\documentclass[english]{article}
\usepackage[T1]{fontenc}
\usepackage[latin9]{inputenc}
\usepackage[active]{srcltx}
\usepackage{color}
\usepackage{float}
\usepackage{calc}
\usepackage{amsmath}
\usepackage{amssymb}
\usepackage{stmaryrd}

\makeatletter

\providecommand{\tabularnewline}{\\}

\pdfoutput=1
\usepackage[T1]{fontenc}
\usepackage[latin9]{inputenc}
\usepackage{color}
\usepackage{float}
\usepackage{graphicx}
\usepackage{graphics}
\usepackage{esint}
\usepackage{enumerate}

\makeatletter

\providecommand{\tabularnewline}{\\}

\usepackage{amsfonts,amssymb}
\usepackage{amsmath}
\usepackage{latexsym}
\usepackage{epsfig}
\usepackage{cite}
\usepackage{graphicx}
\usepackage[colorlinks,linkcolor=blue]{hyperref}
\newcommand{\dt}{t}
\newcommand{\csf}{\Psi}
\newcommand{\G}{G}
\newcommand{\HH}{H}
\newcommand{\be}{\begin{equation}}
\newcommand{\ee}{\end{equation}}
\newcommand{\refb}[1]{(\ref{#1})}
\newcommand{\Ci}{C}

\newcommand{\sectiono}[1]{\section{#1}\setcounter{equation}{0}}
\renewcommand{\theequation}{\thesection.\arabic{equation}}
\newcommand{\ra}{\rangle}
\newcommand{\la}{\langle}
\newcommand{\p}{\partial}
\newcommand{\hp}{{\Phi}}
\newcommand{\hq}{{Q_B}}
\newcommand{\he}{{\eta_0}}
\newcommand{\ha}{{{A}}}
\newcommand{\rrr}{\big\rangle\big\rangle}
\newcommand{\lllb}{\Bigl\langle\Bigl\langle}
\newcommand{\rrrb}{\Bigr\rangle\Bigr\rangle}

\allowdisplaybreaks 

\usepackage{babel}
\usepackage{tcolorbox}

\usepackage{tikz}

\makeatother

\usepackage{babel}
\begin{document}
{}~ \hfill\vbox{\hbox{CTP-SCU/2024011}}\break
\vskip 3.0cm
\centerline{\Large \bf String Scattering and Evolution of Ryu-Takayanagi Surface} 
\vspace*{1.0ex}

\vspace*{10.0ex}
\centerline{\large Xin Jiang, Houwen Wu and Haitang Yang}
\vspace*{7.0ex}
\vspace*{4.0ex}

\centerline{\large \it College of Physics}
\centerline{\large \it Sichuan University}
\centerline{\large \it Chengdu, 610065, China} \vspace*{1.0ex}

\vspace*{4.0ex}

\centerline{domoki@stu.scu.edu.cn, iverwu@scu.edu.cn, hyanga@scu.edu.cn}

\vspace*{10.0ex}
\centerline{\bf Abstract} \bigskip \smallskip
In this paper, our aim is to illustrate that the process of open string scattering corresponds to the evolution of the entanglement wedge, where the  scattering distance is identified as the entanglement wedge cross section. Moreover, open-closed string scattering, specifically the disk-disk interaction, works for the evolution of the reflected entanglement wedge, with the circumference of the waist cross section equating to the reflected entropy. It therefore provides evidence for the deep connections between the string worldsheet and the Ryu-Takayanagi surface.  This connection is not only a coincidence rooted in hyperbolic geometry; it also reflects an additional correspondence between two distinct theories: mutual information and the geometric BV master equation.

\vfill 
\eject
\baselineskip=16pt
\vspace*{10.0ex}
\tableofcontents

\section{Introduction}

The AdS/CFT correspondence plays a central role in modern theoretical
physics. It conjectures that a weakly coupled gravitational theory
in the bulk of AdS$_{d+1}$ is equivalent to a strongly coupled $d$
dimensional conformal field theory (CFT) on the conformally flat boundary
\cite{Maldacena:1997re}. It provides a testable realization for the
holographic principle \cite{tHooft:1993dmi,Susskind:1994vu}. One
of the most successful supports for the AdS/CFT correspondence is
the Ryu and Takayanagi (RT) formula, which asserts the equality of
the entanglement entropy (EE) of the CFT$_{d}$ and the accordingly
defined minimal surface area in the bulk AdS$_{d+1}$ \cite{Ryu:2006bv,Ryu:2006ef,Hubeny:2007xt}.
In their work, the geodesic length (minimal surface area) in AdS$_{3}$
is calculated and found in agreement with the EE of CFT$_{2}$. This
identification is then verified extensively by follow-ups, referring
to a recent review \cite{Rangamani:2016dms} and references therein.
To quantify multipartite entanglement, the EE for a pure state becomes
inadequate, requiring consideration of the entanglement of purification.
Consequently, the bulk dual, represented by the RT surface, needs
to be extended to include the entanglement wedges, entanglement wedge
cross sections (EWCS) and reflected surfaces.

In a recent work \cite{Wang:2021aog}, we studied the connections
between the hyperbolic string vertices of closed string field theory
(CSFT) and the reflected surfaces. This connection arises from the
fact that they are both subsets of hyperbolic geometry. And the critical
length of hyperbolic geometry plays a central role in this connection.
In CSFT, the main aim is to seek all consistent string vertices that
integrate worldsheet correlators of string fields over specific regions
of moduli spaces \cite{Zwiebach:1992ie}. The construction of these
vertices depends on how Riemann surfaces are coordinatized and how
the corresponding parts of moduli spaces are covered. The consistency
requires that the string vertices satisfy the geometric Batalin-Vilkoviski
(BV) master equation. A recent development in constructing string
vertices involves the introduction of hyperbolic geometry, providing
a natural approach to understanding moduli space integration \cite{Moosavian:2017qsp,Moosavian:2017sev,Costello:2019fuh}.
This result is soon generalized to hyperbolic open-closed string vertices
\cite{Cho:2019anu}. Further developments considering hyperbolic string
vertices can be found in the following Refs: \cite{Firat:2021ukc,Ishibashi:2022qcz,Erbin:2022rgx,Firat:2023glo,Firat:2023suh,Erbin:2023hcs,Firat:2023gfn}.
To be specific, the hyperbolic three-string vertices in CSFT are constructed
in three steps \cite{Costello:2019fuh}:
\begin{enumerate}
\item Prepare two right-angled hexagons with side geodesic lengths $L/2$,
$\vartheta$, $L/2$, $\vartheta$, $L/2$, $\vartheta$. 
\item Glue two hexagons together along the geodesics $\vartheta$, giving
the Y-piece $\tilde{\mathcal{V}}_{0,3}\left(L\right)$, which includes
three geodesic boundaries with lengths $\left(L/2\right)\times2=L$. 
\item Graft three flat semi-infinite cylinders with a circumference of $L$
to the geodesic boundaries of $\tilde{\mathcal{V}}_{0,3}\left(L\right)$
to obtain the three-string vertices $\mathcal{V}_{0,3}\left(L\right)$. 
\end{enumerate}
To satisfy the BV master equation, the boundary lengths of three-string
vertices $\mathcal{V}_{0,3}\left(L\right)$ are required to be $L\leq L_{*}=\log\left(3+2\sqrt{2}\right)$.
Furthermore, by gluing two Y-pieces $\tilde{\mathcal{V}}_{0,3}\left(L\right)$
along the geodesic boundaries of length $L$, a new closed geodesic
with length $\delta$ is formed. These two simple geodesics intersect
twice on the X-piece. When $L\leq L_{*}$, the critical length $\delta$
always has a lower bound $\delta>2L_{*}$ which completely agrees
with the lower bound of reflected entropy: $S_{R}>2L_{*}$ \cite{Dutta:2019gen}.
Based on this point, we can build a one-to-one correspondence between
$\delta$ and $S_{R}$. This consideration also works for open string
field theory (OSFT). In OSFT, the critical length can be easily obtained
for a single sheet of $\tilde{\mathcal{V}}_{0,3}\left(L\right)$. 

However, these correspondences only connect the critical length and
the value of reflected entropy. One might wonder whether we can further
understand the relations between the string worldsheet and the evolved
RT surface through these correspondences. Since string vertices are
constructed within the unit Poincar$\acute{\mathrm{e}}$ disk, we
set the anti--de Sitter (AdS) radius $\ell=1$ for simplicity. In
order to establish a connection between the properties of entanglement
entropy and string vertices, we also make use of the \textquotedbl area
law,\textquotedbl{} which is inspired by holographic entanglement
entropy. The entanglement entropy $S_{EE}$ can be expressed as

\begin{equation}
S_{EE}=\frac{L_{o/c}}{4G_{N}^{\left(3\right)}},\label{eq:HOLO}
\end{equation}

\noindent where $L_{o/c}$ represents the geodesic lengths obtained
from the open or closed string vertices, respectively, and $G_{N}^{\left(3\right)}$
is the three-dimensional Newton constant. This relationship allows
us to map geometric quantities from string theory to physical observables
in the dual quantum theory, providing a direct connection between
string theory and entanglement entropy via the holographic principle.

In this paper, our aim is to establish connections between the processes
of string scattering and the evolution of RT surfaces. Specifically,
we focus on specific RT surfaces in multipartite systems: the entanglement
wedge which is surrounded by the RT surfaces, the EWCS and reflected
surface. The EWCS, which serves as the bulk dual of entanglement of
purification, measures entanglement entropy for mixed states \cite{Takayanagi:2017knl}.
When two entangled regions, $A$ and $B$, are far apart, the EWCS
vanishes, separating into two distinct systems. This well-known phase
transition occurs at minimal EWCS $E_{W}^{min}\left(\rho_{AB}\right)=\frac{c}{6}\log\left(3+2\sqrt{2}\right)=\frac{c}{6}L_{*}$
\cite{Takayanagi:2017knl}. \emph{In other words, the phase transition
process is analogous to a scattering process where RT surfaces $\gamma_{C}$
and $\gamma_{D}$ interact to form $\gamma_{A}$ and $\gamma_{D}$}.
In OSFT, we have a similar process, where two open strings interact,
transforming into another pair of open strings \cite{Witten:1985cc}. 

\begin{figure}
\begin{centering}

\tikzset{every picture/.style={line width=0.4pt}} 

\begin{tikzpicture}[x=0.4pt,y=0.4pt,yscale=-1,xscale=1]

\draw   (782.95,193.67) .. controls (782.95,144.59) and (822.74,104.79) .. (871.82,104.79) .. controls (920.91,104.79) and (960.7,144.59) .. (960.7,193.67) .. controls (960.7,242.76) and (920.91,282.55) .. (871.82,282.55) .. controls (822.74,282.55) and (782.95,242.76) .. (782.95,193.67) -- cycle ;
\draw  [draw opacity=0][line width=1.5]  (805.86,135.07) .. controls (820.12,146.51) and (829.86,169.32) .. (829.89,195.59) .. controls (829.92,220.47) and (821.24,242.25) .. (808.25,254.24) -- (785.04,195.65) -- cycle ; \draw  [color={rgb, 255:red, 80; green, 227; blue, 194 }  ,draw opacity=1 ][line width=1.5]  (805.86,135.07) .. controls (820.12,146.51) and (829.86,169.32) .. (829.89,195.59) .. controls (829.92,220.47) and (821.24,242.25) .. (808.25,254.24) ;  
\draw  [draw opacity=0][line width=1.5]  (937.16,134.11) .. controls (924.72,148.7) and (900.44,158.74) .. (872.44,159) .. controls (844.3,159.26) and (819.73,149.57) .. (807.09,135.09) -- (872.01,112.3) -- cycle ; \draw  [color={rgb, 255:red, 80; green, 227; blue, 194 }  ,draw opacity=1 ][line width=1.5]  (937.16,134.11) .. controls (924.72,148.7) and (900.44,158.74) .. (872.44,159) .. controls (844.3,159.26) and (819.73,149.57) .. (807.09,135.09) ;  
\draw  [draw opacity=0][line width=1.5]  (936.01,253.49) .. controls (924.43,240.65) and (916.73,219.44) .. (916.44,195.37) .. controls (916.13,169.87) and (924.22,147.38) .. (936.65,134.56) -- (961.29,194.83) -- cycle ; \draw  [color={rgb, 255:red, 80; green, 227; blue, 194 }  ,draw opacity=1 ][line width=1.5]  (936.01,253.49) .. controls (924.43,240.65) and (916.73,219.44) .. (916.44,195.37) .. controls (916.13,169.87) and (924.22,147.38) .. (936.65,134.56) ;  
\draw  [draw opacity=0][line width=1.5]  (807.86,255.7) .. controls (820.56,241.35) and (845.03,231.77) .. (873.02,232.04) .. controls (899.94,232.3) and (923.4,241.59) .. (936.23,255.27) -- (872.58,278.74) -- cycle ; \draw  [color={rgb, 255:red, 80; green, 227; blue, 194 }  ,draw opacity=1 ][line width=1.5]  (807.86,255.7) .. controls (820.56,241.35) and (845.03,231.77) .. (873.02,232.04) .. controls (899.94,232.3) and (923.4,241.59) .. (936.23,255.27) ;  
\draw  [draw opacity=0][line width=1.5]  (807.32,255.18) .. controls (792.01,239.61) and (782.53,218.11) .. (782.53,194.36) .. controls (782.53,171.38) and (791.4,150.5) .. (805.86,135.07) -- (867.28,194.36) -- cycle ; \draw  [color={rgb, 255:red, 245; green, 166; blue, 35 }  ,draw opacity=1 ][line width=1.5]  (807.32,255.18) .. controls (792.01,239.61) and (782.53,218.11) .. (782.53,194.36) .. controls (782.53,171.38) and (791.4,150.5) .. (805.86,135.07) ;  
\draw  [draw opacity=0][line width=1.5]  (937.16,134.11) .. controls (952.27,149.88) and (961.45,171.5) .. (961.14,195.25) .. controls (960.83,218.23) and (951.69,238.99) .. (937.03,254.23) -- (876.4,194.13) -- cycle ; \draw  [color={rgb, 255:red, 189; green, 16; blue, 224 }  ,draw opacity=1 ][line width=1.5]  (937.16,134.11) .. controls (952.27,149.88) and (961.45,171.5) .. (961.14,195.25) .. controls (960.83,218.23) and (951.69,238.99) .. (937.03,254.23) ;  
\draw   (28.69,193.98) .. controls (28.69,144.9) and (68.48,105.1) .. (117.56,105.1) .. controls (166.65,105.1) and (206.44,144.9) .. (206.44,193.98) .. controls (206.44,243.07) and (166.65,282.86) .. (117.56,282.86) .. controls (68.48,282.86) and (28.69,243.07) .. (28.69,193.98) -- cycle ;
\draw  [draw opacity=0][line width=1.5]  (182.89,134.42) .. controls (170.46,149.01) and (146.18,159.05) .. (118.18,159.31) .. controls (90.04,159.57) and (65.47,149.88) .. (52.82,135.4) -- (117.75,112.61) -- cycle ; \draw  [color={rgb, 255:red, 80; green, 227; blue, 194 }  ,draw opacity=1 ][line width=1.5]  (182.89,134.42) .. controls (170.46,149.01) and (146.18,159.05) .. (118.18,159.31) .. controls (90.04,159.57) and (65.47,149.88) .. (52.82,135.4) ;  
\draw  [draw opacity=0][line width=1.5]  (53.59,256.01) .. controls (66.3,241.66) and (90.76,232.08) .. (118.76,232.35) .. controls (145.68,232.61) and (169.14,241.9) .. (181.96,255.58) -- (118.32,279.05) -- cycle ; \draw  [color={rgb, 255:red, 80; green, 227; blue, 194 }  ,draw opacity=1 ][line width=1.5]  (53.59,256.01) .. controls (66.3,241.66) and (90.76,232.08) .. (118.76,232.35) .. controls (145.68,232.61) and (169.14,241.9) .. (181.96,255.58) ;  
\draw  [draw opacity=0][line width=1.5]  (53.06,255.49) .. controls (37.74,239.92) and (28.27,218.42) .. (28.27,194.67) .. controls (28.27,171.69) and (37.14,150.81) .. (51.6,135.38) -- (113.02,194.67) -- cycle ; \draw  [color={rgb, 255:red, 245; green, 166; blue, 35 }  ,draw opacity=1 ][line width=1.5]  (53.06,255.49) .. controls (37.74,239.92) and (28.27,218.42) .. (28.27,194.67) .. controls (28.27,171.69) and (37.14,150.81) .. (51.6,135.38) ;  
\draw  [draw opacity=0][line width=1.5]  (182.89,134.42) .. controls (198.01,150.19) and (207.19,171.81) .. (206.88,195.56) .. controls (206.57,218.54) and (197.43,239.3) .. (182.77,254.54) -- (122.14,194.44) -- cycle ; \draw  [color={rgb, 255:red, 189; green, 16; blue, 224 }  ,draw opacity=1 ][line width=1.5]  (182.89,134.42) .. controls (198.01,150.19) and (207.19,171.81) .. (206.88,195.56) .. controls (206.57,218.54) and (197.43,239.3) .. (182.77,254.54) ;  
\draw   (392.02,192.98) .. controls (392.02,143.9) and (431.81,104.1) .. (480.9,104.1) .. controls (529.98,104.1) and (569.77,143.9) .. (569.77,192.98) .. controls (569.77,242.07) and (529.98,281.86) .. (480.9,281.86) .. controls (431.81,281.86) and (392.02,242.07) .. (392.02,192.98) -- cycle ;
\draw  [draw opacity=0][line width=1.5]  (414.93,134.38) .. controls (429.19,145.82) and (438.93,168.63) .. (438.96,194.9) .. controls (438.99,219.78) and (430.32,241.56) .. (417.32,253.55) -- (394.11,194.96) -- cycle ; \draw  [color={rgb, 255:red, 80; green, 227; blue, 194 }  ,draw opacity=1 ][line width=1.5]  (414.93,134.38) .. controls (429.19,145.82) and (438.93,168.63) .. (438.96,194.9) .. controls (438.99,219.78) and (430.32,241.56) .. (417.32,253.55) ;  
\draw  [draw opacity=0][line width=1.5]  (545.09,252.8) .. controls (533.51,239.96) and (525.8,218.75) .. (525.51,194.68) .. controls (525.2,169.18) and (533.29,146.69) .. (545.73,133.87) -- (570.36,194.14) -- cycle ; \draw  [color={rgb, 255:red, 80; green, 227; blue, 194 }  ,draw opacity=1 ][line width=1.5]  (545.09,252.8) .. controls (533.51,239.96) and (525.8,218.75) .. (525.51,194.68) .. controls (525.2,169.18) and (533.29,146.69) .. (545.73,133.87) ;  
\draw  [draw opacity=0][line width=1.5]  (416.4,254.49) .. controls (401.08,238.92) and (391.61,217.42) .. (391.61,193.67) .. controls (391.61,170.69) and (400.48,149.81) .. (414.93,134.38) -- (476.35,193.67) -- cycle ; \draw  [color={rgb, 255:red, 245; green, 166; blue, 35 }  ,draw opacity=1 ][line width=1.5]  (416.4,254.49) .. controls (401.08,238.92) and (391.61,217.42) .. (391.61,193.67) .. controls (391.61,170.69) and (400.48,149.81) .. (414.93,134.38) ;  
\draw  [draw opacity=0][line width=1.5]  (546.23,133.42) .. controls (561.34,149.19) and (570.53,170.81) .. (570.21,194.56) .. controls (569.91,217.54) and (560.76,238.3) .. (546.1,253.54) -- (485.47,193.44) -- cycle ; \draw  [color={rgb, 255:red, 189; green, 16; blue, 224 }  ,draw opacity=1 ][line width=1.5]  (546.23,133.42) .. controls (561.34,149.19) and (570.53,170.81) .. (570.21,194.56) .. controls (569.91,217.54) and (560.76,238.3) .. (546.1,253.54) ;  
\draw   (285,176.45) -- (309,176.45) -- (309,170.6) -- (325,182.3) -- (309,194) -- (309,188.15) -- (285,188.15) -- cycle ;
\draw [color={rgb, 255:red, 208; green, 2; blue, 27 }  ,draw opacity=1 ][line width=2.25]  [dash pattern={on 6.75pt off 4.5pt}]  (118.88,160.45) -- (118.88,231.6) ;
\draw    (871.82,165.6) -- (871.82,190.67) ;
\draw [shift={(871.82,193.67)}, rotate = 270] [fill={rgb, 255:red, 0; green, 0; blue, 0 }  ][line width=0.08]  [draw opacity=0] (8.93,-4.29) -- (0,0) -- (8.93,4.29) -- cycle    ;
\draw    (872.23,225.6) -- (872.23,200.27) ;
\draw [shift={(872.23,197.27)}, rotate = 90] [fill={rgb, 255:red, 0; green, 0; blue, 0 }  ][line width=0.08]  [draw opacity=0] (8.93,-4.29) -- (0,0) -- (8.93,4.29) -- cycle    ;
\draw    (908.2,196.35) -- (881.95,196.35) ;
\draw [shift={(911.2,196.35)}, rotate = 180] [fill={rgb, 255:red, 0; green, 0; blue, 0 }  ][line width=0.08]  [draw opacity=0] (8.93,-4.29) -- (0,0) -- (8.93,4.29) -- cycle    ;
\draw    (835.2,196.27) -- (862.23,196.27) ;
\draw [shift={(832.2,196.27)}, rotate = 0] [fill={rgb, 255:red, 0; green, 0; blue, 0 }  ][line width=0.08]  [draw opacity=0] (8.93,-4.29) -- (0,0) -- (8.93,4.29) -- cycle    ;
\draw    (863,150) .. controls (902.8,120.15) and (817.06,103.76) .. (873.34,63.21) ;
\draw [shift={(874.2,62.6)}, rotate = 144.74] [color={rgb, 255:red, 0; green, 0; blue, 0 }  ][line width=0.75]    (10.93,-3.29) .. controls (6.95,-1.4) and (3.31,-0.3) .. (0,0) .. controls (3.31,0.3) and (6.95,1.4) .. (10.93,3.29)   ;
\draw    (871,240) .. controls (856.27,254.53) and (834.42,306.08) .. (903.16,302.66) ;
\draw [shift={(904.2,302.6)}, rotate = 176.73] [color={rgb, 255:red, 0; green, 0; blue, 0 }  ][line width=0.75]    (10.93,-3.29) .. controls (6.95,-1.4) and (3.31,-0.3) .. (0,0) .. controls (3.31,0.3) and (6.95,1.4) .. (10.93,3.29)   ;
\draw    (933,189) .. controls (972.6,159.3) and (968.29,221.34) .. (1011.86,185.71) ;
\draw [shift={(1013.2,184.6)}, rotate = 139.82] [color={rgb, 255:red, 0; green, 0; blue, 0 }  ][line width=0.75]    (10.93,-3.29) .. controls (6.95,-1.4) and (3.31,-0.3) .. (0,0) .. controls (3.31,0.3) and (6.95,1.4) .. (10.93,3.29)   ;
\draw    (806,189) .. controls (783.31,166.71) and (778.25,220.06) .. (753.57,157.55) ;
\draw [shift={(753.2,156.6)}, rotate = 68.66] [color={rgb, 255:red, 0; green, 0; blue, 0 }  ][line width=0.75]    (10.93,-3.29) .. controls (6.95,-1.4) and (3.31,-0.3) .. (0,0) .. controls (3.31,0.3) and (6.95,1.4) .. (10.93,3.29)   ;
\draw  [fill={rgb, 255:red, 74; green, 144; blue, 226 }  ,fill opacity=1 ] (642.2,187.8) -- (658.15,176) -- (658.15,181.9) -- (690.05,181.9) -- (690.05,176) -- (706,187.8) -- (690.05,199.6) -- (690.05,193.7) -- (658.15,193.7) -- (658.15,199.6) -- cycle ;
\draw    (113.16,147.08) .. controls (132.07,107.28) and (175.72,156.58) .. (196.84,93.05) ;
\draw [shift={(197.16,92.08)}, rotate = 107.9] [color={rgb, 255:red, 0; green, 0; blue, 0 }  ][line width=0.75]    (10.93,-3.29) .. controls (6.95,-1.4) and (3.31,-0.3) .. (0,0) .. controls (3.31,0.3) and (6.95,1.4) .. (10.93,3.29)   ;
\draw    (152.16,225.08) .. controls (175.8,200.46) and (218.84,231.13) .. (227.78,271.24) ;
\draw [shift={(228.16,273.08)}, rotate = 258.96] [color={rgb, 255:red, 0; green, 0; blue, 0 }  ][line width=0.75]    (10.93,-3.29) .. controls (6.95,-1.4) and (3.31,-0.3) .. (0,0) .. controls (3.31,0.3) and (6.95,1.4) .. (10.93,3.29)   ;
\draw    (438.6,146.08) .. controls (500.73,119.35) and (528.8,108.3) .. (533.23,62.48) ;
\draw [shift={(533.36,61.08)}, rotate = 94.86] [color={rgb, 255:red, 0; green, 0; blue, 0 }  ][line width=0.75]    (10.93,-3.29) .. controls (6.95,-1.4) and (3.31,-0.3) .. (0,0) .. controls (3.31,0.3) and (6.95,1.4) .. (10.93,3.29)   ;
\draw    (537.8,215.08) .. controls (570.14,221.94) and (578.47,255.69) .. (566.55,285.27) ;
\draw [shift={(565.8,287.08)}, rotate = 293.43] [color={rgb, 255:red, 0; green, 0; blue, 0 }  ][line width=0.75]    (10.93,-3.29) .. controls (6.95,-1.4) and (3.31,-0.3) .. (0,0) .. controls (3.31,0.3) and (6.95,1.4) .. (10.93,3.29)   ;

\draw (3.57,177.25) node [anchor=north west][inner sep=0.75pt]    {$A$};
\draw (217.99,177.25) node [anchor=north west][inner sep=0.75pt]    {$B$};
\draw (366.9,176.25) node [anchor=north west][inner sep=0.75pt]    {$A$};
\draw (581.33,176.25) node [anchor=north west][inner sep=0.75pt]    {$B$};
\draw (265,109) node [anchor=north west][inner sep=0.75pt]   [align=left] {\begin{minipage}[lt]{44.1pt}\setlength\topsep{0pt}
\begin{center}
phase\\transition
\end{center}

\end{minipage}};
\draw (67,185.4) node [anchor=north west][inner sep=0.75pt]  [color={rgb, 255:red, 208; green, 2; blue, 27 }  ,opacity=1 ]  {$\textcolor[rgb]{0.82,0.01,0.11}{E}\textcolor[rgb]{0.82,0.01,0.11}{_{W}}$};
\draw (156,344) node [anchor=north west][inner sep=0.75pt]  [color={rgb, 255:red, 0; green, 0; blue, 0 }  ,opacity=1 ] [align=left] {\begin{minipage}[lt]{164.19pt}\setlength\topsep{0pt}
\textbf{evolution of entanglement wedge}
\begin{center}
\textbf{(RT surface scattering)}
\end{center}

\end{minipage}};
\draw (756,342.52) node [anchor=north west][inner sep=0.75pt]  [color={rgb, 255:red, 0; green, 0; blue, 0 }  ,opacity=1 ] [align=left] {\textbf{open string scattering}};
\draw (897,35) node [anchor=north west][inner sep=0.75pt]   [align=left] {\begin{minipage}[lt]{46.36pt}\setlength\topsep{0pt}
\begin{center}
incoming \\string 1
\end{center}

\end{minipage}};
\draw (942,275) node [anchor=north west][inner sep=0.75pt]   [align=left] {\begin{minipage}[lt]{43.53pt}\setlength\topsep{0pt}
\begin{center}
incoming\\string 2
\end{center}

\end{minipage}};
\draw (994,132) node [anchor=north west][inner sep=0.75pt]   [align=left] {\begin{minipage}[lt]{41.85pt}\setlength\topsep{0pt}
\begin{center}
outgoing\\string 4
\end{center}

\end{minipage}};
\draw (686,100) node [anchor=north west][inner sep=0.75pt]   [align=left] {\begin{minipage}[lt]{44.68pt}\setlength\topsep{0pt}
\begin{center}
outgoing \\string 3
\end{center}

\end{minipage}};
\draw (111,73.25) node [anchor=north west][inner sep=0.75pt]    {$C$};
\draw (113,295.25) node [anchor=north west][inner sep=0.75pt]    {$D$};
\draw (474,294.25) node [anchor=north west][inner sep=0.75pt]    {$D$};
\draw (473,75.25) node [anchor=north west][inner sep=0.75pt]    {$C$};
\draw (163,64) node [anchor=north west][inner sep=0.75pt]   [align=left] {RT surface of C};
\draw (173,282) node [anchor=north west][inner sep=0.75pt]   [align=left] {RT surface of D};
\draw (491,36) node [anchor=north west][inner sep=0.75pt]   [align=left] {RT surface of A};
\draw (512,294) node [anchor=north west][inner sep=0.75pt]   [align=left] {RT surface of B};

\end{tikzpicture}
\par\end{centering}
\caption{\label{fig:rough relation}This picture roughly illustrates the correspondence
between the RT surface scattering and open string scattering.}
\end{figure}
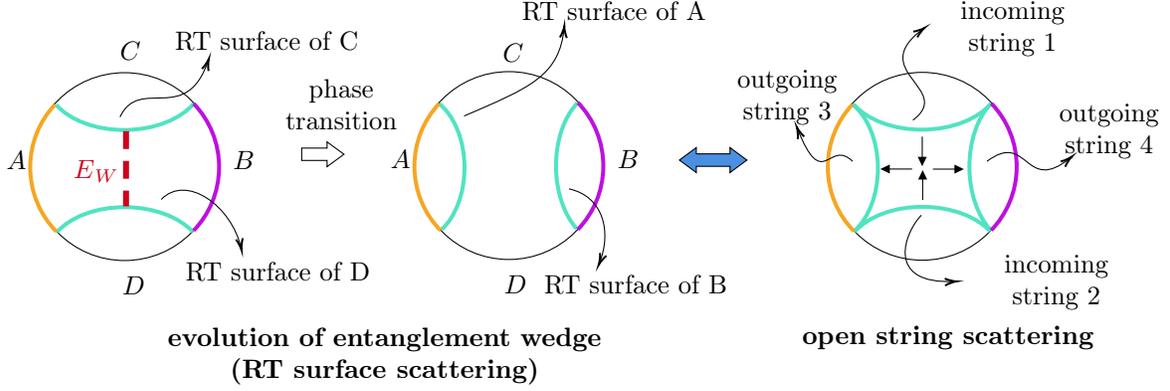

The scattering distance, obtained from hyperbolic string vertices,
is also $L_{*}$. Thus, open string corresponds to the RT surface,
open string scattering corresponds to the scattering of RT surfaces,
and finally, the open string scattering distance is related to the
EWCS by using (\ref{eq:HOLO}), as roughly illustrated in figure (\ref{fig:rough relation}).
It is important to note that this correspondence is not only a coincidence
rooted in hyperbolic geometry. It involves an additional connection
between two distinct theories beyond hyperbolic geometry. In CSFT,
hyperbolic string vertices are governed by two factors: hyperbolic
geometry and the geometric BV master equation. Similarly, the phase
transition in entanglement entropy depends on two components: hyperbolic
geometry and mutual information. The link between hyperbolic string
vertices and the phase transition in entanglement entropy arises from
the relationship between the geometric BV master equation and mutual
information, which are shown as follows: 

\vspace*{2.0ex}
\noindent \begin{center}
\begin{tabular}{ccc}
\cline{1-1} \cline{3-3} 
\multicolumn{1}{|c|}{Phase transition of multipartite entanglement} & \multicolumn{1}{c|}{=} & \multicolumn{1}{c|}{Hyperbolic geometry + Mutual information}\tabularnewline
\cline{1-1} \cline{3-3} 
 &  & \tabularnewline
$\Updownarrow$ &  & $\Updownarrow$\tabularnewline
 &  & \tabularnewline
\cline{1-1} \cline{3-3} 
\multicolumn{1}{|c|}{Hyperbolic string vertices} & \multicolumn{1}{c|}{=} & \multicolumn{1}{c|}{Hyperbolic geometry + Geometric BV equation}\tabularnewline
\cline{1-1} \cline{3-3} 
\end{tabular}
\par\end{center}

\vspace*{2.0ex}

\noindent Both equations determine an equal boundary length for the
X-piece, thereby sharing the same critical length, $L_{*}$.We will
elaborate on this in the following sections.

On the other hand, canonical purification, another purification method,
doubles and glues the original EWCS to yield the reflected entropy
$S_{R}$, also serving as a measure of multipartite entanglement \cite{Dutta:2019gen}.
As expected, $S_{R}\left(A,B\right)=2E_{W}\left(\rho_{AB}\right)$
for bipartite systems, with a transition point $S_{R}^{min}\left(A,B\right)=\frac{c}{3}L_{*}$.
The phase transition process agrees perfectly with disk-disk interactions
of open strings, resulting in a closed string cylinder. However, studying
its relationship with open-closed string field theory poses a challenge.
This is because the recent constructions for hyperbolic open-closed
string vertices rely on flat disks and cylinders (a disk with a bulk
puncture is equivalent to a flat circle with a marking, while an annulus
without a puncture is represented as empty set \cite{Cho:2019anu}),
whereas EWCS exhibits hyperbolic characteristics. To bridge this gap,
we utilize open-closed string duality, revealing that a slice of disk-disk
interaction corresponds to hyperbolic open string scattering. Therefore,
its waist of hyperbolic cylinder can be calculated, measuring $2L_{*}$,
corresponds to the reflected entropy. In short, these observations
reveal deep connections between string scattering and the scattering
of RT surfaces.

This paper is structured as follows: In Section 2, we provide a brief
overview of the entanglement wedge cross section, reflected entropy,
and phase transitions. Section 3 offers a review of the fundamentals
of hyperbolic open and closed string vertices. Section 4 establishes
connections between string vertices and the evolution of the entanglement
wedge and its reflected extension. Section 5 discusses the application
of our correspondence to the black hole information paradox. The final
section includes conclusions and discussions.

\section{EWCS, reflected entropy and their phase transitions}

In this section, we will review the EWCS, reflected entropy and their
phase transitions. It is well known that the traditional entanglement
entropy (EE) serves as a measure of correlation between subsystems.
In the simplest quantum system, pure state, the divided two subsystems
can be denoted as $A$ and $B$. The total Hilbert space is decomposed
into $\mathcal{H}=\mathcal{H}_{A}\varotimes\mathcal{H}_{B}$. By tracing
out the degrees of freedom associated with the region $B$, a reduced
density matrix for the region $A$ can be derived, and denoted as
$\rho_{A}=\mathrm{Tr}_{\mathcal{H}_{B}}\rho$. The entanglement entropy
of region $A$ is then defined using the von Neumann entropy $S_{EE}\left(A\right)=-\mathrm{Tr}_{\mathcal{H}_{A}}\left(\rho_{A}\log\rho_{A}\right)$,
and we also have $S_{EE}\left(A\right)=S_{EE}\left(B\right)$. However,
this method fails to measure the entanglement for mixed states, as
the entanglement entropy of such states is always nonzero whether
they are entangled or not. To solve this problem, one approach is
to purify the mixed states and then introduce the entanglement of
purification $E_{P}$, whose bulk dual is conjectured to be the area
of the entanglement wedge cross section $E_{W}$, specifically $E_{P}=E_{W}/4G_{N}$.
For convenience, hereinafter, we set $4G_{N}=1$. The computation
of $E_{P}$ is typically challenging as it requires minimizing over
all possible purifications. To pass through this difficulty, an alternative
method known as canonical purification has been proposed in \cite{Dutta:2019gen},
defining the reflected entropy $S_{R}$. This method is to copy the
original mixed state as the purification. For a bipartite system,
this identical copy is denoted as $A^{*}B^{*}$. The purified state
and the reflected entropy are then defined as $\left|\sqrt{\rho_{AB}}\right\rangle _{AA^{*}BB^{*}}=\underset{i}{\sum}\sqrt{p_{i}}\left|i\right\rangle _{AB}\otimes\left|i\right\rangle _{A^{*}B^{*}}$
and $S_{R}\left(A:B\right)=S\left(AA^{*}\right)_{\sqrt{\rho_{AB}}}=-\mathrm{Tr}\rho_{AA^{*}}\log\rho_{AA^{*}}$
where $\rho_{AA}=\mathrm{Tr}_{\mathcal{H}_{BB^{*}}}\left|\rho_{AB}\right\rangle \left\langle \rho_{AB}\right|$.
From the definition, it is easy to see $S_{R}\left(A:B\right)=2E_{W}\left(A:B\right)$.
The bulk interpretation of canonical purification is proposed in \cite{Engelhardt:2018kcs},
which is referred to as the reflected surface, also denoted as $S_{R}$
to avoid confusion. Recently, we proposed a much simpler alternative
for purification to measure the entanglements of mixed states \cite{Jiang:2024akz}.

In the rest of this section, we plan to introduce the bulk duals of
the entanglement of purification and the canonical purification, which
are denoted as EWCS and reflected surface separately. Moreover, we
will also present the ingredient feature for these two purifications:
connected and disconnected phase transitions. 

\vspace*{2.0ex}

\subsection{Entanglement wedge cross section (EWCS) \cite{Takayanagi:2017knl}}

\begin{figure}[h]
\begin{centering}

\tikzset{every picture/.style={line width=0.55pt}} 

\begin{tikzpicture}[x=0.55pt,y=0.55pt,yscale=-1,xscale=1]

\draw    (149,202) -- (149,62.45) ;
\draw [shift={(149,60.45)}, rotate = 90] [color={rgb, 255:red, 0; green, 0; blue, 0 }  ][line width=0.75]    (10.93,-3.29) .. controls (6.95,-1.4) and (3.31,-0.3) .. (0,0) .. controls (3.31,0.3) and (6.95,1.4) .. (10.93,3.29)   ;
\draw    (25.21,202.57) -- (278.09,202.57) ;
\draw  [draw opacity=0][line width=1.5]  (93.49,202.46) .. controls (93.49,202.29) and (93.49,202.12) .. (93.49,201.95) .. controls (93.52,171.29) and (118.39,146.46) .. (149.05,146.49) .. controls (179.51,146.52) and (204.21,171.07) .. (204.51,201.45) -- (149,202) -- cycle ; \draw  [color={rgb, 255:red, 80; green, 227; blue, 194 }  ,draw opacity=1 ][line width=1.5]  (93.49,202.46) .. controls (93.49,202.29) and (93.49,202.12) .. (93.49,201.95) .. controls (93.52,171.29) and (118.39,146.46) .. (149.05,146.49) .. controls (179.51,146.52) and (204.21,171.07) .. (204.51,201.45) ;  
\draw  [draw opacity=0][line width=1.5]  (39.88,202.46) .. controls (39.87,202.27) and (39.87,202.09) .. (39.87,201.9) .. controls (39.93,141.64) and (88.83,92.82) .. (149.1,92.87) .. controls (209.36,92.93) and (258.18,141.83) .. (258.13,202.1) .. controls (258.13,202.25) and (258.12,202.41) .. (258.12,202.57) -- (149,202) -- cycle ; \draw  [color={rgb, 255:red, 80; green, 227; blue, 194 }  ,draw opacity=1 ][line width=1.5]  (39.88,202.46) .. controls (39.87,202.27) and (39.87,202.09) .. (39.87,201.9) .. controls (39.93,141.64) and (88.83,92.82) .. (149.1,92.87) .. controls (209.36,92.93) and (258.18,141.83) .. (258.13,202.1) .. controls (258.13,202.25) and (258.12,202.41) .. (258.12,202.57) ;  
\draw [color={rgb, 255:red, 245; green, 166; blue, 35 }  ,draw opacity=1 ][line width=1.5]    (39.88,202.46) -- (93.49,202.46) ;
\draw [color={rgb, 255:red, 144; green, 19; blue, 254 }  ,draw opacity=1 ][line width=1.5]    (204.51,202.57) -- (258.12,202.57) ;
\draw    (199,178) .. controls (238.6,148.3) and (245.95,215.1) .. (297.43,179.12) ;
\draw [shift={(299,178)}, rotate = 143.99] [color={rgb, 255:red, 0; green, 0; blue, 0 }  ][line width=0.75]    (10.93,-3.29) .. controls (6.95,-1.4) and (3.31,-0.3) .. (0,0) .. controls (3.31,0.3) and (6.95,1.4) .. (10.93,3.29)   ;
\draw    (235,131) .. controls (274.4,101.45) and (296.42,108.24) .. (303.77,148.59) ;
\draw [shift={(304.09,150.45)}, rotate = 260.54] [color={rgb, 255:red, 0; green, 0; blue, 0 }  ][line width=0.75]    (10.93,-3.29) .. controls (6.95,-1.4) and (3.31,-0.3) .. (0,0) .. controls (3.31,0.3) and (6.95,1.4) .. (10.93,3.29)   ;
\draw [color={rgb, 255:red, 208; green, 2; blue, 27 }  ,draw opacity=1 ][line width=2.25]  [dash pattern={on 6.75pt off 4.5pt}]  (148.88,92.45) -- (148.88,146.45) ;
\draw   (413,141.7) -- (440.05,141.7) -- (440.05,135.45) -- (458.09,147.95) -- (440.05,160.45) -- (440.05,154.2) -- (413,154.2) -- cycle ;

\draw    (662,202) -- (662,62.45) ;
\draw [shift={(662,60.45)}, rotate = 90] [color={rgb, 255:red, 0; green, 0; blue, 0 }  ][line width=0.75]    (10.93,-3.29) .. controls (6.95,-1.4) and (3.31,-0.3) .. (0,0) .. controls (3.31,0.3) and (6.95,1.4) .. (10.93,3.29)   ;
\draw    (538.21,202.57) -- (791.09,202.57) ;
\draw  [draw opacity=0][line width=1.5]  (552.67,202.68) .. controls (552.67,202.6) and (552.67,202.51) .. (552.67,202.43) .. controls (552.69,171.8) and (564.81,146.98) .. (579.73,146.99) .. controls (594.63,147.01) and (606.7,171.78) .. (606.7,202.36) -- (579.68,202.46) -- cycle ; \draw  [color={rgb, 255:red, 80; green, 227; blue, 194 }  ,draw opacity=1 ][line width=1.5]  (552.67,202.68) .. controls (552.67,202.6) and (552.67,202.51) .. (552.67,202.43) .. controls (552.69,171.8) and (564.81,146.98) .. (579.73,146.99) .. controls (594.63,147.01) and (606.7,171.78) .. (606.7,202.36) ;  
\draw [color={rgb, 255:red, 245; green, 166; blue, 35 }  ,draw opacity=1 ][line width=1.5]    (553.09,202.36) -- (606.7,202.36) ;
\draw [color={rgb, 255:red, 144; green, 19; blue, 254 }  ,draw opacity=1 ][line width=1.5]    (717.51,202.57) -- (771.12,202.57) ;
\draw  [draw opacity=0][line width=1.5]  (717.09,202.89) .. controls (717.09,202.81) and (717.09,202.72) .. (717.09,202.64) .. controls (717.11,172.01) and (729.23,147.19) .. (744.15,147.2) .. controls (759.06,147.22) and (771.12,172) .. (771.12,202.57) -- (744.11,202.67) -- cycle ; \draw  [color={rgb, 255:red, 80; green, 227; blue, 194 }  ,draw opacity=1 ][line width=1.5]  (717.09,202.89) .. controls (717.09,202.81) and (717.09,202.72) .. (717.09,202.64) .. controls (717.11,172.01) and (729.23,147.19) .. (744.15,147.2) .. controls (759.06,147.22) and (771.12,172) .. (771.12,202.57) ;  
\draw    (158.23,102.18) -- (158.23,115.27) ;
\draw [shift={(158.23,118.27)}, rotate = 270] [fill={rgb, 255:red, 0; green, 0; blue, 0 }  ][line width=0.08]  [draw opacity=0] (8.93,-4.29) -- (0,0) -- (8.93,4.29) -- cycle    ;
\draw    (158.23,134.27) -- (158.23,121.27) ;
\draw [shift={(158.23,118.27)}, rotate = 90] [fill={rgb, 255:red, 0; green, 0; blue, 0 }  ][line width=0.08]  [draw opacity=0] (8.93,-4.29) -- (0,0) -- (8.93,4.29) -- cycle    ;

\draw (60,210) node [anchor=north west][inner sep=0.75pt]   [align=left] {A};
\draw (227,208) node [anchor=north west][inner sep=0.75pt]   [align=left] {B};
\draw (298.09,153.85) node [anchor=north west][inner sep=0.75pt]  [color={rgb, 255:red, 80; green, 227; blue, 194 }  ,opacity=1 ]  {$\Gamma _{AB}^{min}$};
\draw (112,112.4) node [anchor=north west][inner sep=0.75pt]  [color={rgb, 255:red, 208; green, 2; blue, 27 }  ,opacity=1 ]  {$\textcolor[rgb]{0.82,0.01,0.11}{E}\textcolor[rgb]{0.82,0.01,0.11}{_{W}}$};
\draw (383,113) node [anchor=north west][inner sep=0.75pt]   [align=left] {Phase transition};
\draw (573,210) node [anchor=north west][inner sep=0.75pt]   [align=left] {A};
\draw (740,208) node [anchor=north west][inner sep=0.75pt]   [align=left] {B};
\draw (698,113) node [anchor=north west][inner sep=0.75pt]   [align=left] {\textcolor[rgb]{0.31,0.89,0.76}{RT surface}};
\draw (535,112) node [anchor=north west][inner sep=0.75pt]   [align=left] {\textcolor[rgb]{0.31,0.89,0.76}{RT surface}};

\end{tikzpicture}
\par\end{centering}
\centering{}\caption{\label{fig:2 intervals}The entanglement wedge cross section $E_{W}\left(A,B\right)$
(red dashed line) is assumed to be a bulk dual of the entanglement
of purification $E_{P}\left(A,B\right)$ for bipartite mixed states.
The essential ingredient is that there is a lower bound $E_{W}^{min}\left(A,B\right)$
for the EWCS, due to the mutual information vanishing when $A$ and
$B$ are distant. This lower bound indicates the phase transition
point between the connected and disconnected entanglement wedges,
transitioning from the left-hand side picture to the right-hand side
picture.}
\end{figure}
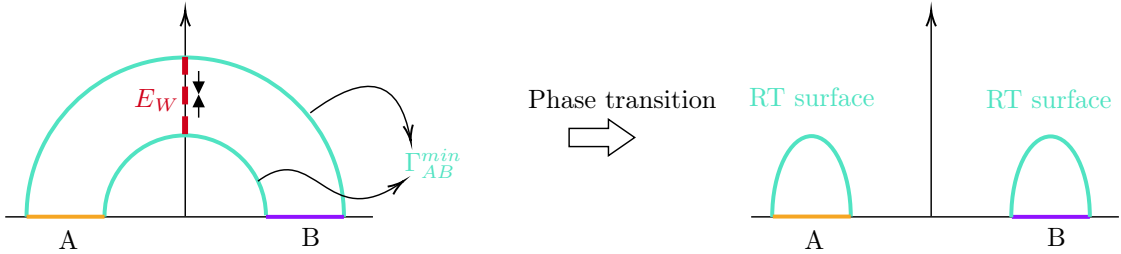

\noindent Now, let us first review the EWCS based on \cite{Jokela:2019ebz}.
Consider two subsystems, $A$ and $B$, on the boundary without any
overlap, see figure (\ref{fig:2 intervals}). The entanglement wedge
$M_{AB}$ is defined by a region whose boundary is given by:

\begin{equation}
\partial M_{AB}=A\cup B\cup\Gamma_{AB}^{min},
\end{equation}

\noindent where $\Gamma_{AB}^{min}$ denotes the area of the minimal
surface corresponding to the union $A\cup B$. Now, let us split the
boundary of the entanglement wedge into two parts:

\begin{equation}
\partial M_{AB}=\tilde{\Gamma}_{A}\cup\tilde{\Gamma}_{B},
\end{equation}

\noindent where $A\subset\tilde{\Gamma}_{A}$ and $B\subset\tilde{\Gamma}_{B}$.
The holographic entanglement entropy is obtained by searching for
the minimal surface $\Sigma_{AB}^{min}$ which satisfies:
\begin{enumerate}
\item $\partial\Sigma_{AB}^{min}=\partial\tilde{\Gamma}_{A}=\partial\tilde{\Gamma}_{B}$,
\item $\Sigma_{AB}^{min}$ is homologous to $\tilde{\Gamma}_{A}$ or $\tilde{\Gamma}_{B}$.
\end{enumerate}
Then the EWCS is defined by the minimized area $\Sigma_{AB}^{min}$
of all possible splits of the entanglement wedge:

\begin{equation}
E_{W}\left(A:B\right)=\underset{\tilde{\Gamma}_{A}\subset\partial M_{AB}}{min}\left(\frac{A\left(\Sigma_{AB}^{min}\right)}{4G_{N}}\right).
\end{equation}

\noindent When two entangled regions, $A$ and $B$, are far apart,
the entanglement wedge cross section vanishes, giving rise to two
separate systems. This phenomenon represents a phase transition between
the disconnected and connected phases. To see it clearly, let us consider
it in a time slice of AdS$_{3}$ and set the AdS radius to be unity.
If we choose $A=\left[a_{1},a_{2}\right]$ and $B=\left[b_{1},b_{2}\right]$
where $a_{1}<a_{2}<b_{1}<b_{2}$. The EWCS is given by:

\begin{equation}
E_{W}\left(A:B\right)=\frac{c}{6}\log\left(1+2z+2\sqrt{z\left(z+1\right)}\right),
\end{equation}

\noindent and the corresponding mutual information is given by

\begin{equation}
I\left(A:B\right)=\frac{c}{3}\log z,
\end{equation}

\noindent where $z$ denotes the cross ratio:

\begin{equation}
z=\frac{\left(a_{2}-a_{1}\right)\left(b_{2}-b_{1}\right)}{\left(b_{1}-a_{2}\right)\left(b_{2}-a_{1}\right)}\geq0.
\end{equation}

\noindent when $z\leq1$, $E_{W}\left(A:B\right)=I\left(A:B\right)=0$.
Therefore, it gives a transition point at $z=1$:

\begin{equation}
E_{W}^{min}\left(A:B\right)=\frac{c}{6}\log\left(3+2\sqrt{2}\right).
\end{equation}

\vspace*{2.0ex}

\subsection{Reflected surface \cite{Dutta:2019gen}}

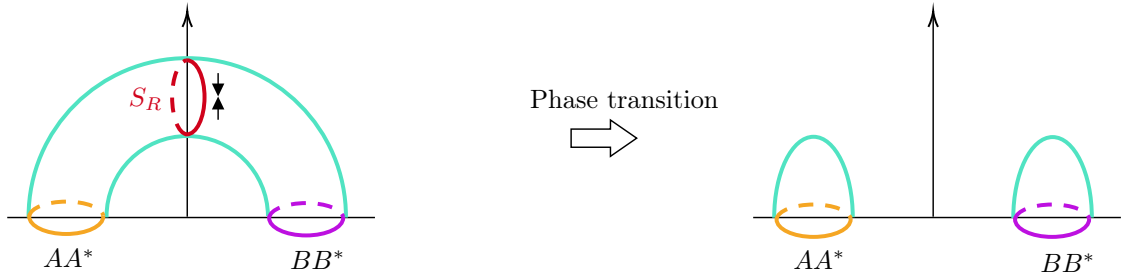
\begin{figure}[h]
\begin{centering}

\tikzset{every picture/.style={line width=0.55pt}} 

\begin{tikzpicture}[x=0.55pt,y=0.55pt,yscale=-1,xscale=1]

\draw    (149,202) -- (149,62.45) ;
\draw [shift={(149,60.45)}, rotate = 90] [color={rgb, 255:red, 0; green, 0; blue, 0 }  ][line width=0.75]    (10.93,-3.29) .. controls (6.95,-1.4) and (3.31,-0.3) .. (0,0) .. controls (3.31,0.3) and (6.95,1.4) .. (10.93,3.29)   ;
\draw    (25.21,202.57) -- (278.09,202.57) ;
\draw  [draw opacity=0][line width=1.5]  (93.49,202.46) .. controls (93.49,202.29) and (93.49,202.12) .. (93.49,201.95) .. controls (93.52,171.29) and (118.39,146.46) .. (149.05,146.49) .. controls (179.51,146.52) and (204.21,171.07) .. (204.51,201.45) -- (149,202) -- cycle ; \draw  [color={rgb, 255:red, 80; green, 227; blue, 194 }  ,draw opacity=1 ][line width=1.5]  (93.49,202.46) .. controls (93.49,202.29) and (93.49,202.12) .. (93.49,201.95) .. controls (93.52,171.29) and (118.39,146.46) .. (149.05,146.49) .. controls (179.51,146.52) and (204.21,171.07) .. (204.51,201.45) ;  
\draw  [draw opacity=0][line width=1.5]  (39.88,202.46) .. controls (39.87,202.27) and (39.87,202.09) .. (39.87,201.9) .. controls (39.93,141.64) and (88.83,92.82) .. (149.1,92.87) .. controls (209.36,92.93) and (258.18,141.83) .. (258.13,202.1) .. controls (258.13,202.25) and (258.12,202.41) .. (258.12,202.57) -- (149,202) -- cycle ; \draw  [color={rgb, 255:red, 80; green, 227; blue, 194 }  ,draw opacity=1 ][line width=1.5]  (39.88,202.46) .. controls (39.87,202.27) and (39.87,202.09) .. (39.87,201.9) .. controls (39.93,141.64) and (88.83,92.82) .. (149.1,92.87) .. controls (209.36,92.93) and (258.18,141.83) .. (258.13,202.1) .. controls (258.13,202.25) and (258.12,202.41) .. (258.12,202.57) ;  
\draw   (413,141.7) -- (440.05,141.7) -- (440.05,135.45) -- (458.09,147.95) -- (440.05,160.45) -- (440.05,154.2) -- (413,154.2) -- cycle ;

\draw    (662,202) -- (662,62.45) ;
\draw [shift={(662,60.45)}, rotate = 90] [color={rgb, 255:red, 0; green, 0; blue, 0 }  ][line width=0.75]    (10.93,-3.29) .. controls (6.95,-1.4) and (3.31,-0.3) .. (0,0) .. controls (3.31,0.3) and (6.95,1.4) .. (10.93,3.29)   ;
\draw    (538.21,202.57) -- (791.09,202.57) ;
\draw  [draw opacity=0][line width=1.5]  (552.67,202.68) .. controls (552.67,202.6) and (552.67,202.51) .. (552.67,202.43) .. controls (552.69,171.8) and (564.81,146.98) .. (579.73,146.99) .. controls (594.63,147.01) and (606.7,171.78) .. (606.7,202.36) -- (579.68,202.46) -- cycle ; \draw  [color={rgb, 255:red, 80; green, 227; blue, 194 }  ,draw opacity=1 ][line width=1.5]  (552.67,202.68) .. controls (552.67,202.6) and (552.67,202.51) .. (552.67,202.43) .. controls (552.69,171.8) and (564.81,146.98) .. (579.73,146.99) .. controls (594.63,147.01) and (606.7,171.78) .. (606.7,202.36) ;  
\draw  [draw opacity=0][line width=1.5]  (717.09,202.89) .. controls (717.09,202.81) and (717.09,202.72) .. (717.09,202.64) .. controls (717.11,172.01) and (729.23,147.19) .. (744.15,147.2) .. controls (759.06,147.22) and (771.12,172) .. (771.12,202.57) -- (744.11,202.67) -- cycle ; \draw  [color={rgb, 255:red, 80; green, 227; blue, 194 }  ,draw opacity=1 ][line width=1.5]  (717.09,202.89) .. controls (717.09,202.81) and (717.09,202.72) .. (717.09,202.64) .. controls (717.11,172.01) and (729.23,147.19) .. (744.15,147.2) .. controls (759.06,147.22) and (771.12,172) .. (771.12,202.57) ;  
\draw    (170.23,103.18) -- (170.23,116.27) ;
\draw [shift={(170.23,119.27)}, rotate = 270] [fill={rgb, 255:red, 0; green, 0; blue, 0 }  ][line width=0.08]  [draw opacity=0] (8.93,-4.29) -- (0,0) -- (8.93,4.29) -- cycle    ;
\draw    (170.23,135.27) -- (170.23,122.27) ;
\draw [shift={(170.23,119.27)}, rotate = 90] [fill={rgb, 255:red, 0; green, 0; blue, 0 }  ][line width=0.08]  [draw opacity=0] (8.93,-4.29) -- (0,0) -- (8.93,4.29) -- cycle    ;

\draw  [draw opacity=0][dash pattern={on 5.63pt off 4.5pt}][line width=1.5]  (40.4,201.53) .. controls (41.1,196.02) and (52.21,191.63) .. (65.82,191.62) .. controls (79.17,191.61) and (90.13,195.81) .. (91.21,201.17) -- (65.83,202.06) -- cycle ; \draw  [color={rgb, 255:red, 245; green, 166; blue, 35 }  ,draw opacity=1 ][dash pattern={on 5.63pt off 4.5pt}][line width=1.5]  (40.4,201.53) .. controls (41.1,196.02) and (52.21,191.63) .. (65.82,191.62) .. controls (79.17,191.61) and (90.13,195.81) .. (91.21,201.17) ;  
\draw  [draw opacity=0][line width=1.5]  (91.21,201.17) .. controls (91.23,201.37) and (91.24,201.58) .. (91.24,201.79) .. controls (91.25,208.47) and (79.85,213.89) .. (65.79,213.9) .. controls (51.72,213.92) and (40.32,208.52) .. (40.31,201.84) .. controls (40.31,201.51) and (40.34,201.17) .. (40.39,200.84) -- (65.77,201.82) -- cycle ; \draw  [color={rgb, 255:red, 245; green, 166; blue, 35 }  ,draw opacity=1 ][line width=1.5]  (91.21,201.17) .. controls (91.23,201.37) and (91.24,201.58) .. (91.24,201.79) .. controls (91.25,208.47) and (79.85,213.89) .. (65.79,213.9) .. controls (51.72,213.92) and (40.32,208.52) .. (40.31,201.84) .. controls (40.31,201.51) and (40.34,201.17) .. (40.39,200.84) ;  

\draw  [draw opacity=0][dash pattern={on 5.63pt off 4.5pt}][line width=1.5]  (205.4,202.53) .. controls (206.1,197.02) and (217.21,192.63) .. (230.82,192.62) .. controls (244.17,192.61) and (255.13,196.81) .. (256.21,202.17) -- (230.83,203.06) -- cycle ; \draw  [color={rgb, 255:red, 189; green, 16; blue, 224 }  ,draw opacity=1 ][dash pattern={on 5.63pt off 4.5pt}][line width=1.5]  (205.4,202.53) .. controls (206.1,197.02) and (217.21,192.63) .. (230.82,192.62) .. controls (244.17,192.61) and (255.13,196.81) .. (256.21,202.17) ;  
\draw  [draw opacity=0][line width=1.5]  (256.21,202.17) .. controls (256.23,202.37) and (256.24,202.58) .. (256.24,202.79) .. controls (256.25,209.47) and (244.85,214.89) .. (230.79,214.9) .. controls (216.72,214.92) and (205.32,209.52) .. (205.31,202.84) .. controls (205.31,202.51) and (205.34,202.17) .. (205.39,201.84) -- (230.77,202.82) -- cycle ; \draw  [color={rgb, 255:red, 189; green, 16; blue, 224 }  ,draw opacity=1 ][line width=1.5]  (256.21,202.17) .. controls (256.23,202.37) and (256.24,202.58) .. (256.24,202.79) .. controls (256.25,209.47) and (244.85,214.89) .. (230.79,214.9) .. controls (216.72,214.92) and (205.32,209.52) .. (205.31,202.84) .. controls (205.31,202.51) and (205.34,202.17) .. (205.39,201.84) ;  

\draw  [draw opacity=0][dash pattern={on 5.63pt off 4.5pt}][line width=1.5]  (148.4,145.22) .. controls (142.9,144.49) and (138.56,133.36) .. (138.6,119.76) .. controls (138.65,106.4) and (142.9,95.46) .. (148.26,94.41) -- (149.04,119.79) -- cycle ; \draw  [color={rgb, 255:red, 208; green, 2; blue, 27 }  ,draw opacity=1 ][dash pattern={on 5.63pt off 4.5pt}][line width=1.5]  (148.4,145.22) .. controls (142.9,144.49) and (138.56,133.36) .. (138.6,119.76) .. controls (138.65,106.4) and (142.9,95.46) .. (148.26,94.41) ;  
\draw  [draw opacity=0][line width=1.5]  (148.26,94.41) .. controls (148.47,94.39) and (148.68,94.38) .. (148.88,94.38) .. controls (155.56,94.4) and (160.93,105.82) .. (160.89,119.88) .. controls (160.84,133.95) and (155.39,145.33) .. (148.72,145.31) .. controls (148.38,145.31) and (148.04,145.28) .. (147.71,145.22) -- (148.8,119.84) -- cycle ; \draw  [color={rgb, 255:red, 208; green, 2; blue, 27 }  ,draw opacity=1 ][line width=1.5]  (148.26,94.41) .. controls (148.47,94.39) and (148.68,94.38) .. (148.88,94.38) .. controls (155.56,94.4) and (160.93,105.82) .. (160.89,119.88) .. controls (160.84,133.95) and (155.39,145.33) .. (148.72,145.31) .. controls (148.38,145.31) and (148.04,145.28) .. (147.71,145.22) ;  

\draw  [draw opacity=0][dash pattern={on 5.63pt off 4.5pt}][line width=1.5]  (554.4,203.53) .. controls (555.1,198.02) and (566.21,193.63) .. (579.82,193.62) .. controls (593.17,193.61) and (604.13,197.81) .. (605.21,203.17) -- (579.83,204.06) -- cycle ; \draw  [color={rgb, 255:red, 245; green, 166; blue, 35 }  ,draw opacity=1 ][dash pattern={on 5.63pt off 4.5pt}][line width=1.5]  (554.4,203.53) .. controls (555.1,198.02) and (566.21,193.63) .. (579.82,193.62) .. controls (593.17,193.61) and (604.13,197.81) .. (605.21,203.17) ;  
\draw  [draw opacity=0][line width=1.5]  (605.21,203.17) .. controls (605.23,203.37) and (605.24,203.58) .. (605.24,203.79) .. controls (605.25,210.47) and (593.85,215.89) .. (579.79,215.9) .. controls (565.72,215.92) and (554.32,210.52) .. (554.31,203.84) .. controls (554.31,203.51) and (554.34,203.17) .. (554.39,202.84) -- (579.77,203.82) -- cycle ; \draw  [color={rgb, 255:red, 245; green, 166; blue, 35 }  ,draw opacity=1 ][line width=1.5]  (605.21,203.17) .. controls (605.23,203.37) and (605.24,203.58) .. (605.24,203.79) .. controls (605.25,210.47) and (593.85,215.89) .. (579.79,215.9) .. controls (565.72,215.92) and (554.32,210.52) .. (554.31,203.84) .. controls (554.31,203.51) and (554.34,203.17) .. (554.39,202.84) ;  

\draw  [draw opacity=0][dash pattern={on 5.63pt off 4.5pt}][line width=1.5]  (718.4,203.53) .. controls (719.1,198.02) and (730.21,193.63) .. (743.82,193.62) .. controls (757.17,193.61) and (768.13,197.81) .. (769.21,203.17) -- (743.83,204.06) -- cycle ; \draw  [color={rgb, 255:red, 189; green, 16; blue, 224 }  ,draw opacity=1 ][dash pattern={on 5.63pt off 4.5pt}][line width=1.5]  (718.4,203.53) .. controls (719.1,198.02) and (730.21,193.63) .. (743.82,193.62) .. controls (757.17,193.61) and (768.13,197.81) .. (769.21,203.17) ;  
\draw  [draw opacity=0][line width=1.5]  (769.21,203.17) .. controls (769.23,203.37) and (769.24,203.58) .. (769.24,203.79) .. controls (769.25,210.47) and (757.85,215.89) .. (743.79,215.9) .. controls (729.72,215.92) and (718.32,210.52) .. (718.31,203.84) .. controls (718.31,203.51) and (718.34,203.17) .. (718.39,202.84) -- (743.77,203.82) -- cycle ; \draw  [color={rgb, 255:red, 189; green, 16; blue, 224 }  ,draw opacity=1 ][line width=1.5]  (769.21,203.17) .. controls (769.23,203.37) and (769.24,203.58) .. (769.24,203.79) .. controls (769.25,210.47) and (757.85,215.89) .. (743.79,215.9) .. controls (729.72,215.92) and (718.32,210.52) .. (718.31,203.84) .. controls (718.31,203.51) and (718.34,203.17) .. (718.39,202.84) ;

\draw (107,111.4) node [anchor=north west][inner sep=0.75pt]  [color={rgb, 255:red, 208; green, 2; blue, 27 }  ,opacity=1 ]  {$\textcolor[rgb]{0.82,0.01,0.11}{S_{R}}$};
\draw (383,113) node [anchor=north west][inner sep=0.75pt]   [align=left] {Phase transition};
\draw (48,222.4) node [anchor=north west][inner sep=0.75pt]    {$AA^{*}$};
\draw (218,223.4) node [anchor=north west][inner sep=0.75pt]    {$BB^{*}$};
\draw (564,223.4) node [anchor=north west][inner sep=0.75pt]    {$AA^{*}$};
\draw (734,225.4) node [anchor=north west][inner sep=0.75pt]    {$BB^{*}$};

\end{tikzpicture}
\par\end{centering}
\centering{}\caption{\label{fig:2 reflected surface}The phase transition of the reflected
entanglement wedge is illustrated in the figure. The red circle, referred
to as the reflected surface, with a length $S_{R}$, represents the
bulk dual of reflected entropy. It comes from the copying and gluing
of entanglement wedges. This surface is obtained through the copying
and gluing of entanglement wedges. The lower bound of the EWCS also
gives a lower bound for the reflected surface.}
\end{figure}

\noindent The bulk dual of the reflected entropy is called reflected
surface, which is also known as $S_{R}$. In the bulk interpretation,
the reflected surface can be obtained by gluing two entanglement wedges
shown in figure (\ref{fig:2 intervals}) along the cyan geodesics,
the result is shown in figure (\ref{fig:2 reflected surface}). As
definition, the reflected surface is a closed simple geodesic, and
it gives:

\begin{equation}
S_{R}\left(A:B\right)=2E_{W}\left(A:B\right).
\end{equation}

\noindent Due to the lower bound of $E_{W}$, there also exists a
lower bound for $S_{R}$:

\begin{equation}
S_{R}^{min}\left(A:B\right)=2E_{W}^{min}\left(A:B\right)=\frac{c}{3}\log\left(3+2\sqrt{2}\right).
\end{equation}

\noindent In the rest of this paper, we will refer to the surface
of this cylinder as the ``reflected entanglement wedge'' for simplicity.

\section{Hyperbolic open and closed string vertices}

In this section, we aim to introduce the geometric BV master equation
and its solution, known as string vertices, which are essential in
open and closed string field theories. A good short review for the
string field theory can be found in \cite{Maccaferri:2023vns}, on
which our review is partially based. In string field theory, the string
vertices play an important role since they are connected to off-shell
amplitudes. These off-shell amplitudes can be derived from integrals
over the moduli space of Riemann surfaces with genus $g$ and $n$
punctures. The correlation function in this integral is related to
the states inserted at the punctures, which are not always $\left(0,0\right)$
conformal primaries. Local coordinate maps are therefore needed to
map the punctured disk onto the Riemann surface, and the different
choices of the coordinate maps represent different field redefinitions.
The complete amplitudes resulted from summing all Feynman diagrams
can be obtained by gluing these string vertices and propagators. Therefore,
the correct integration over the moduli space, without any overlapping
or missing regions, is crucial for obtaining these amplitudes. And,
this correct integration leads to the string vertices $\mathcal{V}$
satisfying the geometric BV master equation:

\begin{equation}
\partial\mathcal{V}+\hbar\triangle\mathcal{V}+\frac{1}{2}\left\{ \mathcal{V},\mathcal{V}\right\} =0,\label{eq:BV equation}
\end{equation}
where $\mathcal{V}$ is a formal sum of string vertices associated
to various moduli spaces $\mathcal{V}_{g,n}$, $\partial$ denotes
the boundary operator acting on the moduli space, $\triangle$ denotes
to remove disks/semidisks of two marked points on one Riemann surface,
and then twist-sewing the boundaries of these two disks/semidisks.
$\left\{ \;,\;\right\} $ stands for removing two disks/semidisks
on two input Riemann surfaces respectively, and then twist-sewing
them together \cite{Sen:1993kb,Sen:1994kx}. 

In the rest of this section, let us briefly review the recent development
that construct string vertices using hyperbolic geometry.

\vspace*{2.0ex}

At first, we introduce the Ultra-Parallel Theorem:

\vspace*{2.0ex}

\noindent \textbf{Theorem} \textbf{1} (Ultra-parallel theorem \cite{Buser})
(Theorem 1.1.6)): Any pair of ultra-parallel geodesics in the Poincar$\acute{\mathrm{e}}$
disk has a unique perpendicular geodesics.

\vspace*{2.0ex}

\begin{figure}[h]
\begin{centering}

\tikzset{every picture/.style={line width=0.55pt}} 

\begin{tikzpicture}[x=0.55pt,y=0.55pt,yscale=-1,xscale=1]

\draw  [line width=1.5]  (258,132.91) .. controls (258,96.41) and (287.59,66.82) .. (324.09,66.82) .. controls (360.59,66.82) and (390.18,96.41) .. (390.18,132.91) .. controls (390.18,169.41) and (360.59,199) .. (324.09,199) .. controls (287.59,199) and (258,169.41) .. (258,132.91) -- cycle ;
\draw  [draw opacity=0][line width=1.5]  (310.91,197.65) .. controls (307.2,185.24) and (317.2,171.23) .. (333.3,166.33) .. controls (349.07,161.53) and (364.85,167.25) .. (369.08,179.12) -- (340.15,188.84) -- cycle ; \draw  [color={rgb, 255:red, 80; green, 227; blue, 194 }  ,draw opacity=1 ][line width=1.5]  (310.91,197.65) .. controls (307.2,185.24) and (317.2,171.23) .. (333.3,166.33) .. controls (349.07,161.53) and (364.85,167.25) .. (369.08,179.12) ;  
\draw  [draw opacity=0][line width=1.5]  (336.49,68.37) .. controls (340.16,80.53) and (329.69,94.51) .. (313.04,99.63) .. controls (297.21,104.5) and (281.49,99.61) .. (276.73,88.68) -- (306.26,77.58) -- cycle ; \draw  [color={rgb, 255:red, 80; green, 227; blue, 194 }  ,draw opacity=1 ][line width=1.5]  (336.49,68.37) .. controls (340.16,80.53) and (329.69,94.51) .. (313.04,99.63) .. controls (297.21,104.5) and (281.49,99.61) .. (276.73,88.68) ;  
\draw [color={rgb, 255:red, 208; green, 2; blue, 27 }  ,draw opacity=1 ][line width=1.5]    (313.93,100.25) -- (334.25,165.57) ;
\draw  [color={rgb, 255:red, 208; green, 2; blue, 27 }  ,draw opacity=1 ][line width=1.5]  (313.93,100.25) -- (321.33,97.72) -- (324.03,105.06) -- (316.63,107.59) -- cycle ;
\draw  [color={rgb, 255:red, 208; green, 2; blue, 27 }  ,draw opacity=1 ][line width=1.5]  (331.93,158.1) -- (339.45,155.96) -- (341.77,163.43) -- (334.25,165.57) -- cycle ;
\draw    (362,164) .. controls (382.79,116.97) and (397.7,157.67) .. (430.02,110.94) ;
\draw [shift={(431,109.5)}, rotate = 123.96] [color={rgb, 255:red, 0; green, 0; blue, 0 }  ][line width=0.75]    (10.93,-3.29) .. controls (6.95,-1.4) and (3.31,-0.3) .. (0,0) .. controls (3.31,0.3) and (6.95,1.4) .. (10.93,3.29)   ;
\draw    (340,86) .. controls (382.79,110.38) and (357.26,68.92) .. (425.96,93.13) ;
\draw [shift={(427,93.5)}, rotate = 199.65] [color={rgb, 255:red, 0; green, 0; blue, 0 }  ][line width=0.75]    (10.93,-3.29) .. controls (6.95,-1.4) and (3.31,-0.3) .. (0,0) .. controls (3.31,0.3) and (6.95,1.4) .. (10.93,3.29)   ;

\draw (435,89) node [anchor=north west][inner sep=0.75pt]   [align=left] {Any two geodesics};

\end{tikzpicture}
\par\end{centering}
\centering{}\caption{\label{fig:orthogonal} The red line denotes the shortest geodesic
between any two cyan boundary-anchored geodesics. This red geodesic
is perpendicular to the two cyan geodesics.}
\end{figure}
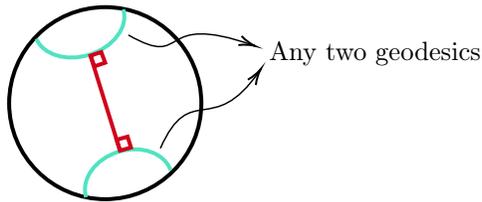

\noindent Following the Ultra-Parallel Theorem, it shows that there
is a unique perpendicular line connecting any pair of boundary-anchored
geodesics, see figure (\ref{fig:orthogonal}). 

\vspace*{2.0ex}

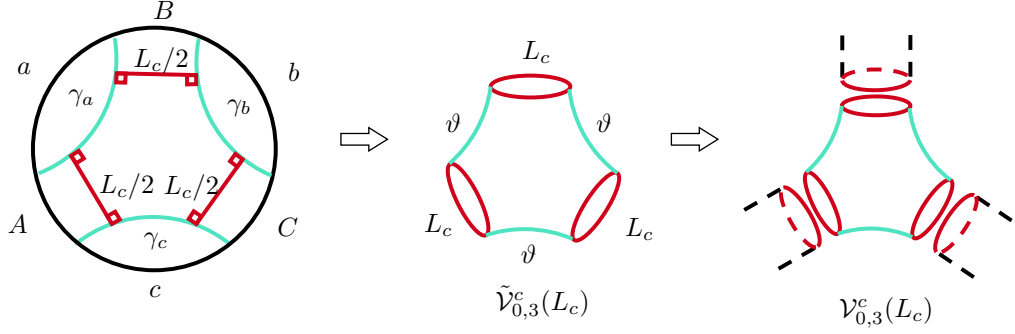
\begin{figure}[h]
\begin{centering}

\tikzset{every picture/.style={line width=0.5pt}} 

\begin{tikzpicture}[x=0.5pt,y=0.5pt,yscale=-1,xscale=1]

\draw  [draw opacity=0][line width=1.5]  (81.04,215.33) .. controls (97.34,202.12) and (118.17,194.32) .. (140.79,194.61) .. controls (162.65,194.88) and (182.65,202.68) .. (198.38,215.49) -- (139.6,287.68) -- cycle ; \draw  [color={rgb, 255:red, 80; green, 227; blue, 194 }  ,draw opacity=1 ][line width=1.5]  (81.04,215.33) .. controls (97.34,202.12) and (118.17,194.32) .. (140.79,194.61) .. controls (162.65,194.88) and (182.65,202.68) .. (198.38,215.49) ;  
\draw  [line width=1.5]  (49.01,143.13) .. controls (49.01,92.41) and (90.13,51.3) .. (140.85,51.3) .. controls (191.57,51.3) and (232.69,92.41) .. (232.69,143.13) .. controls (232.69,193.86) and (191.57,234.97) .. (140.85,234.97) .. controls (90.13,234.97) and (49.01,193.86) .. (49.01,143.13) -- cycle ;

\draw  [draw opacity=0][line width=1.5]  (229.46,163.29) .. controls (210.14,155.11) and (193.46,140.39) .. (183.08,120.3) .. controls (173.04,100.88) and (170.51,79.56) .. (174.43,59.65) -- (265.78,77.58) -- cycle ; \draw  [color={rgb, 255:red, 80; green, 227; blue, 194 }  ,draw opacity=1 ][line width=1.5]  (229.46,163.29) .. controls (210.14,155.11) and (193.46,140.39) .. (183.08,120.3) .. controls (173.04,100.88) and (170.51,79.56) .. (174.43,59.65) ;  
\draw  [draw opacity=0][line width=1.5]  (110.82,59.18) .. controls (114.4,79.86) and (111.06,101.85) .. (99.78,121.46) .. controls (88.88,140.41) and (72.32,154.07) .. (53.47,161.55) -- (19.09,75.04) -- cycle ; \draw  [color={rgb, 255:red, 80; green, 227; blue, 194 }  ,draw opacity=1 ][line width=1.5]  (110.82,59.18) .. controls (114.4,79.86) and (111.06,101.85) .. (99.78,121.46) .. controls (88.88,140.41) and (72.32,154.07) .. (53.47,161.55) ;  

\draw [color={rgb, 255:red, 208; green, 2; blue, 27 }  ,draw opacity=1 ][line width=1.5]    (172.45,87.17) -- (112.29,85.86) ;
\draw  [color={rgb, 255:red, 208; green, 2; blue, 27 }  ,draw opacity=1 ][line width=1.5]  (172.45,87.17) -- (172.5,94.04) -- (165.62,94.25) -- (165.58,87.37) -- cycle ;
\draw  [color={rgb, 255:red, 208; green, 2; blue, 27 }  ,draw opacity=1 ][line width=1.5]  (119.17,86.02) -- (118.86,92.89) -- (111.98,92.73) -- (112.29,85.86) -- cycle ;

\draw [color={rgb, 255:red, 208; green, 2; blue, 27 }  ,draw opacity=1 ][line width=1.5]    (77.88,148.28) -- (108.76,199.93) ;
\draw  [color={rgb, 255:red, 208; green, 2; blue, 27 }  ,draw opacity=1 ][line width=1.5]  (77.88,148.28) -- (83.68,144.59) -- (87.51,150.31) -- (81.71,154) -- cycle ;
\draw  [color={rgb, 255:red, 208; green, 2; blue, 27 }  ,draw opacity=1 ][line width=1.5]  (105.23,194.02) -- (111.22,190.64) -- (114.74,196.54) -- (108.75,199.93) -- cycle ;

\draw [color={rgb, 255:red, 208; green, 2; blue, 27 }  ,draw opacity=1 ][line width=1.5]    (172.61,199.92) -- (207.63,151) ;
\draw  [color={rgb, 255:red, 208; green, 2; blue, 27 }  ,draw opacity=1 ][line width=1.5]  (172.61,199.92) -- (166.9,196.08) -- (170.62,190.29) -- (176.32,194.13) -- cycle ;
\draw  [color={rgb, 255:red, 208; green, 2; blue, 27 }  ,draw opacity=1 ][line width=1.5]  (203.63,156.59) -- (198.13,152.45) -- (202.14,146.86) -- (207.63,151) -- cycle ;

\draw   (282,131.4) -- (303.12,131.4) -- (303.12,127) -- (317.2,135.8) -- (303.12,144.6) -- (303.12,140.2) -- (282,140.2) -- cycle ;
\draw  [color={rgb, 255:red, 208; green, 2; blue, 27 }  ,draw opacity=1 ][line width=1.5]  (395.29,95.86) .. controls (395.29,91.29) and (408.78,87.58) .. (425.41,87.58) .. controls (442.04,87.58) and (455.52,91.29) .. (455.52,95.86) .. controls (455.52,100.44) and (442.04,104.14) .. (425.41,104.14) .. controls (408.78,104.14) and (395.29,100.44) .. (395.29,95.86) -- cycle ;
\draw  [color={rgb, 255:red, 208; green, 2; blue, 27 }  ,draw opacity=1 ][line width=1.5]  (364.68,154.59) .. controls (368.77,152.54) and (378.13,162.93) .. (385.58,177.8) .. controls (393.03,192.67) and (395.76,206.39) .. (391.67,208.44) .. controls (387.58,210.48) and (378.23,200.09) .. (370.78,185.22) .. controls (363.32,170.35) and (360.6,156.64) .. (364.68,154.59) -- cycle ;
\draw  [color={rgb, 255:red, 208; green, 2; blue, 27 }  ,draw opacity=1 ][line width=1.5]  (491.08,161.39) .. controls (494.93,163.85) and (490.78,177.21) .. (481.8,191.21) .. controls (472.82,205.21) and (462.43,214.56) .. (458.58,212.1) .. controls (454.73,209.63) and (458.88,196.28) .. (467.86,182.27) .. controls (476.84,168.27) and (487.23,158.92) .. (491.08,161.39) -- cycle ;
\draw  [draw opacity=0][line width=1.5]  (491.08,161.39) .. controls (480.32,153.26) and (471.16,142.65) .. (464.56,129.86) .. controls (459.24,119.56) and (456.02,108.73) .. (454.77,97.86) -- (547.26,87.14) -- cycle ; \draw  [color={rgb, 255:red, 80; green, 227; blue, 194 }  ,draw opacity=1 ][line width=1.5]  (491.08,161.39) .. controls (480.32,153.26) and (471.16,142.65) .. (464.56,129.86) .. controls (459.24,119.56) and (456.02,108.73) .. (454.77,97.86) ;  
\draw  [draw opacity=0][line width=1.5]  (395.29,95.86) .. controls (393.77,107.66) and (389.94,119.36) .. (383.64,130.32) .. controls (378.32,139.57) and (371.65,147.56) .. (364.03,154.18) -- (302.96,83.9) -- cycle ; \draw  [color={rgb, 255:red, 80; green, 227; blue, 194 }  ,draw opacity=1 ][line width=1.5]  (395.29,95.86) .. controls (393.77,107.66) and (389.94,119.36) .. (383.64,130.32) .. controls (378.32,139.57) and (371.65,147.56) .. (364.03,154.18) ;  
\draw  [draw opacity=0][line width=1.5]  (391.67,208.44) .. controls (401.18,205.32) and (411.35,203.69) .. (421.91,203.82) .. controls (434.47,203.98) and (446.41,206.62) .. (457.28,211.27) -- (420.72,296.9) -- cycle ; \draw  [color={rgb, 255:red, 80; green, 227; blue, 194 }  ,draw opacity=1 ][line width=1.5]  (391.67,208.44) .. controls (401.18,205.32) and (411.35,203.69) .. (421.91,203.82) .. controls (434.47,203.98) and (446.41,206.62) .. (457.28,211.27) ;  

\draw   (532,131.4) -- (553.12,131.4) -- (553.12,127) -- (567.2,135.8) -- (553.12,144.6) -- (553.12,140.2) -- (532,140.2) -- cycle ;
\draw  [color={rgb, 255:red, 208; green, 2; blue, 27 }  ,draw opacity=1 ][line width=1.5]  (660.97,110.47) .. controls (660.97,106.54) and (672.57,103.35) .. (686.88,103.35) .. controls (701.2,103.35) and (712.8,106.54) .. (712.8,110.47) .. controls (712.8,114.41) and (701.2,117.6) .. (686.88,117.6) .. controls (672.57,117.6) and (660.97,114.41) .. (660.97,110.47) -- cycle ;
\draw  [color={rgb, 255:red, 208; green, 2; blue, 27 }  ,draw opacity=1 ][line width=1.5]  (634.63,161.01) .. controls (638.14,159.25) and (646.19,168.19) .. (652.61,180.99) .. controls (659.02,193.79) and (661.37,205.59) .. (657.85,207.35) .. controls (654.33,209.11) and (646.28,200.17) .. (639.87,187.37) .. controls (633.46,174.58) and (631.11,162.78) .. (634.63,161.01) -- cycle ;
\draw  [color={rgb, 255:red, 208; green, 2; blue, 27 }  ,draw opacity=1 ][line width=1.5]  (743.4,166.86) .. controls (746.71,168.99) and (743.14,180.48) .. (735.41,192.53) .. controls (727.69,204.58) and (718.74,212.62) .. (715.43,210.5) .. controls (712.12,208.38) and (715.69,196.89) .. (723.42,184.84) .. controls (731.14,172.79) and (740.09,164.74) .. (743.4,166.86) -- cycle ;
\draw  [draw opacity=0][line width=1.5]  (743.4,166.86) .. controls (734.14,159.87) and (726.26,150.73) .. (720.57,139.73) .. controls (715.99,130.87) and (713.23,121.55) .. (712.15,112.19) -- (791.74,102.97) -- cycle ; \draw  [color={rgb, 255:red, 80; green, 227; blue, 194 }  ,draw opacity=1 ][line width=1.5]  (743.4,166.86) .. controls (734.14,159.87) and (726.26,150.73) .. (720.57,139.73) .. controls (715.99,130.87) and (713.23,121.55) .. (712.15,112.19) ;  
\draw  [draw opacity=0][line width=1.5]  (660.97,110.47) .. controls (659.66,120.63) and (656.36,130.7) .. (650.94,140.13) .. controls (646.36,148.09) and (640.62,154.96) .. (634.06,160.66) -- (581.5,100.18) -- cycle ; \draw  [color={rgb, 255:red, 80; green, 227; blue, 194 }  ,draw opacity=1 ][line width=1.5]  (660.97,110.47) .. controls (659.66,120.63) and (656.36,130.7) .. (650.94,140.13) .. controls (646.36,148.09) and (640.62,154.96) .. (634.06,160.66) ;  
\draw  [draw opacity=0][line width=1.5]  (657.85,207.35) .. controls (666.03,204.67) and (674.78,203.26) .. (683.87,203.38) .. controls (694.68,203.52) and (704.96,205.79) .. (714.31,209.79) -- (682.85,283.48) -- cycle ; \draw  [color={rgb, 255:red, 80; green, 227; blue, 194 }  ,draw opacity=1 ][line width=1.5]  (657.85,207.35) .. controls (666.03,204.67) and (674.78,203.26) .. (683.87,203.38) .. controls (694.68,203.52) and (704.96,205.79) .. (714.31,209.79) ;  
\draw [line width=1.5]  [dash pattern={on 5.63pt off 4.5pt}]  (661.28,57.16) -- (661.28,91.33) ;
\draw [line width=1.5]  [dash pattern={on 5.63pt off 4.5pt}]  (712.8,56.89) -- (712.8,91.06) ;
\draw  [draw opacity=0][line width=1.5]  (712.8,91.06) .. controls (711.65,95.08) and (700.67,98.4) .. (687.23,98.64) .. controls (673.05,98.9) and (661.47,95.63) .. (661.28,91.33) -- (687.09,90.79) -- cycle ; \draw  [color={rgb, 255:red, 208; green, 2; blue, 27 }  ,draw opacity=1 ][line width=1.5]  (712.8,91.06) .. controls (711.65,95.08) and (700.67,98.4) .. (687.23,98.64) .. controls (673.05,98.9) and (661.47,95.63) .. (661.28,91.33) ;  
\draw  [draw opacity=0][dash pattern={on 5.63pt off 4.5pt}][line width=1.5]  (661.39,89.92) .. controls (662.63,85.93) and (673.69,82.87) .. (687.13,82.94) .. controls (701.31,83.02) and (712.81,86.56) .. (712.9,90.86) -- (687.09,90.79) -- cycle ; \draw  [color={rgb, 255:red, 208; green, 2; blue, 27 }  ,draw opacity=1 ][dash pattern={on 5.63pt off 4.5pt}][line width=1.5]  (661.39,89.92) .. controls (662.63,85.93) and (673.69,82.87) .. (687.13,82.94) .. controls (701.31,83.02) and (712.81,86.56) .. (712.9,90.86) ;  

\draw [line width=1.5]  [dash pattern={on 5.63pt off 4.5pt}]  (613.05,233.83) -- (642.67,216.79) ;
\draw [line width=1.5]  [dash pattern={on 5.63pt off 4.5pt}]  (587.12,189.31) -- (616.73,172.26) ;
\draw  [draw opacity=0][line width=1.5]  (616.73,172.26) .. controls (620.78,171.26) and (629.14,179.12) .. (636.06,190.65) .. controls (643.36,202.81) and (646.3,214.47) .. (642.67,216.79) -- (629.33,194.69) -- cycle ; \draw  [color={rgb, 255:red, 208; green, 2; blue, 27 }  ,draw opacity=1 ][line width=1.5]  (616.73,172.26) .. controls (620.78,171.26) and (629.14,179.12) .. (636.06,190.65) .. controls (643.36,202.81) and (646.3,214.47) .. (642.67,216.79) ;  
\draw  [draw opacity=0][dash pattern={on 5.63pt off 4.5pt}][line width=1.5]  (641.39,217.4) .. controls (637.31,218.31) and (629.14,210.25) .. (622.5,198.57) .. controls (615.49,186.23) and (612.83,174.5) .. (616.51,172.27) -- (629.33,194.69) -- cycle ; \draw  [color={rgb, 255:red, 208; green, 2; blue, 27 }  ,draw opacity=1 ][dash pattern={on 5.63pt off 4.5pt}][line width=1.5]  (641.39,217.4) .. controls (637.31,218.31) and (629.14,210.25) .. (622.5,198.57) .. controls (615.49,186.23) and (612.83,174.5) .. (616.51,172.27) ;  

\draw [line width=1.5]  [dash pattern={on 5.63pt off 4.5pt}]  (789.36,197.34) -- (761.03,178.24) ;
\draw [line width=1.5]  [dash pattern={on 5.63pt off 4.5pt}]  (760.79,240.21) -- (732.45,221.11) ;
\draw  [draw opacity=0][line width=1.5]  (732.45,221.11) .. controls (729.77,217.92) and (733.15,206.95) .. (740.46,195.67) .. controls (748.17,183.77) and (757.35,176) .. (761.03,178.24) -- (747.05,199.94) -- cycle ; \draw  [color={rgb, 255:red, 208; green, 2; blue, 27 }  ,draw opacity=1 ][line width=1.5]  (732.45,221.11) .. controls (729.77,217.92) and (733.15,206.95) .. (740.46,195.67) .. controls (748.17,183.77) and (757.35,176) .. (761.03,178.24) ;  
\draw  [draw opacity=0][dash pattern={on 5.63pt off 4.5pt}][line width=1.5]  (762.14,179.12) .. controls (764.75,182.37) and (761.11,193.26) .. (753.54,204.36) .. controls (745.54,216.08) and (736.18,223.64) .. (732.56,221.31) -- (747.05,199.94) -- cycle ; \draw  [color={rgb, 255:red, 208; green, 2; blue, 27 }  ,draw opacity=1 ][dash pattern={on 5.63pt off 4.5pt}][line width=1.5]  (762.14,179.12) .. controls (764.75,182.37) and (761.11,193.26) .. (753.54,204.36) .. controls (745.54,216.08) and (736.18,223.64) .. (732.56,221.31) ;

\draw (28,191.4) node [anchor=north west][inner sep=0.75pt]    {$A$};
\draw (138,29.4) node [anchor=north west][inner sep=0.75pt]    {$B$};
\draw (232,193.4) node [anchor=north west][inner sep=0.75pt]    {$C$};
\draw (35,75.4) node [anchor=north west][inner sep=0.75pt]    {$a$};
\draw (240,75.4) node [anchor=north west][inner sep=0.75pt]    {$b$};
\draw (135,244.4) node [anchor=north west][inner sep=0.75pt]    {$c$};
\draw (123,64.4) node [anchor=north west][inner sep=0.75pt]    {$L_{c} /2$};
\draw (97,159.4) node [anchor=north west][inner sep=0.75pt]    {$L_{c} /2$};
\draw (146,160.4) node [anchor=north west][inner sep=0.75pt]    {$L_{c} /2$};
\draw (417,59.4) node [anchor=north west][inner sep=0.75pt]    {$L_{c}$};
\draw (343,190.4) node [anchor=north west][inner sep=0.75pt]    {$L_{c}$};
\draw (495,193.4) node [anchor=north west][inner sep=0.75pt]    {$L_{c}$};
\draw (74,96.4) node [anchor=north west][inner sep=0.75pt]    {$\gamma _{a}$};
\draw (194,103.4) node [anchor=north west][inner sep=0.75pt]    {$\gamma _{b}$};
\draw (131,204.4) node [anchor=north west][inner sep=0.75pt]    {$\gamma _{c}$};
\draw (359,114.4) node [anchor=north west][inner sep=0.75pt]    {$\vartheta $};
\draw (473,114.4) node [anchor=north west][inner sep=0.75pt]    {$\vartheta $};
\draw (416,213.4) node [anchor=north west][inner sep=0.75pt]    {$\vartheta $};
\draw (397,244.4) node [anchor=north west][inner sep=0.75pt]    {$\tilde{\mathcal{V}}_{0,3}^{c}( L_{c})$};
\draw (657,253.4) node [anchor=north west][inner sep=0.75pt]    {$\mathcal{V}_{0,3}^{c}( L_{c})$};

\end{tikzpicture}
\par\end{centering}
\centering{}\caption{\label{fig:Y-pants} The construction for the closed string vertices.}
\end{figure}

\noindent \textbf{Closed string vertices:} To construct the hyperbolic
closed string vertices, we need to construct a Y-piece. The first
step is to prepare a right-angled hexagon with side lengths $L_{c}/2$,
$\vartheta$, $L_{c}/2$, $\vartheta$, $L_{c}/2$, $\vartheta$,
see figure (\ref{fig:Y-pants}). This hexagon can be formed by plotting
three boundary-anchored cyan geodesics $\gamma_{a}=\gamma_{b}=\gamma_{c}$
in the Poincar$\acute{\mathrm{e}}$ disk, where it is essential that
the boundary regions have equal distances: $a=b=c$ and $A=B=C$.
From the ultra-parallel theorem, we can construct three unique perpendicular
geodesics with lengths $L_{c}/2$, which are defined as the boundaries
of the right-angled hexagon. The hexagon, as required, is enclosed
by the cyan and red geodesics. Subsequently, we can copy this hexagon
and then glue them along the cyan geodesics. The result is a pair-of-pants
of boundary width $L_{c}$, known as a Y-piece or also called $\tilde{\mathcal{V}}_{0,3}^{c}\left(L_{c}\right)$.
By grafting three semi-infinite cylinders of width $L_{c}$, we obtain
a hyperbolic three-string vertex $\mathcal{V}_{0,3}^{c}\left(L_{c}\right)$
in CSFT. The semi-infinite cylinders can be conformally mapped to
the punctured disks, yielding a Riemann surface with markings.

\vspace*{2.0ex}

\begin{figure}[h]
\begin{centering}

\tikzset{every picture/.style={line width=0.5pt}} 

\begin{tikzpicture}[x=0.5pt,y=0.5pt,yscale=-1,xscale=1]

\draw  [draw opacity=0][line width=1.5]  (80.04,215.33) .. controls (96.34,202.12) and (117.17,194.32) .. (139.79,194.61) .. controls (161.65,194.88) and (181.65,202.68) .. (197.38,215.49) -- (138.6,287.68) -- cycle ; \draw  [color={rgb, 255:red, 80; green, 227; blue, 194 }  ,draw opacity=1 ][line width=1.5]  (80.04,215.33) .. controls (96.34,202.12) and (117.17,194.32) .. (139.79,194.61) .. controls (161.65,194.88) and (181.65,202.68) .. (197.38,215.49) ;  
\draw  [line width=1.5]  (48.01,143.13) .. controls (48.01,92.41) and (89.13,51.3) .. (139.85,51.3) .. controls (190.57,51.3) and (231.69,92.41) .. (231.69,143.13) .. controls (231.69,193.86) and (190.57,234.97) .. (139.85,234.97) .. controls (89.13,234.97) and (48.01,193.86) .. (48.01,143.13) -- cycle ;
\draw  [draw opacity=0][line width=1.5]  (229.46,163.29) .. controls (210.14,155.11) and (193.46,140.39) .. (183.08,120.3) .. controls (173.04,100.88) and (170.51,79.56) .. (174.43,59.65) -- (265.78,77.58) -- cycle ; \draw  [color={rgb, 255:red, 80; green, 227; blue, 194 }  ,draw opacity=1 ][line width=1.5]  (229.46,163.29) .. controls (210.14,155.11) and (193.46,140.39) .. (183.08,120.3) .. controls (173.04,100.88) and (170.51,79.56) .. (174.43,59.65) ;  
\draw  [draw opacity=0][line width=1.5]  (110.82,59.18) .. controls (114.4,79.86) and (111.06,101.85) .. (99.78,121.46) .. controls (88.88,140.41) and (72.32,154.07) .. (53.47,161.55) -- (19.09,75.04) -- cycle ; \draw  [color={rgb, 255:red, 80; green, 227; blue, 194 }  ,draw opacity=1 ][line width=1.5]  (110.82,59.18) .. controls (114.4,79.86) and (111.06,101.85) .. (99.78,121.46) .. controls (88.88,140.41) and (72.32,154.07) .. (53.47,161.55) ;  
\draw [color={rgb, 255:red, 208; green, 2; blue, 27 }  ,draw opacity=1 ][line width=1.5]    (172.45,87.17) -- (112.29,85.86) ;
\draw  [color={rgb, 255:red, 208; green, 2; blue, 27 }  ,draw opacity=1 ][line width=1.5]  (172.45,87.17) -- (172.5,94.04) -- (165.62,94.25) -- (165.58,87.37) -- cycle ;
\draw  [color={rgb, 255:red, 208; green, 2; blue, 27 }  ,draw opacity=1 ][line width=1.5]  (119.17,86.02) -- (118.86,92.89) -- (111.98,92.73) -- (112.29,85.86) -- cycle ;

\draw [color={rgb, 255:red, 208; green, 2; blue, 27 }  ,draw opacity=1 ][line width=1.5]    (77.88,148.28) -- (108.76,199.93) ;
\draw  [color={rgb, 255:red, 208; green, 2; blue, 27 }  ,draw opacity=1 ][line width=1.5]  (77.88,148.28) -- (83.68,144.59) -- (87.51,150.31) -- (81.71,154) -- cycle ;
\draw  [color={rgb, 255:red, 208; green, 2; blue, 27 }  ,draw opacity=1 ][line width=1.5]  (105.23,194.02) -- (111.22,190.64) -- (114.74,196.54) -- (108.75,199.93) -- cycle ;

\draw [color={rgb, 255:red, 208; green, 2; blue, 27 }  ,draw opacity=1 ][line width=1.5]    (172.61,199.92) -- (207.63,151) ;
\draw  [color={rgb, 255:red, 208; green, 2; blue, 27 }  ,draw opacity=1 ][line width=1.5]  (172.61,199.92) -- (166.9,196.08) -- (170.62,190.29) -- (176.32,194.13) -- cycle ;
\draw  [color={rgb, 255:red, 208; green, 2; blue, 27 }  ,draw opacity=1 ][line width=1.5]  (203.63,156.59) -- (198.13,152.45) -- (202.14,146.86) -- (207.63,151) -- cycle ;

\draw   (282,131.4) -- (303.12,131.4) -- (303.12,127) -- (317.2,135.8) -- (303.12,144.6) -- (303.12,140.2) -- (282,140.2) -- cycle ;
\draw   (532,131.4) -- (553.12,131.4) -- (553.12,127) -- (567.2,135.8) -- (553.12,144.6) -- (553.12,140.2) -- (532,140.2) -- cycle ;
\draw  [draw opacity=0][line width=1.5]  (395.76,202.93) .. controls (405.48,199.64) and (415.92,197.92) .. (426.76,198.06) .. controls (438.34,198.21) and (449.39,200.46) .. (459.57,204.45) -- (425.58,291.13) -- cycle ; \draw  [color={rgb, 255:red, 80; green, 227; blue, 194 }  ,draw opacity=1 ][line width=1.5]  (395.76,202.93) .. controls (405.48,199.64) and (415.92,197.92) .. (426.76,198.06) .. controls (438.34,198.21) and (449.39,200.46) .. (459.57,204.45) ;  
\draw  [draw opacity=0][line width=1.5]  (493.04,153.15) .. controls (483.54,145.35) and (475.43,135.53) .. (469.43,123.92) .. controls (463.82,113.06) and (460.56,101.62) .. (459.45,90.17) -- (552.13,81.19) -- cycle ; \draw  [color={rgb, 255:red, 80; green, 227; blue, 194 }  ,draw opacity=1 ][line width=1.5]  (493.04,153.15) .. controls (483.54,145.35) and (475.43,135.53) .. (469.43,123.92) .. controls (463.82,113.06) and (460.56,101.62) .. (459.45,90.17) ;  
\draw  [draw opacity=0][line width=1.5]  (399.29,88.86) .. controls (397.92,101.21) and (394.04,113.48) .. (387.45,124.93) .. controls (381.52,135.23) and (373.93,143.97) .. (365.21,151) -- (306.76,78.52) -- cycle ; \draw  [color={rgb, 255:red, 80; green, 227; blue, 194 }  ,draw opacity=1 ][line width=1.5]  (399.29,88.86) .. controls (397.92,101.21) and (394.04,113.48) .. (387.45,124.93) .. controls (381.52,135.23) and (373.93,143.97) .. (365.21,151) ;  
\draw [color={rgb, 255:red, 208; green, 2; blue, 27 }  ,draw opacity=1 ][line width=1.5]    (459.45,90.17) -- (399.29,88.86) ;
\draw  [color={rgb, 255:red, 208; green, 2; blue, 27 }  ,draw opacity=1 ][line width=1.5]  (459.45,90.17) -- (459.5,97.04) -- (452.62,97.25) -- (452.58,90.37) -- cycle ;
\draw  [color={rgb, 255:red, 208; green, 2; blue, 27 }  ,draw opacity=1 ][line width=1.5]  (406.17,89.02) -- (405.86,95.89) -- (398.98,95.73) -- (399.29,88.86) -- cycle ;

\draw [color={rgb, 255:red, 208; green, 2; blue, 27 }  ,draw opacity=1 ][line width=1.5]    (364.88,151.28) -- (395.76,202.93) ;
\draw  [color={rgb, 255:red, 208; green, 2; blue, 27 }  ,draw opacity=1 ][line width=1.5]  (364.88,151.28) -- (370.68,147.59) -- (374.51,153.31) -- (368.71,157) -- cycle ;
\draw  [color={rgb, 255:red, 208; green, 2; blue, 27 }  ,draw opacity=1 ][line width=1.5]  (392.23,197.02) -- (398.22,193.64) -- (401.74,199.54) -- (395.75,202.93) -- cycle ;

\draw [color={rgb, 255:red, 208; green, 2; blue, 27 }  ,draw opacity=1 ][line width=1.5]    (459.61,202.92) -- (494.63,154) ;
\draw  [color={rgb, 255:red, 208; green, 2; blue, 27 }  ,draw opacity=1 ][line width=1.5]  (459.61,202.92) -- (453.9,199.08) -- (457.62,193.29) -- (463.32,197.13) -- cycle ;
\draw  [color={rgb, 255:red, 208; green, 2; blue, 27 }  ,draw opacity=1 ][line width=1.5]  (490.63,159.59) -- (485.13,155.45) -- (489.14,149.86) -- (494.63,154) -- cycle ;

\draw  [draw opacity=0][line width=1.5]  (657.94,213.19) .. controls (666.86,210.18) and (676.44,208.59) .. (686.39,208.72) .. controls (697.01,208.86) and (707.15,210.93) .. (716.49,214.59) -- (685.3,294.12) -- cycle ; \draw  [color={rgb, 255:red, 80; green, 227; blue, 194 }  ,draw opacity=1 ][line width=1.5]  (657.94,213.19) .. controls (666.86,210.18) and (676.44,208.59) .. (686.39,208.72) .. controls (697.01,208.86) and (707.15,210.93) .. (716.49,214.59) ;  
\draw  [draw opacity=0][line width=1.5]  (747.2,167.52) .. controls (738.48,160.36) and (731.04,151.35) .. (725.54,140.69) .. controls (720.39,130.74) and (717.4,120.24) .. (716.38,109.73) -- (801.42,101.49) -- cycle ; \draw  [color={rgb, 255:red, 80; green, 227; blue, 194 }  ,draw opacity=1 ][line width=1.5]  (747.2,167.52) .. controls (738.48,160.36) and (731.04,151.35) .. (725.54,140.69) .. controls (720.39,130.74) and (717.4,120.24) .. (716.38,109.73) ;  
\draw  [draw opacity=0][line width=1.5]  (661.18,108.53) .. controls (659.92,119.86) and (656.36,131.12) .. (650.32,141.63) .. controls (644.88,151.07) and (637.91,159.09) .. (629.91,165.54) -- (576.28,99.04) -- cycle ; \draw  [color={rgb, 255:red, 80; green, 227; blue, 194 }  ,draw opacity=1 ][line width=1.5]  (661.18,108.53) .. controls (659.92,119.86) and (656.36,131.12) .. (650.32,141.63) .. controls (644.88,151.07) and (637.91,159.09) .. (629.91,165.54) ;  
\draw [color={rgb, 255:red, 208; green, 2; blue, 27 }  ,draw opacity=1 ][line width=1.5]    (716.38,109.73) -- (661.18,108.53) ;
\draw  [color={rgb, 255:red, 208; green, 2; blue, 27 }  ,draw opacity=1 ][line width=1.5]  (716.38,109.73) -- (716.42,116.04) -- (710.11,116.22) -- (710.07,109.91) -- cycle ;
\draw  [color={rgb, 255:red, 208; green, 2; blue, 27 }  ,draw opacity=1 ][line width=1.5]  (667.49,108.67) -- (667.21,114.98) -- (660.9,114.83) -- (661.18,108.53) -- cycle ;

\draw [color={rgb, 255:red, 208; green, 2; blue, 27 }  ,draw opacity=1 ][line width=1.5]    (629.61,165.8) -- (657.94,213.19) ;
\draw  [color={rgb, 255:red, 208; green, 2; blue, 27 }  ,draw opacity=1 ][line width=1.5]  (629.61,165.8) -- (634.93,162.41) -- (638.44,167.66) -- (633.12,171.05) -- cycle ;
\draw  [color={rgb, 255:red, 208; green, 2; blue, 27 }  ,draw opacity=1 ][line width=1.5]  (654.7,207.77) -- (660.2,204.66) -- (663.43,210.09) -- (657.94,213.19) -- cycle ;

\draw [color={rgb, 255:red, 208; green, 2; blue, 27 }  ,draw opacity=1 ][line width=1.5]    (716.52,213.18) -- (748.66,168.29) ;
\draw  [color={rgb, 255:red, 208; green, 2; blue, 27 }  ,draw opacity=1 ][line width=1.5]  (716.52,213.18) -- (711.29,209.66) -- (714.7,204.34) -- (719.93,207.87) -- cycle ;
\draw  [color={rgb, 255:red, 208; green, 2; blue, 27 }  ,draw opacity=1 ][line width=1.5]  (744.99,173.42) -- (739.94,169.63) -- (743.62,164.5) -- (748.66,168.29) -- cycle ;

\draw [color={rgb, 255:red, 208; green, 2; blue, 27 }  ,draw opacity=1 ][line width=1.5]    (715.47,100.55) -- (660.27,99.35) ;
\draw [line width=1.5]  [dash pattern={on 5.63pt off 4.5pt}]  (660.27,69.75) -- (660.27,99.35) ;
\draw [line width=1.5]  [dash pattern={on 5.63pt off 4.5pt}]  (715.47,70.95) -- (715.47,100.55) ;

\draw [color={rgb, 255:red, 208; green, 2; blue, 27 }  ,draw opacity=1 ][line width=1.5]    (621.96,172.87) -- (650.37,220.21) ;
\draw [line width=1.5]  [dash pattern={on 5.63pt off 4.5pt}]  (625.32,235.99) -- (650.37,220.21) ;
\draw [line width=1.5]  [dash pattern={on 5.63pt off 4.5pt}]  (596.91,188.65) -- (621.96,172.87) ;

\draw [color={rgb, 255:red, 208; green, 2; blue, 27 }  ,draw opacity=1 ][line width=1.5]    (723.98,219.79) -- (755.56,174.5) ;
\draw [line width=1.5]  [dash pattern={on 5.63pt off 4.5pt}]  (780.21,190.91) -- (755.56,174.5) ;
\draw [line width=1.5]  [dash pattern={on 5.63pt off 4.5pt}]  (748.62,236.2) -- (723.98,219.79) ;

\draw (420,62.4) node [anchor=north west][inner sep=0.75pt]    {$L_{o}$};
\draw (351,183.4) node [anchor=north west][inner sep=0.75pt]    {$L_{o}$};
\draw (490,183.4) node [anchor=north west][inner sep=0.75pt]    {$L_{o}$};
\draw (28,191.4) node [anchor=north west][inner sep=0.75pt]    {$A$};
\draw (138,29.4) node [anchor=north west][inner sep=0.75pt]    {$B$};
\draw (232,193.4) node [anchor=north west][inner sep=0.75pt]    {$C$};
\draw (35,75.4) node [anchor=north west][inner sep=0.75pt]    {$a$};
\draw (240,75.4) node [anchor=north west][inner sep=0.75pt]    {$b$};
\draw (135,244.4) node [anchor=north west][inner sep=0.75pt]    {$c$};
\draw (134,63.4) node [anchor=north west][inner sep=0.75pt]    {$L_{o}$};
\draw (97,159.4) node [anchor=north west][inner sep=0.75pt]    {$L_{o}$};
\draw (161,161.4) node [anchor=north west][inner sep=0.75pt]    {$L_{o}$};
\draw (74,96.4) node [anchor=north west][inner sep=0.75pt]    {$\gamma _{a}$};
\draw (194,103.4) node [anchor=north west][inner sep=0.75pt]    {$\gamma _{b}$};
\draw (131,204.4) node [anchor=north west][inner sep=0.75pt]    {$\gamma _{c}$};
\draw (360,104.4) node [anchor=north west][inner sep=0.75pt]    {$\vartheta $};
\draw (482,104.4) node [anchor=north west][inner sep=0.75pt]    {$\vartheta $};
\draw (422,213.4) node [anchor=north west][inner sep=0.75pt]    {$\vartheta $};
\draw (399,243.4) node [anchor=north west][inner sep=0.75pt]    {$\tilde{\mathcal{V}}_{0,3}^{o}( L_{o})$};
\draw (659,252.4) node [anchor=north west][inner sep=0.75pt]    {$\mathcal{V}_{0,3}^{o}( L_{o})$};

\end{tikzpicture}
\par\end{centering}
\centering{}\caption{\label{fig:open string vertices}The construction for the open string
vertices. Note that we can also view geodesics of lengths $\vartheta$
as the boundaries of open string vertices $\tilde{\mathcal{V}}_{0,3}^{o}\left(\vartheta\right)$.
Grafting three semi-infinite strips onto the boundaries of $\tilde{\mathcal{V}}_{0,3}^{o}\left(\vartheta\right)$
with a width of $\vartheta$ results in $\mathcal{V}_{0,3}^{o}\left(\vartheta\right)$.
In this paper, for simplicity, we only consider $L_{o}$ as the boundary.}
\end{figure}
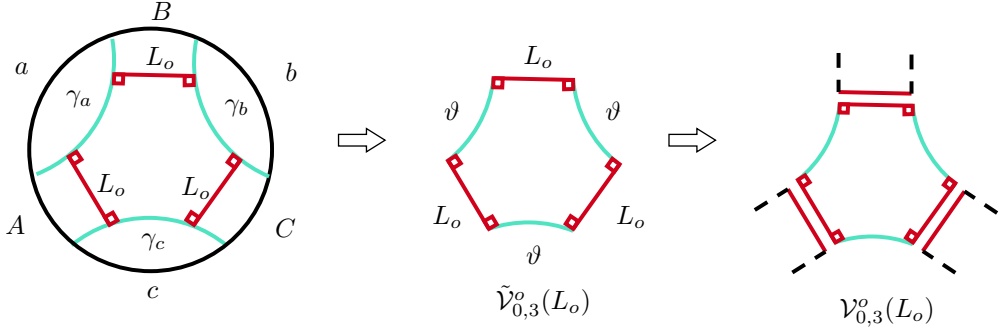

\noindent \textbf{Open string vertices:} In figure (\ref{fig:Y-pants}),
if we do not glue two hexagons together, the illustration represents
the open string vertices $\tilde{\mathcal{V}}_{0,3}^{o}\left(L_{o}\right)$
of OSFT with boundary lengths $L_{o}=L_{c}/2$. The open string vertices
$\mathcal{V}_{0,3}^{o}\left(L_{o}\right)$ is then obtained by grafting
three semi-infinite strips on the boundaries of $\tilde{\mathcal{V}}_{0,3}^{o}\left(L_{o}\right)$
of width $L_{o}$, see figure (\ref{fig:open string vertices}). The
open string boundary conditions apply to the three cyan geodesics.

\vspace*{2.0ex}

\noindent To satisfy the BV equation and construct the hyperbolic
vertices, we need to introduce two essential ingredients: the \textbf{Collar
theorem} and \textbf{systole}.

\vspace*{2.0ex}

\begin{figure}[h]
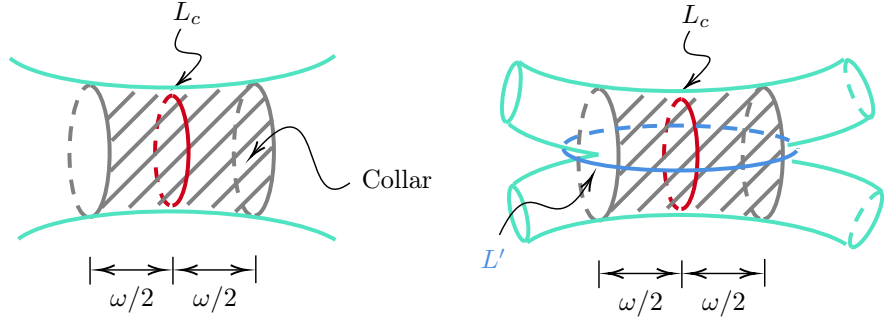

\begin{centering}

\tikzset{every picture/.style={line width=0.55pt}} 


\par\end{centering}
\centering{}\caption{\label{fig:collar} The left-side picture denotes the collar (shaded
region) for the red geodesic of length $L_{c}$. Due to the Collar
theorem, if we require the closed blue geodesic of length $L^{\prime}$
that emerges from sewing between two pairs of pants to be larger than
$L_{c}$, we will obtain constraints for $\omega$ and $L_{c}$: $L^{\prime}>\omega\protect\geq L_{c}$.}
\end{figure}

\noindent \textbf{Theorem} \textbf{2} (Collar theorem \cite{Buser})
(Theorem 4.1.1)): Let $\sigma_{i}$ be simple closed geodesics on
a hyperbolic surface $S$, the collars:

\begin{equation}
\mathcal{C}\left(\sigma_{i}\right)=\left\{ p\in S|d\left(p,\sigma_{i}\right)\leq\frac{1}{2}\omega_{i}\right\} ,
\end{equation}

\noindent of widths $\omega_{i}$

\begin{equation}
\sinh\left(\frac{1}{2}\omega_{i}\right)\sinh\left(\frac{1}{2}L_{c}\left(\sigma_{i}\right)\right)=1,\label{eq:collar}
\end{equation}

\noindent are pairwise disjoint. We illustrate the simplest example
in the left-hand side picture of figure (\ref{fig:collar}). Based
on the Collar theorem, for the open string worldsheet, the half-collars
of the red boundaries with width $L_{o}$ do not overlap with each
other in figure (\ref{fig:open string vertices}). 

\vspace*{2.0ex}

\noindent From equation (\ref{eq:collar}), a critical length can
be defined for $\omega_{*}=L_{*}$, such that
\begin{equation}
L_{*}=2\mathrm{arc}\sinh\left(1\right)=2\log\left(1+\sqrt{2}\right)=\log\left(3+2\sqrt{2}\right).
\end{equation}

\noindent Now, let us explain why the BV equation requires the Collar
theorem. The gluing and cutting of string vertices require the geodesic
boundaries to be minimal and equal. Based on the Collar theorem, if
$L_{c}\leq L_{*}$, then the collar $\omega\geq L_{c}$. It ensures
that the new closed geodesic (with length $L^{\prime}>\omega$) created
by the sewing of two vertices is always larger than $L_{c}$ \cite{Buser,Costello:2019fuh},
as shown in the right-hand side picture of figure (\ref{fig:collar})
for example. Moreover, it also guarantees that any two simple geodesics
(of lengths $\leq L_{*}$) do not intersect with each other and give
the fundamental building block.

\vspace*{2.0ex}

Secondly, let us introduce the surfaces with systole:

\vspace*{2.0ex}

\noindent \textbf{Systole: }The systole $sys\left[\Sigma\right]$
of a surface $\Sigma$ denotes the length of the shortest non-contractible
closed geodesic that is not a boundary component.

\vspace*{2.0ex}

\begin{figure}[h]
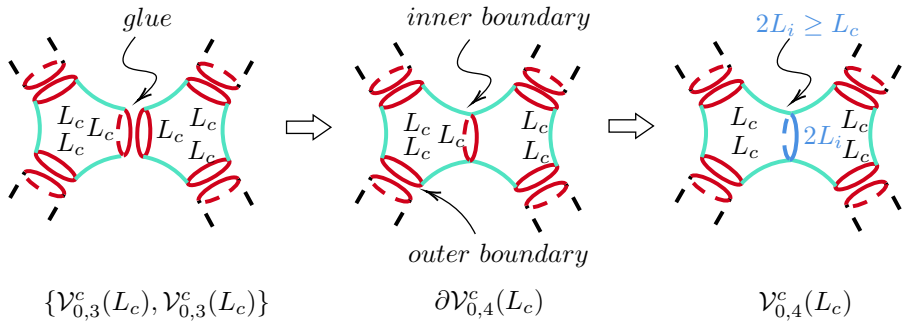

\begin{centering}

\tikzset{every picture/.style={line width=0.5pt}} 


\par\end{centering}
\centering{}\caption{\label{fig:closed BV} The gluing of two closed string vertices occurs
when all outer and inner boundaries have equal lengths. It denotes
the boundary of vertices $\partial\mathcal{V}_{0,4}^{c}\left(L_{c}\right)$
in the moduli space. The construction requires the systole $sys\left[\tilde{\mathcal{V}}\left(L_{c}\right)\right]\protect\geq L_{c}$
for the vertices $\mathcal{V}_{0,4}^{c}\left(L_{c}\right)$, and it
can be roughly seen as $2L_{i}\protect\geq L_{c}$ in the last picture.}
\end{figure}

\begin{figure}[h]
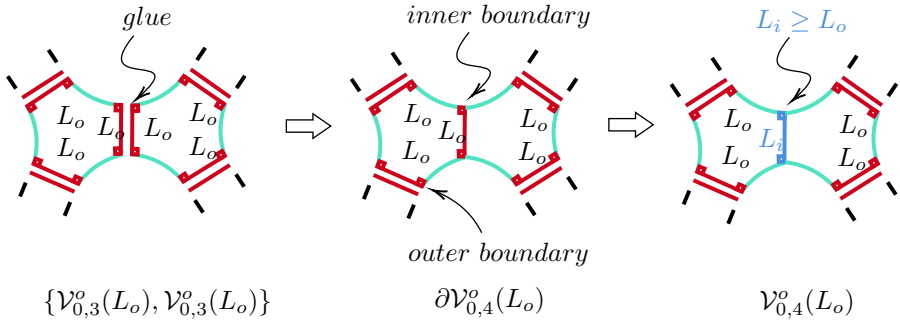

\begin{centering}

\tikzset{every picture/.style={line width=0.5pt}} 


\par\end{centering}
\centering{}\caption{\label{fig:open BV}The gluing of two open string vertices occurs
when all outer and inner boundaries have equal lengths. It denotes
the boundary of vertices $\partial\mathcal{V}_{0,4}^{o}\left(L_{o}\right)$
in the moduli space. The construction requires the systole $sys\left[\tilde{\mathcal{V}}\left(L_{o}\right)\right]\protect\geq L_{o}$
for the vertices $\mathcal{V}_{0,4}^{o}\left(L_{o}\right)$, and it
can be roughly seen as $L_{i}\protect\geq L_{o}$ in the last picture.}
\end{figure}

\noindent The construction of hyperbolic string vertices $\tilde{\mathcal{V}}_{g,n}\left(L\right)$
requires $sys\left[\Sigma\right]\geq L$, where $L$ is the length
of the boundary geodesics. In other words, there is no geodesic of
length less than $L$ on the surface of string vertices. This result
guarantees that $\partial\mathcal{V}_{g,n}\left(L\right)$ is the
boundary of $\mathcal{V}_{g,n}\left(L\right)$ in the moduli space,
as seen in the two pictures on the right-hand side of figures (\ref{fig:closed BV})
and (\ref{fig:open BV}). 

\vspace*{2.0ex}

Using these results, K. Costello and B. Zwiebach \cite{Costello:2019fuh}
established the following results in CSFT for the closed string interaction:
\begin{enumerate}
\item The sets of closed string vertices $\mathcal{V}_{g,n}^{c}\left(L_{c}\right)$
of width $L_{c}\leq L_{*}$ and $\mathrm{sys}\left[\tilde{\Sigma}_{c}\right]\geq L_{c}$
satisfy the quantum geometric master equation.
\item The closed string vertices $\mathcal{V}_{g,n}^{c}\left(L_{c}\right)$
of width $L_{c}>0$ and $\mathrm{sys}\left[\tilde{\Sigma}_{c}\right]\geq L_{c}$
fulfill the classical geometric master equation (corresponding to
$\hbar\rightarrow0$ in the BV equation), and they corresponds to
classical hyperbolic CSFT.
\end{enumerate}
In OSFT, there are similar results:
\begin{enumerate}
\item The sets of open string vertices $\mathcal{V}_{g,n}^{o}\left(L_{o}\right)$
of width $L_{o}\leq\frac{L_{*}}{2}$ and $\mathrm{sys}\left[\tilde{\Sigma}_{o}\right]\geq L_{o}$
meet the quantum geometric master equation.
\item The open string vertices $\mathcal{V}_{g,n}^{o}\left(L_{o}\right)$
of width $L_{o}>0$ and $\mathrm{sys}\left[\tilde{\Sigma}_{o}\right]\geq L_{o}$
satisfy the classical geometric master equation.
\end{enumerate}
The hyperbolic open string vertices can be defined by analogy to the
closed string vertices:

\begin{equation}
\tilde{\mathcal{V}}_{g,n}^{o}\left(L_{o}\right)\equiv\left\{ \tilde{\Sigma}_{o}\in\mathcal{M}_{g,n,L_{o}}|\mathrm{sys}\left[\tilde{\Sigma}_{o}\right]\geq L_{o}\right\} ,
\end{equation}

\noindent where $\mathcal{M}_{g,n,L_{o}}$ denotes the moduli space
of genus $g$ surfaces with $n$ open string legs of length $L_{o}$.
The open string vertices $\mathcal{V}_{g,n}^{o}\left(L_{o}\right)$
are constructed by grafting flat $n$ semi-infinite strips onto the
punctures.

It is important to note that the BV master equation requires that
both open string and closed string vertices possess boundaries of
the same geodesic lengths. To illustrate this concept, let us refer
to figures (\ref{fig:closed BV}) and (\ref{fig:open BV}). We start
with $\mathcal{V}_{0,3}^{c}\left(L_{c}\right)$/$\mathcal{V}_{0,3}^{o}\left(L_{o}\right)$
and its copy, where the three outer boundaries of $\mathcal{V}_{0,3}^{c}\left(L_{c}\right)$/$\mathcal{V}_{0,3}^{o}\left(L_{o}\right)$
have lengths $L_{c}$/$L_{o}$. By attaching two $\mathcal{V}_{0,3}^{c}\left(L_{c}\right)$/$\mathcal{V}_{0,3}^{o}\left(L_{o}\right)$
together, we create the boundary of the moduli space $\partial\mathcal{V}_{0,4}^{c}\left(L_{c}\right)/\partial\mathcal{V}_{0,4}^{o}\left(L_{o}\right)$,
indicating that all inner and outer boundaries have the same length
$L_{c}$/$L_{o}$. If we were to relax the length of the inner boundary
to an arbitrary value $2L_{i}$/$L_{i}$ that satisfies the systole
requirements, the result is $\mathcal{V}_{0,4}^{c}\left(L_{c}\right)$/$\mathcal{V}_{0,4}^{o}\left(L_{o}\right)$. 

\section{Connections between string vertices and entanglement wedge evolution}

In this section, we plan to figure out the relations between phase
transition of entanglement wedge and string vertices.

\begin{figure}[h]
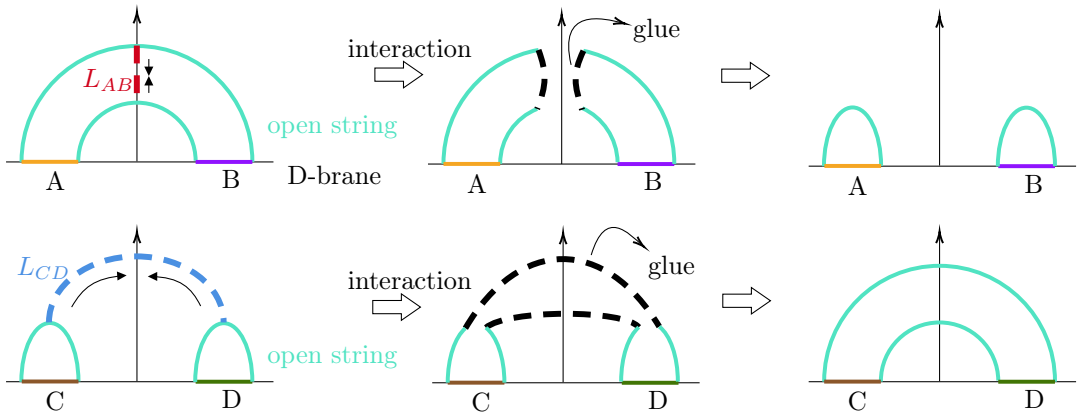

\begin{centering}

\tikzset{every picture/.style={line width=0.40pt}} 


\par\end{centering}
\caption{\label{fig:4 open}These pictures demonstrate the two types of open
string interactions. When two open strings interact with each other,
they break at the mid-point and then glue to the other half of the
string. Here, $L_{AB}$ and $L_{CD}$ denote the distances between
the two open strings.}
\end{figure}

Let us first revisit figure (\ref{fig:2 intervals}), which illustrates
the entanglement wedge bounded by the RT surfaces between $A$ and
$B$. When the EWCS reaches its minimal value $E_{W}^{min}\left(A,B\right)$,
the entanglement wedge vanishes and transforms into two separate RT
surfaces. This process is very similar to the four open strings scattering:
when two open strings interact, they give another pair of open strings.
Moreover, in the theory of hyperbolic open string vertices, this minimal
value can be understood as a scattering distance, indicating that
interaction occurs only when the width reaches a critical length.
In figure (\ref{fig:4 open}), we show two types of interactions between
open strings: the top-down interaction introduces the critical length
$L_{AB}$, and the left-right interaction provides the critical length
$L_{CD}$. 

\begin{figure}[h]
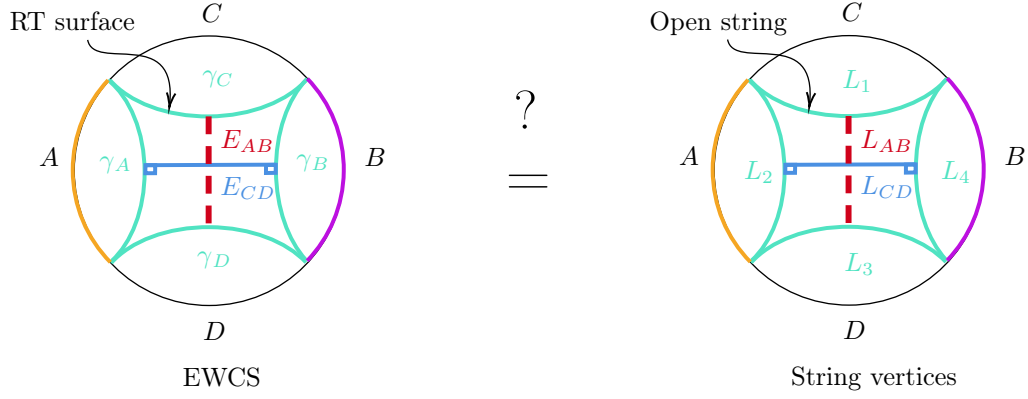

\begin{centering}

\tikzset{every picture/.style={line width=0.5pt}} 

%
\par\end{centering}
\caption{\label{fig:4 corres}The left-hand side image depicts the quadrilateral
resulting from the phase transition of the EWCS, where $\gamma_{i}$
represents the RT surface, and $E_{AB/CD}$ denotes the geodesic length
between the RT surfaces. The right-hand side image illustrates the
quadrilateral formed by open string vertices, where $L_{i}$ represents
the boundary of the string vertices, and $L_{AB/CD}$ is the scattering
distance (geodesic length between open strings).}
\end{figure}

By plotting these processes involving RT surfaces and open strings
in the Poincar$\acute{\mathrm{e}}$ disk, we obtain two similar configurations:
hyperbolic ideal quadrilaterals with mutually perpendicular geodesics,
as shown in figure (\ref{fig:4 corres}). Our goal is to verify whether
these two configurations are indeed identical, specifically:

\begin{equation}
E_{AB}^{min}\stackrel{?}{=}L_{AB}^{min},\qquad\mathrm{and}\qquad E_{CD}^{min}\stackrel{?}{=}L_{CD}^{min}.
\end{equation}

\noindent Note the EWCS is defined as $E_{W}\left(A,B\right)=E_{AB}/4G_{N}^{\left(3\right)}$.
However, this equivalence cannot be obtained based only on hyperbolic
geometry, as it provides the geodesics but not their specific lengths.
To establish a connection between the lengths in the two theories,
a deeper investigation into their correspondence is required. \emph{Fortunately,
we find that mutual information is related to the geometric BV master
equation, bridging this gap.}

\subsection{Correspondence between mutual information and the geometric BV equation}

\subsubsection{Mutual information}

Let us first consider the mutual information

\begin{equation}
I\left(A:B\right)=S_{A}+S_{B}-S_{A\cup B},
\end{equation}

\noindent where $S_{A}$, $S_{B}$, and $S_{A\cup B}$ denote the
entanglement entropies of regions $A$, $B$, and the combined region
$A\cup B$, respectively. If we represent the remaining regions of
the system as $C$ and $D$, the mutual information can also be expressed
as

\begin{equation}
I\left(A:B\right)=S_{A}+S_{B}-\left(S_{C}+S_{D}\right).
\end{equation}

\noindent The sub-additivity condition guarantees that $I\left(A:B\right)\geq0$,
which implies:

\begin{equation}
S_{A}+S_{B}\geq S_{C}+S_{D}.
\end{equation}

\noindent Conversely, $I\left(C:D\right)$ becomes relevant when

\begin{equation}
S_{A}+S_{B}\leq S_{C}+S_{D}.
\end{equation}

\noindent A phase transition occurs at the point where

\begin{equation}
S_{A}+S_{B}=S_{C}+S_{D}.\label{eq:mutual eq}
\end{equation}

\begin{figure}[H]
\begin{centering}

\tikzset{every picture/.style={line width=0.55pt}} 

\begin{tikzpicture}[x=0.55pt,y=0.55pt,yscale=-1,xscale=1]

\draw   (71.19,148.55) .. controls (71.19,92.21) and (116.87,46.53) .. (173.22,46.53) .. controls (229.56,46.53) and (275.24,92.21) .. (275.24,148.55) .. controls (275.24,204.9) and (229.56,250.58) .. (173.22,250.58) .. controls (116.87,250.58) and (71.19,204.9) .. (71.19,148.55) -- cycle ;
\draw  [draw opacity=0][dash pattern={on 5.63pt off 4.5pt}][line width=1.5]  (97.49,81.28) .. controls (113.86,94.42) and (125.04,120.6) .. (125.08,150.76) .. controls (125.11,179.31) and (115.15,204.32) .. (100.23,218.09) -- (73.59,150.82) -- cycle ; \draw  [color={rgb, 255:red, 80; green, 227; blue, 194 }  ,draw opacity=1 ][dash pattern={on 5.63pt off 4.5pt}][line width=1.5]  (97.49,81.28) .. controls (113.86,94.42) and (125.04,120.6) .. (125.08,150.76) .. controls (125.11,179.31) and (115.15,204.32) .. (100.23,218.09) ;  
\draw  [draw opacity=0][line width=2.25]  (248.21,80.18) .. controls (233.94,96.93) and (206.07,108.45) .. (173.93,108.75) .. controls (141.62,109.05) and (113.42,97.93) .. (98.9,81.3) -- (173.43,55.14) -- cycle ; \draw  [color={rgb, 255:red, 80; green, 227; blue, 194 }  ,draw opacity=1 ][line width=2.25]  (248.21,80.18) .. controls (233.94,96.93) and (206.07,108.45) .. (173.93,108.75) .. controls (141.62,109.05) and (113.42,97.93) .. (98.9,81.3) ;  
\draw  [draw opacity=0][dash pattern={on 5.63pt off 4.5pt}][line width=1.5]  (246.9,217.22) .. controls (233.61,202.49) and (224.76,178.14) .. (224.43,150.5) .. controls (224.08,121.23) and (233.36,95.42) .. (247.64,80.69) -- (275.92,149.88) -- cycle ; \draw  [color={rgb, 255:red, 80; green, 227; blue, 194 }  ,draw opacity=1 ][dash pattern={on 5.63pt off 4.5pt}][line width=1.5]  (246.9,217.22) .. controls (233.61,202.49) and (224.76,178.14) .. (224.43,150.5) .. controls (224.08,121.23) and (233.36,95.42) .. (247.64,80.69) ;  
\draw  [draw opacity=0][line width=2.25]  (99.78,219.76) .. controls (114.37,203.29) and (142.45,192.29) .. (174.59,192.6) .. controls (205.49,192.89) and (232.43,203.56) .. (247.15,219.27) -- (174.08,246.21) -- cycle ; \draw  [color={rgb, 255:red, 80; green, 227; blue, 194 }  ,draw opacity=1 ][line width=2.25]  (99.78,219.76) .. controls (114.37,203.29) and (142.45,192.29) .. (174.59,192.6) .. controls (205.49,192.89) and (232.43,203.56) .. (247.15,219.27) ;  
\draw [color={rgb, 255:red, 208; green, 2; blue, 27 }  ,draw opacity=1 ][line width=2.25]    (173.51,108.74) -- (173.51,192.74) ;
\draw  [draw opacity=0][line width=2.25]  (99.17,219.16) .. controls (81.59,201.29) and (70.71,176.61) .. (70.71,149.35) .. controls (70.71,122.96) and (80.9,99) .. (97.49,81.28) -- (168,149.35) -- cycle ; \draw  [color={rgb, 255:red, 245; green, 166; blue, 35 }  ,draw opacity=1 ][line width=2.25]  (99.17,219.16) .. controls (81.59,201.29) and (70.71,176.61) .. (70.71,149.35) .. controls (70.71,122.96) and (80.9,99) .. (97.49,81.28) ;  
\draw  [draw opacity=0][line width=2.25]  (248.21,80.18) .. controls (265.56,98.28) and (276.11,123.11) .. (275.74,150.37) .. controls (275.4,176.74) and (264.89,200.57) .. (248.07,218.07) -- (178.47,149.08) -- cycle ; \draw  [color={rgb, 255:red, 245; green, 166; blue, 35 }  ,draw opacity=1 ][line width=2.25]  (248.21,80.18) .. controls (265.56,98.28) and (276.11,123.11) .. (275.74,150.37) .. controls (275.4,176.74) and (264.89,200.57) .. (248.07,218.07) ;  
\draw   (450.19,148.83) .. controls (450.19,92.49) and (495.87,46.81) .. (552.22,46.81) .. controls (608.56,46.81) and (654.24,92.49) .. (654.24,148.83) .. controls (654.24,205.18) and (608.56,250.86) .. (552.22,250.86) .. controls (495.87,250.86) and (450.19,205.18) .. (450.19,148.83) -- cycle ;
\draw  [draw opacity=0][line width=2.25]  (476.49,81.56) .. controls (492.86,94.7) and (504.04,120.88) .. (504.08,151.04) .. controls (504.11,179.59) and (494.15,204.6) .. (479.23,218.37) -- (452.59,151.1) -- cycle ; \draw  [color={rgb, 255:red, 80; green, 227; blue, 194 }  ,draw opacity=1 ][line width=2.25]  (476.49,81.56) .. controls (492.86,94.7) and (504.04,120.88) .. (504.08,151.04) .. controls (504.11,179.59) and (494.15,204.6) .. (479.23,218.37) ;  
\draw  [draw opacity=0][dash pattern={on 5.63pt off 4.5pt}][line width=1.5]  (627.21,80.46) .. controls (612.94,97.21) and (585.07,108.73) .. (552.93,109.03) .. controls (520.62,109.33) and (492.42,98.21) .. (477.9,81.58) -- (552.43,55.42) -- cycle ; \draw  [color={rgb, 255:red, 80; green, 227; blue, 194 }  ,draw opacity=1 ][dash pattern={on 5.63pt off 4.5pt}][line width=1.5]  (627.21,80.46) .. controls (612.94,97.21) and (585.07,108.73) .. (552.93,109.03) .. controls (520.62,109.33) and (492.42,98.21) .. (477.9,81.58) ;  
\draw  [draw opacity=0][line width=2.25]  (625.9,217.5) .. controls (612.61,202.77) and (603.76,178.42) .. (603.43,150.78) .. controls (603.08,121.51) and (612.36,95.7) .. (626.64,80.97) -- (654.92,150.16) -- cycle ; \draw  [color={rgb, 255:red, 80; green, 227; blue, 194 }  ,draw opacity=1 ][line width=2.25]  (625.9,217.5) .. controls (612.61,202.77) and (603.76,178.42) .. (603.43,150.78) .. controls (603.08,121.51) and (612.36,95.7) .. (626.64,80.97) ;  
\draw  [draw opacity=0][dash pattern={on 5.63pt off 4.5pt}][line width=1.5]  (478.78,220.04) .. controls (493.37,203.57) and (521.45,192.57) .. (553.59,192.88) .. controls (584.49,193.17) and (611.43,203.84) .. (626.15,219.55) -- (553.08,246.49) -- cycle ; \draw  [color={rgb, 255:red, 80; green, 227; blue, 194 }  ,draw opacity=1 ][dash pattern={on 5.63pt off 4.5pt}][line width=1.5]  (478.78,220.04) .. controls (493.37,203.57) and (521.45,192.57) .. (553.59,192.88) .. controls (584.49,193.17) and (611.43,203.84) .. (626.15,219.55) ;  
\draw [color={rgb, 255:red, 74; green, 144; blue, 226 }  ,draw opacity=1 ][line width=2.25]    (603.16,145.57) -- (504.72,145.96) ;
\draw  [draw opacity=0][line width=2.25]  (474.76,82.55) .. controls (493.99,60.07) and (521.59,45.99) .. (552.24,46.01) .. controls (581.75,46.03) and (608.4,59.11) .. (627.48,80.16) -- (552.16,158.97) -- cycle ; \draw  [color={rgb, 255:red, 245; green, 166; blue, 35 }  ,draw opacity=1 ][line width=2.25]  (474.76,82.55) .. controls (493.99,60.07) and (521.59,45.99) .. (552.24,46.01) .. controls (581.75,46.03) and (608.4,59.11) .. (627.48,80.16) ;  
\draw  [draw opacity=0][line width=2.25]  (626.65,219.04) .. controls (607.89,238.74) and (581.26,250.99) .. (551.77,250.86) .. controls (522.77,250.73) and (496.64,238.66) .. (478.1,219.37) -- (552.21,149.46) -- cycle ; \draw  [color={rgb, 255:red, 245; green, 166; blue, 35 }  ,draw opacity=1 ][line width=2.25]  (626.65,219.04) .. controls (607.89,238.74) and (581.26,250.99) .. (551.77,250.86) .. controls (522.77,250.73) and (496.64,238.66) .. (478.1,219.37) ;  

\draw (181.56,134.09) node [anchor=north west][inner sep=0.75pt]    {$\textcolor[rgb]{0.82,0.01,0.11}{E_{AB}}$};
\draw (36.39,133.47) node [anchor=north west][inner sep=0.75pt]    {$\textcolor[rgb]{0.82,0.01,0.11}{A}$};
\draw (298.39,133.47) node [anchor=north west][inner sep=0.75pt]    {$\textcolor[rgb]{0.82,0.01,0.11}{B}$};
\draw (166.39,21.47) node [anchor=north west][inner sep=0.75pt]    {$\textcolor[rgb]{0.29,0.56,0.89}{C}$};
\draw (167.39,262.23) node [anchor=north west][inner sep=0.75pt]    {$\textcolor[rgb]{0.29,0.56,0.89}{D}$};
\draw (165,76.4) node [anchor=north west][inner sep=0.75pt]    {$\textcolor[rgb]{0.29,0.56,0.89}{\gamma }\textcolor[rgb]{0.29,0.56,0.89}{_{C}}$};
\draw (164,205.4) node [anchor=north west][inner sep=0.75pt]    {$\textcolor[rgb]{0.29,0.56,0.89}{\gamma }\textcolor[rgb]{0.29,0.56,0.89}{_{D}}$};
\draw (91,134.4) node [anchor=north west][inner sep=0.75pt]    {$\textcolor[rgb]{0.82,0.01,0.11}{\gamma }\textcolor[rgb]{0.82,0.01,0.11}{_{A}}$};
\draw (242,133.4) node [anchor=north west][inner sep=0.75pt]    {$\textcolor[rgb]{0.82,0.01,0.11}{\gamma }\textcolor[rgb]{0.82,0.01,0.11}{_{B}}$};
\draw (105,295.4) node [anchor=north west][inner sep=0.75pt]    {$\textcolor[rgb]{0.82,0.01,0.11}{S}\textcolor[rgb]{0.82,0.01,0.11}{_{A}} +\textcolor[rgb]{0.82,0.01,0.11}{S}\textcolor[rgb]{0.82,0.01,0.11}{_{B}}  >\textcolor[rgb]{0.29,0.56,0.89}{S}\textcolor[rgb]{0.29,0.56,0.89}{_{C}} +\textcolor[rgb]{0.29,0.56,0.89}{S}\textcolor[rgb]{0.29,0.56,0.89}{_{D}}$};
\draw (540,157.51) node [anchor=north west][inner sep=0.75pt]  [color={rgb, 255:red, 74; green, 144; blue, 226 }  ,opacity=1 ]  {$E_{CD}$};
\draw (415.39,133.75) node [anchor=north west][inner sep=0.75pt]    {$\textcolor[rgb]{0.82,0.01,0.11}{A}$};
\draw (677.39,133.75) node [anchor=north west][inner sep=0.75pt]    {$\textcolor[rgb]{0.82,0.01,0.11}{B}$};
\draw (545.39,21.75) node [anchor=north west][inner sep=0.75pt]    {$\textcolor[rgb]{0.29,0.56,0.89}{C}$};
\draw (546.39,262.51) node [anchor=north west][inner sep=0.75pt]    {$\textcolor[rgb]{0.29,0.56,0.89}{D}$};
\draw (544,76.68) node [anchor=north west][inner sep=0.75pt]    {$\textcolor[rgb]{0.29,0.56,0.89}{\gamma }\textcolor[rgb]{0.29,0.56,0.89}{_{C}}$};
\draw (543,205.68) node [anchor=north west][inner sep=0.75pt]    {$\textcolor[rgb]{0.29,0.56,0.89}{\gamma }\textcolor[rgb]{0.29,0.56,0.89}{_{D}}$};
\draw (471,132.68) node [anchor=north west][inner sep=0.75pt]    {$\textcolor[rgb]{0.82,0.01,0.11}{\gamma }\textcolor[rgb]{0.82,0.01,0.11}{_{A}}$};
\draw (621,133.68) node [anchor=north west][inner sep=0.75pt]    {$\textcolor[rgb]{0.82,0.01,0.11}{\gamma }\textcolor[rgb]{0.82,0.01,0.11}{_{B}}$};
\draw (484,295.68) node [anchor=north west][inner sep=0.75pt]    {$\textcolor[rgb]{0.82,0.01,0.11}{S}\textcolor[rgb]{0.82,0.01,0.11}{_{A}} +\textcolor[rgb]{0.82,0.01,0.11}{S}\textcolor[rgb]{0.82,0.01,0.11}{_{B}} < \textcolor[rgb]{0.29,0.56,0.89}{S}\textcolor[rgb]{0.29,0.56,0.89}{_{C}} +\textcolor[rgb]{0.29,0.56,0.89}{S}\textcolor[rgb]{0.29,0.56,0.89}{_{D}}$};

\end{tikzpicture}
\par\end{centering}
\caption{\label{fig:phase transition}The left-hand side image depicts the
entanglement wedge enclosed by regions $A$, $B$, $\gamma_{C}$ and
$\gamma_{D}$, which exists only when $S_{A}+S_{B}\protect\geq S_{C}+S_{D}$.
The corresponding EWCS is given by $E_{W}\left(A:B\right)=E_{AB}/4G_{N}^{\left(3\right)}$.
In contrast, the right-hand side image shows the entanglement wedge
when $S_{A}+S_{B}\protect\leq S_{C}+S_{D}$, enclosed by regions $C$,
$D$, $\gamma_{A}$ and $\gamma_{B}$. The corresponding EWCS is expressed
as $E_{W}\left(C:D\right)=E_{CD}/4G_{N}^{\left(3\right)}$.}
\end{figure}
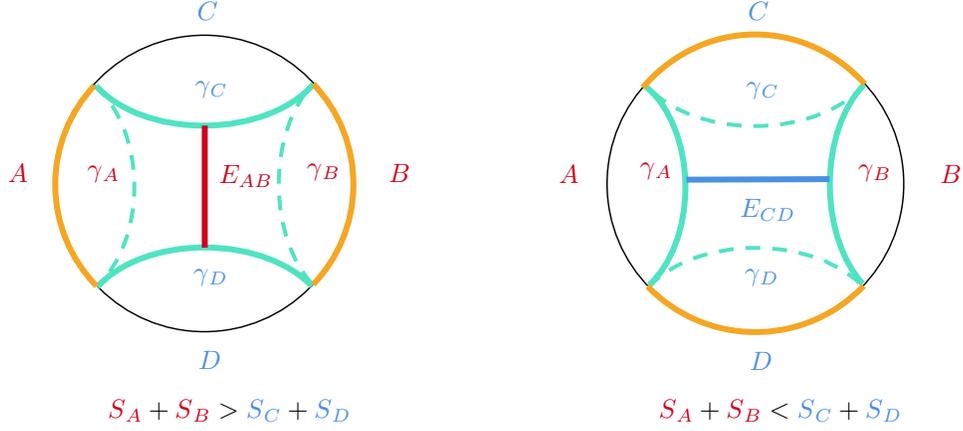

\noindent This phase transition is illustrated in figure (\ref{fig:phase transition}).

\noindent To see how this equivalence (\ref{eq:mutual eq}) imposes
an additional constraint on the hyperbolic ideal quadrilaterals shown
in the left-hand side of figure \textcolor{blue}{(\ref{fig:4 corres})},
we recall the bipartite mixed state in the holographic framework,
where the entanglement entropy $S_{i}$ is given by the geodesic length
of $\gamma_{i}$ in the hyperbolic Poincar$\acute{\mathrm{e}}$ disk:

\begin{equation}
S_{EE}=\frac{\mathrm{Area}\left(\gamma_{i}\right)}{4G_{N}^{\left(3\right)}}.
\end{equation}

\begin{figure}[H]
\begin{centering}

\tikzset{every picture/.style={line width=0.55pt}} 

\begin{tikzpicture}[x=0.55pt,y=0.55pt,yscale=-1,xscale=1]

\draw  [draw opacity=0][line width=1.5]  (408.21,62.06) .. controls (428.64,78.46) and (442.59,111.13) .. (442.63,148.77) .. controls (442.68,184.4) and (430.25,215.61) .. (411.63,232.79) -- (378.38,148.85) -- cycle ; \draw  [color={rgb, 255:red, 80; green, 227; blue, 194 }  ,draw opacity=1 ][line width=1.5]  (408.21,62.06) .. controls (428.64,78.46) and (442.59,111.13) .. (442.63,148.77) .. controls (442.68,184.4) and (430.25,215.61) .. (411.63,232.79) ;  
\draw  [draw opacity=0][line width=1.5]  (595.58,61.33) .. controls (577.77,82.23) and (542.99,96.62) .. (502.88,96.99) .. controls (462.56,97.36) and (427.36,83.48) .. (409.25,62.73) -- (502.26,30.09) -- cycle ; \draw  [color={rgb, 255:red, 80; green, 227; blue, 194 }  ,draw opacity=1 ][line width=1.5]  (595.58,61.33) .. controls (577.77,82.23) and (542.99,96.62) .. (502.88,96.99) .. controls (462.56,97.36) and (427.36,83.48) .. (409.25,62.73) ;  
\draw  [draw opacity=0][line width=1.5]  (594.67,231.7) .. controls (578.08,213.32) and (567.04,182.94) .. (566.62,148.45) .. controls (566.18,111.92) and (577.77,79.7) .. (595.58,61.33) -- (630.87,147.67) -- cycle ; \draw  [color={rgb, 255:red, 80; green, 227; blue, 194 }  ,draw opacity=1 ][line width=1.5]  (594.67,231.7) .. controls (578.08,213.32) and (567.04,182.94) .. (566.62,148.45) .. controls (566.18,111.92) and (577.77,79.7) .. (595.58,61.33) ;  
\draw  [draw opacity=0][line width=1.5]  (411.07,234.88) .. controls (429.27,214.32) and (464.32,200.59) .. (504.43,200.98) .. controls (542.98,201.35) and (576.6,214.66) .. (594.97,234.26) -- (503.79,267.88) -- cycle ; \draw  [color={rgb, 255:red, 80; green, 227; blue, 194 }  ,draw opacity=1 ][line width=1.5]  (411.07,234.88) .. controls (429.27,214.32) and (464.32,200.59) .. (504.43,200.98) .. controls (542.98,201.35) and (576.6,214.66) .. (594.97,234.26) ;  
\draw [color={rgb, 255:red, 74; green, 144; blue, 226 }  ,draw opacity=1 ][line width=1.5]    (566.29,141.94) -- (443.44,142.43) ;
\draw [color={rgb, 255:red, 208; green, 2; blue, 27 }  ,draw opacity=1 ][line width=2.25]  [dash pattern={on 6.75pt off 4.5pt}]  (503.08,96.33) -- (503.08,201.16) ;
\draw  [draw opacity=0][line width=1.5]  (73.45,100.49) .. controls (78.59,114.56) and (81.53,130.76) .. (81.55,148) .. controls (81.57,162.65) and (79.48,176.55) .. (75.72,189.04) -- (17.3,148.08) -- cycle ; \draw  [color={rgb, 255:red, 80; green, 227; blue, 194 }  ,draw opacity=1 ][line width=1.5]  (73.45,100.49) .. controls (78.59,114.56) and (81.53,130.76) .. (81.55,148) .. controls (81.57,162.65) and (79.48,176.55) .. (75.72,189.04) ;  
\draw  [draw opacity=0][line width=1.5]  (198.15,85.44) .. controls (181.86,92.09) and (162.54,96.03) .. (141.8,96.22) .. controls (119.78,96.43) and (99.3,92.38) .. (82.28,85.29) -- (141.18,29.32) -- cycle ; \draw  [color={rgb, 255:red, 80; green, 227; blue, 194 }  ,draw opacity=1 ][line width=1.5]  (198.15,85.44) .. controls (181.86,92.09) and (162.54,96.03) .. (141.8,96.22) .. controls (119.78,96.43) and (99.3,92.38) .. (82.28,85.29) ;  
\draw  [draw opacity=0][line width=1.5]  (212.03,190.83) .. controls (208.04,177.75) and (205.73,163.13) .. (205.54,147.68) .. controls (205.32,129.25) and (208.16,111.92) .. (213.32,96.91) -- (269.79,146.91) -- cycle ; \draw  [color={rgb, 255:red, 80; green, 227; blue, 194 }  ,draw opacity=1 ][line width=1.5]  (212.03,190.83) .. controls (208.04,177.75) and (205.73,163.13) .. (205.54,147.68) .. controls (205.32,129.25) and (208.16,111.92) .. (213.32,96.91) ;  
\draw  [draw opacity=0][line width=1.5]  (95.32,206.98) .. controls (109.76,202.5) and (126.09,200.05) .. (143.35,200.22) .. controls (163.49,200.41) and (182.29,204.13) .. (198.26,210.44) -- (142.71,267.12) -- cycle ; \draw  [color={rgb, 255:red, 80; green, 227; blue, 194 }  ,draw opacity=1 ][line width=1.5]  (95.32,206.98) .. controls (109.76,202.5) and (126.09,200.05) .. (143.35,200.22) .. controls (163.49,200.41) and (182.29,204.13) .. (198.26,210.44) ;  
\draw [color={rgb, 255:red, 74; green, 144; blue, 226 }  ,draw opacity=1 ][line width=1.5]    (205.21,141.17) -- (82.36,141.66) ;
\draw [color={rgb, 255:red, 208; green, 2; blue, 27 }  ,draw opacity=1 ][line width=2.25]  [dash pattern={on 6.75pt off 4.5pt}]  (142,95.56) -- (142,200.4) ;
\draw [color={rgb, 255:red, 189; green, 16; blue, 224 }  ,draw opacity=1 ][line width=1.5]    (198.15,85.44) -- (213.32,96.91) ;
\draw [color={rgb, 255:red, 189; green, 16; blue, 224 }  ,draw opacity=1 ][line width=1.5]    (73.45,100.49) -- (82.28,85.29) ;
\draw [color={rgb, 255:red, 189; green, 16; blue, 224 }  ,draw opacity=1 ][line width=1.5]    (75.72,189.04) -- (95.32,206.98) ;
\draw [color={rgb, 255:red, 189; green, 16; blue, 224 }  ,draw opacity=1 ][line width=1.5]    (212.03,190.83) -- (198.26,210.44) ;
\draw   (305,138.52) -- (332.22,138.52) -- (332.22,133) -- (350.36,144.04) -- (332.22,155.08) -- (332.22,149.56) -- (305,149.56) -- cycle ;

\draw (486.19,109.66) node [anchor=north west][inner sep=0.75pt]    {$a$};
\draw (531.81,122.72) node [anchor=north west][inner sep=0.75pt]    {$b$};
\draw (459.19,122.66) node [anchor=north west][inner sep=0.75pt]    {$b$};
\draw (484.19,164.66) node [anchor=north west][inner sep=0.75pt]    {$a$};
\draw (532.19,63.66) node [anchor=north west][inner sep=0.75pt]    {$\beta _{1}$};
\draw (452.19,62.66) node [anchor=north west][inner sep=0.75pt]    {$\beta _{2}$};
\draw (590.19,104.66) node [anchor=north west][inner sep=0.75pt]    {$\alpha _{1}$};
\draw (400.19,104.66) node [anchor=north west][inner sep=0.75pt]    {$\alpha _{2}$};
\draw (400.19,174.66) node [anchor=north west][inner sep=0.75pt]    {$\alpha _{3}$};
\draw (589.19,174.66) node [anchor=north west][inner sep=0.75pt]    {$\alpha _{4}$};
\draw (457.19,223.66) node [anchor=north west][inner sep=0.75pt]    {$\beta _{3}$};
\draw (535.19,223.66) node [anchor=north west][inner sep=0.75pt]    {$\beta _{4}$};
\draw (125.11,108.89) node [anchor=north west][inner sep=0.75pt]    {$a$};
\draw (170.73,121.96) node [anchor=north west][inner sep=0.75pt]    {$b$};
\draw (98.11,121.89) node [anchor=north west][inner sep=0.75pt]    {$c$};
\draw (123.11,163.89) node [anchor=north west][inner sep=0.75pt]    {$d$};
\draw (165.11,62.89) node [anchor=north west][inner sep=0.75pt]    {$\beta _{1}$};
\draw (110.11,62.89) node [anchor=north west][inner sep=0.75pt]    {$\beta _{2}$};
\draw (225.11,115.89) node [anchor=north west][inner sep=0.75pt]    {$\alpha _{1}$};
\draw (45.11,112.89) node [anchor=north west][inner sep=0.75pt]    {$\alpha _{2}$};
\draw (44.11,154.89) node [anchor=north west][inner sep=0.75pt]    {$\alpha _{3}$};
\draw (225.11,153.89) node [anchor=north west][inner sep=0.75pt]    {$\alpha _{4}$};
\draw (111.11,213.89) node [anchor=north west][inner sep=0.75pt]    {$\beta _{3}$};
\draw (162.11,214.89) node [anchor=north west][inner sep=0.75pt]    {$\beta _{4}$};
\draw (209.11,65.89) node [anchor=north west][inner sep=0.75pt]  [color={rgb, 255:red, 189; green, 16; blue, 224 }  ,opacity=1 ]  {$\sigma _{1}$};
\draw (214.11,204.89) node [anchor=north west][inner sep=0.75pt]  [color={rgb, 255:red, 189; green, 16; blue, 224 }  ,opacity=1 ]  {$\sigma _{4}$};
\draw (54.11,64.89) node [anchor=north west][inner sep=0.75pt]  [color={rgb, 255:red, 189; green, 16; blue, 224 }  ,opacity=1 ]  {$\sigma _{2}$};
\draw (53.11,203.89) node [anchor=north west][inner sep=0.75pt]  [color={rgb, 255:red, 189; green, 16; blue, 224 }  ,opacity=1 ]  {$\sigma _{3}$};
\draw (302.11,109.89) node [anchor=north west][inner sep=0.75pt]  [color={rgb, 255:red, 189; green, 16; blue, 224 }  ,opacity=1 ]  {$\sigma _{i}\rightarrow 0$};

\end{tikzpicture}
\par\end{centering}
\caption{\label{fig:quadrilateral}Hyperbolic ideal quadrilaterals can be constructed
by gluing together four pentagons and setting the four boundary parameters
$\sigma_{i}$ to zero.}
\end{figure}
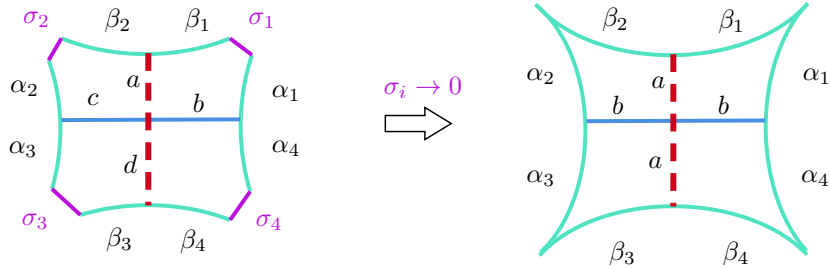

\noindent From this holographic derivation, the equation $S_{A}+S_{B}=S_{C}+S_{D}$
translates to $\gamma_{A}+\gamma_{B}=\gamma_{C}+\gamma_{D}$. To calculate
the transition point determined by this equation, we must consider
a configuration involving four right-angled pentagons, as shown in
figure (\ref{fig:quadrilateral}). 

\noindent For the pentagon in the upper right corner, we have the
relation (\cite{Buser}, Theorem 2.3.4 (i)):

\begin{equation}
\cosh\sigma_{1}=\sinh a\sinh b.
\end{equation}

\noindent As $\sigma_{1}\rightarrow0$, this simplifies to:

\begin{equation}
\sinh a\sinh b=1.
\end{equation}

\noindent Similarly, as $\sigma_{2},\sigma_{3},\sigma_{4}\rightarrow0$,
we obtain:

\begin{eqnarray}
\sinh a\sinh c & = & 1,\nonumber \\
\sinh c\sinh d & = & 1,\nonumber \\
\sinh d\sinh b & = & 1.
\end{eqnarray}

\noindent For the hyperbolic quadrilateral, we thus have:

\begin{equation}
b=c,\qquad a=d.
\end{equation}

\noindent When $\sigma_{i}\rightarrow0$, the right-angled pentagons
reduce to trirectangles. For the trirectangles in the upper right
corner, we have (\cite{Buser}, Theorem 2.3.1 (iv)):

\begin{equation}
\tanh\beta_{1}=\frac{\cosh a}{\coth b},\qquad\tanh\alpha_{1}=\frac{\cosh b}{\coth a}.
\end{equation}

\noindent Similarly, we find:

\begin{eqnarray}
\tanh\beta_{2}=\frac{\cosh a}{\coth b}, & \qquad & \tanh\alpha_{2}=\frac{\cosh b}{\coth a},\nonumber \\
\tanh\beta_{3}=\frac{\cosh a}{\coth b}, & \qquad & \tanh\alpha_{3}=\frac{\cosh b}{\coth a},\nonumber \\
\tanh\beta_{4}=\frac{\cosh a}{\coth b}, & \qquad & \tanh\alpha_{4}=\frac{\cosh b}{\coth a}.\label{eq:alpha beta}
\end{eqnarray}

\noindent TTherefore, the transition point where $\gamma_{A}+\gamma_{B}=\gamma_{C}+\gamma_{D}$
requires

\begin{equation}
\left(\alpha_{2}+\alpha_{3}\right)+\left(\alpha_{1}+\alpha_{4}\right)=\left(\beta_{2}+\beta_{1}\right)+\left(\beta_{3}+\beta_{4}\right).\label{eq:mutual hyper}
\end{equation}

\noindent For $a,b\geq0$, using (\ref{eq:alpha beta}) and (\ref{eq:mutual hyper}),
we obtain:

\begin{equation}
a=b=\frac{1}{2}\log\left(3+2\sqrt{2}\right),\qquad\alpha_{i}=\beta_{i}
\end{equation}

In summary, the transition point of mutual information imposes an
additional constraint on the hyperbolic geometry, such that

\begin{equation}
\gamma_{A}=\gamma_{B}=\gamma_{C}=\gamma_{D},\qquad\mathrm{and}\qquad E_{AB}^{min}=E_{CD}^{min}=\log\left(3+2\sqrt{2}\right),\label{eq:mutual constraint}
\end{equation}

\noindent in the quadrilateral. This result corresponds exactly to
the minimal EWCS \cite{Takayanagi:2017knl}.

\subsubsection{Geometric BV master equation}

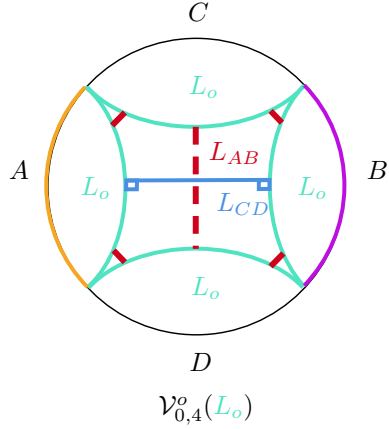
\begin{figure}[h]
\begin{centering}

\tikzset{every picture/.style={line width=0.55pt}} 

\begin{tikzpicture}[x=0.55pt,y=0.55pt,yscale=-1,xscale=1]

\draw   (183.86,133.05) .. controls (183.86,76.7) and (229.54,31.03) .. (285.89,31.03) .. controls (342.24,31.03) and (387.92,76.7) .. (387.92,133.05) .. controls (387.92,189.4) and (342.24,235.08) .. (285.89,235.08) .. controls (229.54,235.08) and (183.86,189.4) .. (183.86,133.05) -- cycle ;
\draw  [draw opacity=0][line width=1.5]  (209.49,64.45) .. controls (225.86,77.59) and (237.04,103.77) .. (237.08,133.93) .. controls (237.11,162.48) and (227.15,187.49) .. (212.23,201.25) -- (185.59,133.99) -- cycle ; \draw  [color={rgb, 255:red, 80; green, 227; blue, 194 }  ,draw opacity=1 ][line width=1.5]  (209.49,64.45) .. controls (225.86,77.59) and (237.04,103.77) .. (237.08,133.93) .. controls (237.11,162.48) and (227.15,187.49) .. (212.23,201.25) ;  
\draw  [draw opacity=0][line width=1.5]  (360.21,63.35) .. controls (345.94,80.1) and (318.07,91.62) .. (285.93,91.92) .. controls (253.62,92.22) and (225.42,81.09) .. (210.9,64.47) -- (285.43,38.31) -- cycle ; \draw  [color={rgb, 255:red, 80; green, 227; blue, 194 }  ,draw opacity=1 ][line width=1.5]  (360.21,63.35) .. controls (345.94,80.1) and (318.07,91.62) .. (285.93,91.92) .. controls (253.62,92.22) and (225.42,81.09) .. (210.9,64.47) ;  
\draw  [draw opacity=0][line width=1.5]  (358.9,200.39) .. controls (345.61,185.65) and (336.76,161.31) .. (336.43,133.67) .. controls (336.08,104.4) and (345.36,78.59) .. (359.64,63.86) -- (387.92,133.05) -- cycle ; \draw  [color={rgb, 255:red, 80; green, 227; blue, 194 }  ,draw opacity=1 ][line width=1.5]  (358.9,200.39) .. controls (345.61,185.65) and (336.76,161.31) .. (336.43,133.67) .. controls (336.08,104.4) and (345.36,78.59) .. (359.64,63.86) ;  
\draw  [draw opacity=0][line width=1.5]  (211.78,202.93) .. controls (226.37,186.45) and (254.45,175.46) .. (286.59,175.77) .. controls (317.49,176.06) and (344.43,186.73) .. (359.15,202.43) -- (286.08,229.37) -- cycle ; \draw  [color={rgb, 255:red, 80; green, 227; blue, 194 }  ,draw opacity=1 ][line width=1.5]  (211.78,202.93) .. controls (226.37,186.45) and (254.45,175.46) .. (286.59,175.77) .. controls (317.49,176.06) and (344.43,186.73) .. (359.15,202.43) ;  
\draw [color={rgb, 255:red, 74; green, 144; blue, 226 }  ,draw opacity=1 ][line width=1.5]    (336.16,128.45) -- (237.72,128.85) ;
\draw  [color={rgb, 255:red, 74; green, 144; blue, 226 }  ,draw opacity=1 ][line width=1.5]  (336.28,128.59) -- (336.24,134.65) -- (328.88,134.77) -- (328.92,128.71) -- cycle ;
\draw  [color={rgb, 255:red, 74; green, 144; blue, 226 }  ,draw opacity=1 ][line width=1.5]  (245.12,128.82) -- (245,134.91) -- (237.6,134.94) -- (237.72,128.85) -- cycle ;

\draw [color={rgb, 255:red, 208; green, 2; blue, 27 }  ,draw opacity=1 ][line width=2.25]  [dash pattern={on 6.75pt off 4.5pt}]  (285.51,91.91) -- (285.51,175.91) ;
\draw  [draw opacity=0][line width=1.5]  (211.17,202.33) .. controls (193.59,184.46) and (182.71,159.78) .. (182.71,132.51) .. controls (182.71,106.13) and (192.9,82.17) .. (209.49,64.45) -- (280,132.51) -- cycle ; \draw  [color={rgb, 255:red, 245; green, 166; blue, 35 }  ,draw opacity=1 ][line width=1.5]  (211.17,202.33) .. controls (193.59,184.46) and (182.71,159.78) .. (182.71,132.51) .. controls (182.71,106.13) and (192.9,82.17) .. (209.49,64.45) ;  
\draw  [draw opacity=0][line width=1.5]  (360.21,63.35) .. controls (377.56,81.45) and (388.11,106.27) .. (387.74,133.53) .. controls (387.4,159.91) and (376.89,183.74) .. (360.07,201.24) -- (290.47,132.24) -- cycle ; \draw  [color={rgb, 255:red, 189; green, 16; blue, 224 }  ,draw opacity=1 ][line width=1.5]  (360.21,63.35) .. controls (377.56,81.45) and (388.11,106.27) .. (387.74,133.53) .. controls (387.4,159.91) and (376.89,183.74) .. (360.07,201.24) ;  
\draw [color={rgb, 255:red, 208; green, 2; blue, 27 }  ,draw opacity=1 ][line width=2.25]    (228,91) -- (236.76,82.08) ;
\draw [color={rgb, 255:red, 208; green, 2; blue, 27 }  ,draw opacity=1 ][line width=2.25]    (337,188) -- (345.76,179.08) ;
\draw [color={rgb, 255:red, 208; green, 2; blue, 27 }  ,draw opacity=1 ][line width=2.25]    (229,177) -- (236.76,185.08) ;
\draw [color={rgb, 255:red, 208; green, 2; blue, 27 }  ,draw opacity=1 ][line width=2.25]    (337,81) -- (344.76,89.08) ;

\draw (278.93,55.01) node [anchor=north west][inner sep=0.75pt]    {$\textcolor[rgb]{0.31,0.89,0.76}{L_{o}}$};
\draw (204.56,124.26) node [anchor=north west][inner sep=0.75pt]    {$\textcolor[rgb]{0.31,0.89,0.76}{L_{o}}$};
\draw (280.56,193.26) node [anchor=north west][inner sep=0.75pt]    {$\textcolor[rgb]{0.31,0.89,0.76}{L_{o}}$};
\draw (353.56,124.26) node [anchor=north west][inner sep=0.75pt]    {$\textcolor[rgb]{0.31,0.89,0.76}{L_{o}}$};
\draw (292.56,100.26) node [anchor=north west][inner sep=0.75pt]    {$\textcolor[rgb]{0.82,0.01,0.11}{L}\textcolor[rgb]{0.82,0.01,0.11}{_{AB}}$};
\draw (298,135.4) node [anchor=north west][inner sep=0.75pt]  [color={rgb, 255:red, 74; green, 144; blue, 226 }  ,opacity=1 ]  {$L_{CD}$};
\draw (155.39,113.64) node [anchor=north west][inner sep=0.75pt]    {$A$};
\draw (401.39,113.64) node [anchor=north west][inner sep=0.75pt]    {$B$};
\draw (258,274.4) node [anchor=north west][inner sep=0.75pt]    {$\mathcal{V}_{0,4}^{o}(\textcolor[rgb]{0.31,0.89,0.76}{L_{o}})$};
\draw (278.39,4.64) node [anchor=north west][inner sep=0.75pt]    {$C$};
\draw (279.39,245.4) node [anchor=north west][inner sep=0.75pt]    {$D$};

\end{tikzpicture}
\par\end{centering}
\centering{}\caption{\label{fig:connection}Based on the hyperbolic four-string vertices,
there exists a minimal value for $L_{AB}$ or $L_{CD}$. It gives
the scattering distance. It also indicates the transition point between
the top-down cyan geodesics and the left-hand cyan geodesics.}
\end{figure}

On the other hand, let us examine how the geometric BV master equation
imposes an additional constraint on the hyperbolic quadrilateral shown
on the right-hand side of figure (\ref{fig:4 corres}), leading to
the determination of the critical length. We begin by transforming
the four open strings scattering (shown in figure (\ref{fig:4 open}))
into hyperbolic four-string vertices. The entire process can be seen
as the joining and splitting of two segments of three-open string
vertices, namely $\left\{ \mathcal{V}_{0,3}^{o}\left(L_{o}\right),\mathcal{V}_{0,3}^{o}\left(L_{o}\right)\right\} $,
or $\mathcal{V}_{0,4}^{o}\left(L_{o}\right)$. To match this process
with the phase transition picture of the the EWCS, $\mathcal{V}_{0,4}^{o}\left(L_{o}\right)$
should have a large $L_{o}$. Therefore, our goal is to calculate
$L_{AB}$ and $L_{CD}$ for the string vertices $\mathcal{V}_{0,4}^{o}\left(L_{o}\right)$,
as illustrated in figure (\ref{fig:connection}).

\noindent Now, let us revisit the requirements of the geometric BV
master equation, as derived in the previous section:
\begin{center}
\noindent\fbox{\begin{minipage}[t]{1\columnwidth - 2\fboxsep - 2\fboxrule}%
The open string vertices $\mathcal{V}_{g,n}^{o}\left(L_{o}\right)$
of width $L_{o}>0$ and $\mathrm{sys}\left[\tilde{\mathcal{V}}\left(L_{o}\right)\right]\geq L_{o}$
satisfy the classical geometric master equation.%
\end{minipage}}
\par\end{center}

\noindent To satisfy the classical BV equation (where there are no
loops in the correspondence), the four cyan boundaries must all be
of equal length, namely $L_{1}=L_{2}=L_{3}=L_{4}=L_{o}$. In hyperbolic
geometry, this requirement gives $L_{AB}=L_{CD}$. Recall the hexagon:
when the four outer red boundaries approach zero, the two inner geodesics
$L_{AB}$ and $L_{CD}$ satisfy the following equality (\cite{Buser},
Theorem 2.3.1 (i)):

\begin{equation}
\sinh\left(\frac{L_{AB}}{2}\right)\sinh\left(\frac{L_{CD}}{2}\right)=1.\label{eq:D1D2}
\end{equation}

\noindent This equation provides the minimal critical length, which
is given by $L_{AB}^{min}=L_{CD}^{min}=L_{*}$:

\begin{equation}
L_{*}=2\mathrm{arc}\sinh\left(1\right)=2\log\left(1+\sqrt{2}\right)=\log\left(3+2\sqrt{2}\right).
\end{equation}

Thus, the geometric BV master equation imposes an additional constraint
on the hyperbolic quadrilateral, resulting in:

\begin{equation}
L_{1}=L_{2}=L_{3}=L_{4}=L_{o},\qquad\mathrm{and}\qquad L_{AB}^{min}=L_{CD}^{min}=L_{*}=\log\left(3+2\sqrt{2}\right),\label{eq:BV constraint}
\end{equation}

\noindent in the quadrilateral. 

\vspace*{2.0ex}

Finally, by comparing the results from mutual information (\ref{eq:mutual constraint})
and the geometric BV master equation (\ref{eq:BV constraint}), we
can confirm that the hyperbolic quadrilaterals in both theories, as
shown in figure (\ref{fig:4 corres}), are identical. Thus, we arrive
at the following:

\begin{equation}
E_{AB}^{min}=L_{AB}^{min}=L_{*},\qquad\mathrm{and}\qquad E_{CD}^{min}=L_{CD}^{min}=L_{*}.
\end{equation}

\noindent This result implies that the correspondence between the
EWCS and string vertices is not only a coincidence arising from hyperbolic
geometry but is obtained by the extra relation between mutual information
and geometric BV master equation of two theories. Thus, the EWCS and
string vertices are directly related based on the observation:

\begin{equation}
E_{W}^{min}\left(A:B/C:D\right)=\frac{E_{AB/CD}^{min}}{4G_{N}^{\left(3\right)}}=\frac{L_{*}}{4G_{N}^{\left(3\right)}}=\frac{c}{6}\log\left(3+2\sqrt{2}\right),
\end{equation}

\noindent where the last equivalence follows from the fact that $c=3/2G_{N}^{(3)}$.
Therefore, the following connections can be established:
\begin{itemize}
\item Open string scattering corresponds to the evolution of the entanglement
wedge.
\item The open string scattering distance $L_{AB/CD}^{min}$ relates to
the EWCS $E_{W}^{min}\left(A:B/C:D\right)$.
\end{itemize}

\subsection{Open-closed string scattering and reflected entanglement wedge evolution}

\begin{figure}[h]
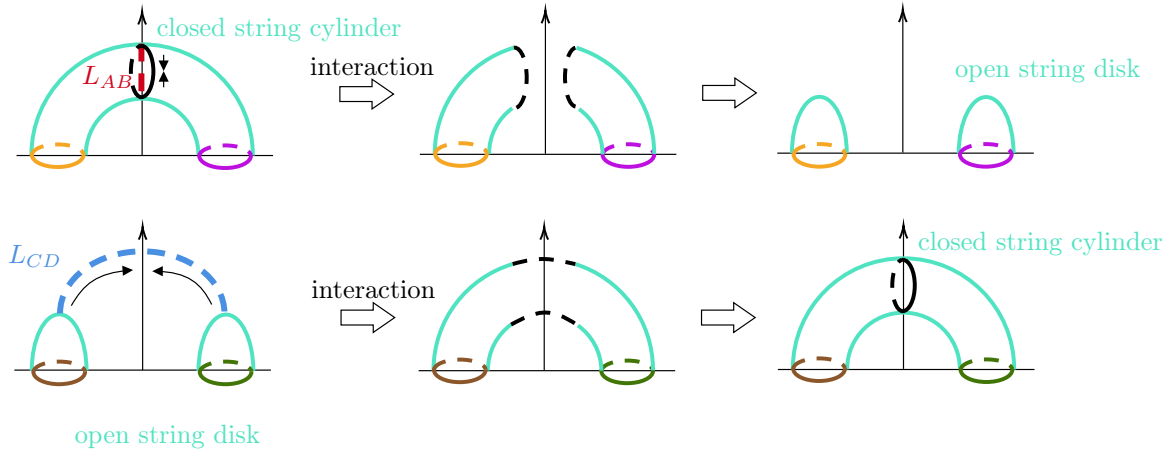

\begin{centering}

\tikzset{every picture/.style={line width=0.45pt}} 


\par\end{centering}
\caption{\label{fig:4 open-close}These pictures illustrate the disk-disk interactions.
In the first row, the center of the closed string cylinder contracts
and then breaks into two open string disks. The second row shows how
the two open string disks interact with each other and transform into
a closed string cylinder.}
\end{figure}

In addition to the connection between entanglement wedge and open
string interaction, there is also a connection to the reflected entropy.
As expected, the closed string vertices enter the story. In this subsection,
we wish to include the reflected entropy in our framework. Let us
recall figure (\ref{fig:2 reflected surface}), which demonstrates
the reflected geometry obtained through copying and gluing the original
entanglement wedge. The EWCS is thus doubled and called the reflected
surface. When the reflected surface reaches its minimal value $S_{R}^{min}\left(A,B\right)$,
the corresponding reflected geometry vanishes and transforms into
two separate RT surfaces. This process can be seen as the interaction
between two open string disks. During this process, the two open string
disks interact with each other, transforming into the closed string
cylinder, as shown in figure (\ref{fig:4 open-close}).

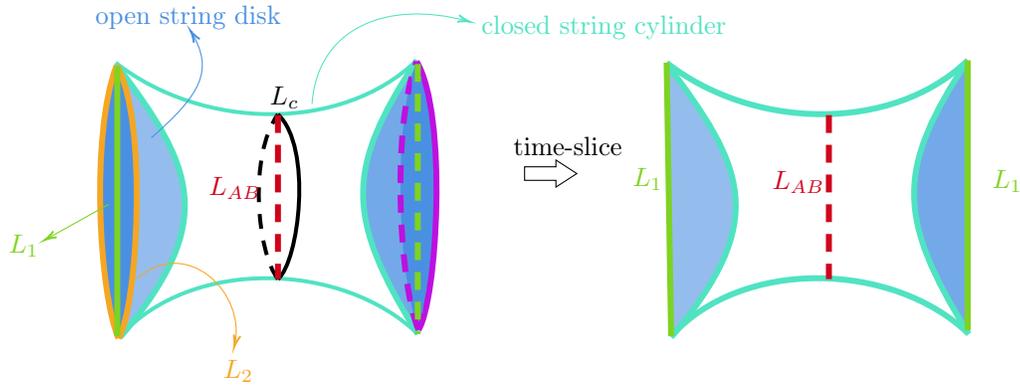
\begin{figure}[h]
\begin{centering}

\tikzset{every picture/.style={line width=0.40pt}} 

\begin{tikzpicture}[x=0.40pt,y=0.40pt,yscale=-1,xscale=1]

\draw [color={rgb, 255:red, 80; green, 227; blue, 194 }  ,draw opacity=1 ][line width=1.5]    (165.31,115.87) .. controls (246.76,179.71) and (378.76,182.71) .. (450.03,113.63) ;
\draw [color={rgb, 255:red, 80; green, 227; blue, 194 }  ,draw opacity=1 ][line width=1.5]    (169.56,374.71) .. controls (236.51,300.08) and (379.76,303.71) .. (451,372.61) ;
\draw [color={rgb, 255:red, 80; green, 227; blue, 194 }  ,draw opacity=1 ][fill={rgb, 255:red, 74; green, 144; blue, 226 }  ,fill opacity=0.59 ][line width=2.25]    (165.31,116.87) .. controls (238.76,213.71) and (263.51,265.08) .. (169.56,375.71) ;
\draw [color={rgb, 255:red, 74; green, 144; blue, 226 }  ,draw opacity=1 ][line width=3.75]    (166.24,133.22) -- (167.5,360.46) ;
\draw [color={rgb, 255:red, 80; green, 227; blue, 194 }  ,draw opacity=1 ][fill={rgb, 255:red, 74; green, 144; blue, 226 }  ,fill opacity=0.77 ][line width=2.25]    (451,372.61) .. controls (381.76,261.71) and (382.76,215.71) .. (450.03,113.63) ;
\draw [color={rgb, 255:red, 189; green, 16; blue, 224 }  ,draw opacity=1 ][fill={rgb, 255:red, 74; green, 144; blue, 226 }  ,fill opacity=1 ][line width=2.25]    (449.9,114.72) .. controls (479.67,164.99) and (469.31,340.92) .. (451.68,368.72) ;
\draw [color={rgb, 255:red, 189; green, 16; blue, 224 }  ,draw opacity=1 ][fill={rgb, 255:red, 74; green, 144; blue, 226 }  ,fill opacity=1 ][line width=2.25]  [dash pattern={on 6.75pt off 4.5pt}]  (451.68,368.72) .. controls (433.56,326.51) and (425.39,178.4) .. (449.9,114.72) ;

\draw [color={rgb, 255:red, 74; green, 144; blue, 226 }  ,draw opacity=1 ][line width=3.75]    (450,129) -- (451.24,352.71) ;
\draw [line width=1.5]  [dash pattern={on 5.63pt off 4.5pt}]  (318.76,320.71) .. controls (298.76,294.71) and (290.76,203.71) .. (318.76,164.71) ;
\draw [line width=1.5]    (318.76,164.71) .. controls (351.76,195.71) and (338.76,303.71) .. (318.76,320.71) ;
\draw [color={rgb, 255:red, 74; green, 144; blue, 226 }  ,draw opacity=1 ]   (199,187) .. controls (245.06,146.57) and (258.83,123.65) .. (237.25,87.61) ;
\draw [shift={(236.24,85.95)}, rotate = 58.13] [color={rgb, 255:red, 74; green, 144; blue, 226 }  ,draw opacity=1 ][line width=0.75]    (10.93,-3.29) .. controls (6.95,-1.4) and (3.31,-0.3) .. (0,0) .. controls (3.31,0.3) and (6.95,1.4) .. (10.93,3.29)   ;
\draw [color={rgb, 255:red, 80; green, 227; blue, 194 }  ,draw opacity=1 ]   (351,155) .. controls (355.74,84.31) and (420.11,55.24) .. (498.58,76.63) ;
\draw [shift={(499.76,76.95)}, rotate = 195.56] [color={rgb, 255:red, 80; green, 227; blue, 194 }  ,draw opacity=1 ][line width=0.75]    (10.93,-3.29) .. controls (6.95,-1.4) and (3.31,-0.3) .. (0,0) .. controls (3.31,0.3) and (6.95,1.4) .. (10.93,3.29)   ;
\draw [color={rgb, 255:red, 245; green, 166; blue, 35 }  ,draw opacity=1 ][fill={rgb, 255:red, 74; green, 144; blue, 226 }  ,fill opacity=1 ][line width=2.25]    (167.95,376.72) .. controls (149.54,333.84) and (141.25,183.39) .. (166.15,118.71) ;
\draw [color={rgb, 255:red, 245; green, 166; blue, 35 }  ,draw opacity=1 ][fill={rgb, 255:red, 74; green, 144; blue, 226 }  ,fill opacity=1 ][line width=2.25]    (166.15,118.71) .. controls (196.38,169.78) and (185.86,348.48) .. (167.95,376.72) ;
\draw [color={rgb, 255:red, 126; green, 211; blue, 33 }  ,draw opacity=1 ][line width=2.25]    (166.87,119.22) -- (166.87,374.71) ;
\draw [color={rgb, 255:red, 208; green, 2; blue, 27 }  ,draw opacity=1 ][line width=2.25]  [dash pattern={on 6.75pt off 4.5pt}]  (318.76,165.48) -- (318.76,320.71) ;
\draw [color={rgb, 255:red, 126; green, 211; blue, 33 }  ,draw opacity=1 ][line width=2.25]  [dash pattern={on 6.75pt off 4.5pt}]  (451,117.11) -- (451,372.61) ;
\draw [color={rgb, 255:red, 80; green, 227; blue, 194 }  ,draw opacity=1 ][line width=2.25]    (686.31,115.87) .. controls (767.76,179.71) and (899.76,182.71) .. (971.03,113.63) ;
\draw [color={rgb, 255:red, 80; green, 227; blue, 194 }  ,draw opacity=1 ][line width=2.25]    (690.56,374.71) .. controls (757.51,300.08) and (900.76,303.71) .. (972,372.61) ;
\draw [color={rgb, 255:red, 80; green, 227; blue, 194 }  ,draw opacity=1 ][fill={rgb, 255:red, 74; green, 144; blue, 226 }  ,fill opacity=0.59 ][line width=2.25]    (686.31,116.87) .. controls (759.76,213.71) and (784.51,265.08) .. (690.56,375.71) ;
\draw [color={rgb, 255:red, 80; green, 227; blue, 194 }  ,draw opacity=1 ][fill={rgb, 255:red, 74; green, 144; blue, 226 }  ,fill opacity=0.77 ][line width=2.25]    (972,372.61) .. controls (902.76,261.71) and (903.76,215.71) .. (971.03,113.63) ;
\draw [color={rgb, 255:red, 126; green, 211; blue, 33 }  ,draw opacity=1 ][line width=2.25]    (686.31,115.87) -- (690.56,374.71) ;
\draw [color={rgb, 255:red, 208; green, 2; blue, 27 }  ,draw opacity=1 ][line width=2.25]  [dash pattern={on 6.75pt off 4.5pt}]  (839.76,165.48) -- (839.76,320.71) ;
\draw [color={rgb, 255:red, 126; green, 211; blue, 33 }  ,draw opacity=1 ][line width=2.25]    (971.03,113.63) -- (971.03,369.13) ;
\draw   (552,214.4) -- (581.64,214.4) -- (581.64,208) -- (601.4,220.8) -- (581.64,233.6) -- (581.64,227.2) -- (552,227.2) -- cycle ;
\draw [color={rgb, 255:red, 245; green, 166; blue, 35 }  ,draw opacity=1 ]   (186,320) .. controls (232.29,279.36) and (289.25,333.5) .. (277.77,383.1) ;
\draw [shift={(277.4,384.6)}, rotate = 284.57] [color={rgb, 255:red, 245; green, 166; blue, 35 }  ,draw opacity=1 ][line width=0.75]    (10.93,-3.29) .. controls (6.95,-1.4) and (3.31,-0.3) .. (0,0) .. controls (3.31,0.3) and (6.95,1.4) .. (10.93,3.29)   ;
\draw [color={rgb, 255:red, 126; green, 211; blue, 33 }  ,draw opacity=1 ]   (158.87,249.84) -- (100.15,282.63) ;
\draw [shift={(98.4,283.6)}, rotate = 330.83] [color={rgb, 255:red, 126; green, 211; blue, 33 }  ,draw opacity=1 ][line width=0.75]    (10.93,-3.29) .. controls (6.95,-1.4) and (3.31,-0.3) .. (0,0) .. controls (3.31,0.3) and (6.95,1.4) .. (10.93,3.29)   ;

\draw (144,60) node [anchor=north west][inner sep=0.75pt]  [color={rgb, 255:red, 74; green, 144; blue, 226 }  ,opacity=1 ] [align=left] {open string disk};
\draw (509,69) node [anchor=north west][inner sep=0.75pt]  [color={rgb, 255:red, 80; green, 227; blue, 194 }  ,opacity=1 ] [align=left] {closed string cylinder};
\draw (539,184) node [anchor=north west][inner sep=0.75pt]   [align=left] {time-slice};
\draw (265,394.4) node [anchor=north west][inner sep=0.75pt]    {$\textcolor[rgb]{0.96,0.65,0.14}{L}\textcolor[rgb]{0.96,0.65,0.14}{_{2}}$};
\draw (62,276.4) node [anchor=north west][inner sep=0.75pt]    {$\textcolor[rgb]{0.49,0.83,0.13}{L}\textcolor[rgb]{0.49,0.83,0.13}{_{1}}$};
\draw (251,221.4) node [anchor=north west][inner sep=0.75pt]  [color={rgb, 255:red, 208; green, 2; blue, 27 }  ,opacity=1 ]  {$\textcolor[rgb]{0.82,0.01,0.11}{L}\textcolor[rgb]{0.82,0.01,0.11}{_{AB}}$};
\draw (784,216.4) node [anchor=north west][inner sep=0.75pt]  [color={rgb, 255:red, 208; green, 2; blue, 27 }  ,opacity=1 ]  {$\textcolor[rgb]{0.82,0.01,0.11}{L}\textcolor[rgb]{0.82,0.01,0.11}{_{AB}}$};
\draw (309,136.4) node [anchor=north west][inner sep=0.75pt]    {$L_{c}$};
\draw (652,213.4) node [anchor=north west][inner sep=0.75pt]    {$\textcolor[rgb]{0.49,0.83,0.13}{L}\textcolor[rgb]{0.49,0.83,0.13}{_{1}}$};
\draw (992,215.4) node [anchor=north west][inner sep=0.75pt]    {$\textcolor[rgb]{0.49,0.83,0.13}{L}\textcolor[rgb]{0.49,0.83,0.13}{_{1}}$};

\end{tikzpicture}
\par\end{centering}
\caption{\label{fig:open-cloesd vertex}The entire process of disk-disk interaction
is illustrated in the left-hand side picture. Owing to open-closed
string duality, taking a slice of the left-hand side picture results
in the right-hand side, representing open string scattering. The orange
and purple circles on the boundary have lengths $L_{2}$, and the
black circle in the middle has a length $L_{c}$. If they are perfect
circles, the green and red lines, with lengths $L_{1}$ and $L_{AB}$
respectively, represent their diameters.}

\end{figure}

\noindent The entire process, which can be seen as the joining of
two segments of open string disks and yielding a closed string cylinder,
is illustrated in figure (\ref{fig:open-cloesd vertex}).

However, there are no well-defined open-closed string vertices to
describe this process. The reason is that, in the recent construction
of hyperbolic open-closed string vertices, the open string disk without
boundary punctures is flat. The interaction between two flat disks
results in a flat cylinder. This implies that it is difficult to establish
a clear connection between this process and the hyperbolic entanglement
wedge. To solve this problem, we can utilize the open-closed string
duality. This duality reveals that a closed string cylinder can be
identified as a one-loop of open string. In other words, we can take
the time slice of this picture, corresponding to open string scattering.
This open string scattering can be described by the previous hyperbolic
open string vertices, as shown in (\ref{fig:connection}).

Now, let us calculate the length $L_{c}$ and compare it with the
reflected entropy $S_{R}$. Before calculation, it is essential to
clarify a point. In figure (\ref{fig:open-cloesd vertex}), if the
waist cross section\footnote{In this subsection, the term 'waist cross section' refers to the cross section of the narrowest part of a hyperbolic cylinder.}
of a cylinder is a perfect circle, we will have:

\begin{equation}
L_{2}=\pi L_{1},\qquad L_{c}=\pi L_{AB}.
\end{equation}

\noindent However, if we consider this process in the Poincar$\acute{\mathrm{e}}$
disk (see figure (\ref{fig:open-cloesd limit})), there is a problem:
the corresponding boundary entangling regions $L_{2}/2$ (in reflected
entropy, this region of length $L_{2}$ is obtained by gluing two
entanglement wedges of boundary lengths $L_{2}/2$, see figure (\ref{fig:2 reflected surface}))
and $L_{1}$ are not equal. It implies that the result is inconsistent
with multipartite entanglement: when we take a slice of the reflected
entanglement wedge, it can be seen as a new entanglement wedge. The
corresponding EWCS and reflected entropy must vanish simultaneously.
In other words, they must have the same entangling regions. However,
if $L_{2}/2<L_{1}$, the EWCS $E_{W}\left(A,B\right)$ will vanish
before the reflected entropy $S_{R}\left(A,B\right)$, which corresponds
to different entangling regions. Therefore, to solve this problem,
the waist cross section of the cylinder cannot be a perfect circle,
and there is a requirement: $L_{2}/2\rightarrow L_{1}$. Then, we
will obtain the following result:

\begin{equation}
L_{c}=2L_{AB}\leq2L_{*},
\end{equation}

\noindent where $L_{CD}\geq L_{*}$, or vice versa. This result is
precisely equivalent to the minimal reflected entropy (\ref{fig:2 reflected surface})
\cite{Dutta:2019gen}. In other words, there will always be one nonvanishing
EWCS between $AB$ or $CD$, and its size is greater than

\begin{equation}
S_{R}^{min}\left(A:B/C:D\right)=\frac{c}{3}\log\left(3+2\sqrt{2}\right)=\frac{2L_{*}}{4G_{N}^{\left(3\right)}},
\end{equation}

In short, we present the following connections in this subsection:
\begin{itemize}
\item Disk-disk string scattering corresponds to the evolution of reflected
entanglement wedge.
\item Circumference of disk-disk scattering waist cross section $L_{c}$
corresponds to the reflected surface/entropy $S_{R}$.
\end{itemize}
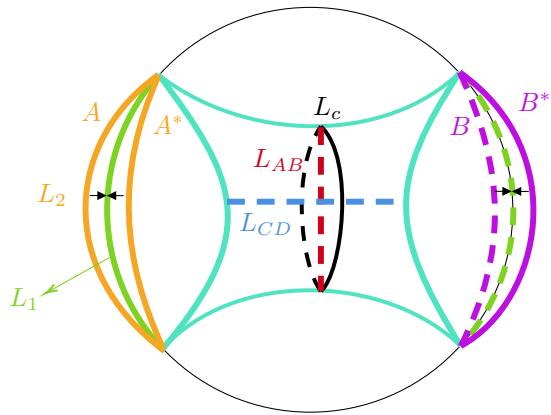
\begin{figure}[h]
\begin{centering}

\tikzset{every picture/.style={line width=0.40pt}} 

\begin{tikzpicture}[x=0.40pt,y=0.40pt,yscale=-1,xscale=1]

\draw   (176.13,238.65) .. controls (176.13,132.86) and (261.89,47.1) .. (367.68,47.1) .. controls (473.47,47.1) and (559.23,132.86) .. (559.23,238.65) .. controls (559.23,344.44) and (473.47,430.2) .. (367.68,430.2) .. controls (261.89,430.2) and (176.13,344.44) .. (176.13,238.65) -- cycle ;
\draw  [draw opacity=0][line width=2.25]  (229.81,372.35) .. controls (196.28,337.77) and (175.63,290.62) .. (175.63,238.65) .. controls (175.63,189.58) and (194.03,144.81) .. (224.31,110.87) -- (367.68,238.65) -- cycle ; \draw  [color={rgb, 255:red, 126; green, 211; blue, 33 }  ,draw opacity=1 ][line width=2.25]  (229.81,372.35) .. controls (196.28,337.77) and (175.63,290.62) .. (175.63,238.65) .. controls (175.63,189.58) and (194.03,144.81) .. (224.31,110.87) ;  
\draw  [draw opacity=0][dash pattern={on 6.75pt off 4.5pt}][line width=2.25]  (509.03,108.63) .. controls (540.55,142.88) and (559.78,188.63) .. (559.73,238.85) .. controls (559.68,288.41) and (540.86,333.57) .. (510,367.61) -- (367.68,238.65) -- cycle ; \draw  [color={rgb, 255:red, 126; green, 211; blue, 33 }  ,draw opacity=1 ][dash pattern={on 6.75pt off 4.5pt}][line width=2.25]  (509.03,108.63) .. controls (540.55,142.88) and (559.78,188.63) .. (559.73,238.85) .. controls (559.68,288.41) and (540.86,333.57) .. (510,367.61) ;  
\draw [color={rgb, 255:red, 80; green, 227; blue, 194 }  ,draw opacity=1 ][line width=1.5]    (224.31,110.87) .. controls (305.76,174.71) and (437.76,177.71) .. (509.03,108.63) ;
\draw [color={rgb, 255:red, 80; green, 227; blue, 194 }  ,draw opacity=1 ][line width=1.5]    (228.56,369.71) .. controls (295.51,295.08) and (438.76,298.71) .. (510,367.61) ;
\draw [color={rgb, 255:red, 245; green, 166; blue, 35 }  ,draw opacity=1 ][line width=2.25]    (229.81,372.35) .. controls (124.76,293.71) and (137.76,171.71) .. (224.31,110.87) ;
\draw [color={rgb, 255:red, 189; green, 16; blue, 224 }  ,draw opacity=1 ][line width=2.25]    (509.03,108.63) .. controls (615.76,183.71) and (586.76,325.71) .. (510,367.61) ;
\draw [color={rgb, 255:red, 80; green, 227; blue, 194 }  ,draw opacity=1 ][fill={rgb, 255:red, 74; green, 144; blue, 226 }  ,fill opacity=0 ][line width=2.25]    (224.31,110.87) .. controls (297.76,207.71) and (322.51,259.08) .. (228.56,369.71) ;
\draw [color={rgb, 255:red, 80; green, 227; blue, 194 }  ,draw opacity=1 ][fill={rgb, 255:red, 74; green, 144; blue, 226 }  ,fill opacity=0 ][line width=2.25]    (510,367.61) .. controls (440.76,256.71) and (441.76,210.71) .. (509.03,108.63) ;
\draw [line width=1.5]  [dash pattern={on 5.63pt off 4.5pt}]  (377.76,315.71) .. controls (357.76,289.71) and (349.76,198.71) .. (377.76,159.71) ;
\draw [line width=1.5]    (377.76,159.71) .. controls (410.76,190.71) and (397.76,298.71) .. (377.76,315.71) ;
\draw [color={rgb, 255:red, 208; green, 2; blue, 27 }  ,draw opacity=1 ][line width=2.25]  [dash pattern={on 6.75pt off 4.5pt}]  (377.76,160.48) -- (377.76,315.71) ;
\draw [color={rgb, 255:red, 245; green, 166; blue, 35 }  ,draw opacity=1 ][line width=2.25]    (228.56,369.71) .. controls (190.2,287.6) and (182.2,204.6) .. (224.31,111.87) ;
\draw [color={rgb, 255:red, 189; green, 16; blue, 224 }  ,draw opacity=1 ][line width=2.25]  [dash pattern={on 6.75pt off 4.5pt}]  (509.03,108.63) .. controls (552.2,201.6) and (554.2,279.6) .. (510,367.61) ;
\draw [color={rgb, 255:red, 126; green, 211; blue, 33 }  ,draw opacity=1 ]   (177.87,283.84) -- (119.15,316.63) ;
\draw [shift={(117.4,317.6)}, rotate = 330.83] [color={rgb, 255:red, 126; green, 211; blue, 33 }  ,draw opacity=1 ][line width=0.75]    (10.93,-3.29) .. controls (6.95,-1.4) and (3.31,-0.3) .. (0,0) .. controls (3.31,0.3) and (6.95,1.4) .. (10.93,3.29)   ;
\draw [color={rgb, 255:red, 74; green, 144; blue, 226 }  ,draw opacity=1 ][line width=2.25]  [dash pattern={on 6.75pt off 4.5pt}]  (289,231) -- (457.76,231) ;
\draw    (159.18,225.15) -- (172.27,225.21) ;
\draw [shift={(175.27,225.23)}, rotate = 180.27] [fill={rgb, 255:red, 0; green, 0; blue, 0 }  ][line width=0.08]  [draw opacity=0] (8.93,-4.29) -- (0,0) -- (8.93,4.29) -- cycle    ;
\draw    (191.27,225.3) -- (178.27,225.24) ;
\draw [shift={(175.27,225.23)}, rotate = 0.27] [fill={rgb, 255:red, 0; green, 0; blue, 0 }  ][line width=0.08]  [draw opacity=0] (8.93,-4.29) -- (0,0) -- (8.93,4.29) -- cycle    ;

\draw    (542.18,221.15) -- (555.27,221.21) ;
\draw [shift={(558.27,221.23)}, rotate = 180.27] [fill={rgb, 255:red, 0; green, 0; blue, 0 }  ][line width=0.08]  [draw opacity=0] (8.93,-4.29) -- (0,0) -- (8.93,4.29) -- cycle    ;
\draw    (574.27,221.3) -- (561.27,221.24) ;
\draw [shift={(558.27,221.23)}, rotate = 0.27] [fill={rgb, 255:red, 0; green, 0; blue, 0 }  ][line width=0.08]  [draw opacity=0] (8.93,-4.29) -- (0,0) -- (8.93,4.29) -- cycle    ;

\draw (107,211.4) node [anchor=north west][inner sep=0.75pt]    {$\textcolor[rgb]{0.96,0.65,0.14}{L}\textcolor[rgb]{0.96,0.65,0.14}{_{2}}$};
\draw (311,178.4) node [anchor=north west][inner sep=0.75pt]    {$\textcolor[rgb]{0.82,0.01,0.11}{L}\textcolor[rgb]{0.82,0.01,0.11}{_{AB}}$};
\draw (81,310.4) node [anchor=north west][inner sep=0.75pt]    {$\textcolor[rgb]{0.49,0.83,0.13}{L}\textcolor[rgb]{0.49,0.83,0.13}{_{1}}$};
\draw (298,239.4) node [anchor=north west][inner sep=0.75pt]    {$\textcolor[rgb]{0.29,0.56,0.89}{L}\textcolor[rgb]{0.29,0.56,0.89}{_{CD}}$};
\draw (368,129.4) node [anchor=north west][inner sep=0.75pt]    {$L_{c}$};
\draw (149,135.4) node [anchor=north west][inner sep=0.75pt]    {$\textcolor[rgb]{0.96,0.65,0.14}{A}$};
\draw (216,147.4) node [anchor=north west][inner sep=0.75pt]    {$\textcolor[rgb]{0.96,0.65,0.14}{A^{*}}$};
\draw (497,148.4) node [anchor=north west][inner sep=0.75pt]    {$\textcolor[rgb]{0.74,0.06,0.88}{B}$};
\draw (562,125.4) node [anchor=north west][inner sep=0.75pt]    {$\textcolor[rgb]{0.74,0.06,0.88}{B^{*}}$};

\end{tikzpicture}
\par\end{centering}
\caption{\label{fig:open-cloesd limit} Disk-disk scattering in the Poincar$\acute{\mathrm{e}}$
disk. To establish a connection between entanglement entropy, the
EWCS, and the reflected surface in this scenario, should be the bulk
duals for the bipartite entanglement entropy corresponding to the
same entangling regions $A$ and $B$. This implies that $L_{1}$
and $L_{2}/2$ must have the same value.}
\end{figure}

\section{Black hole information paradox revisited}

The black hole information paradox is a central issue in modern theoretical
physics. In recent work, Ryu and Takayanagi established a relationship
between the entanglement entropy (von Neumann entropy) of CFT$_{2}$
and the extremal surface in the bulk of AdS$_{3}$. Inspired by this
result, it is possible to compute the von Neumann entropy of gravitational
systems using the quantum extremal surface. This approach can be applied
to the von Neumann entropies of evaporating black holes and their
corresponding Hawking radiation, with results found to be consistent
with the Page curve. However, the corresponding quantum state for
the entanglement entropy of Hawking radiation remains unclear, which
remains one of the central challenges of the information paradox \cite{Almheiri:2020cfm}.

Susskind and Uglum proposed that black hole entropy could be derived
from the closed string worldsheet, where the closed string sphere
is punctured twice by the black hole horizon, as illustrated in the
left image of figure (\ref{fig:SU}) \cite{Susskind:1994sm}. This
picture can also be realized from the open string perspective, where
the open string endpoints are frozen on the horizon. This framework
provides a statistical interpretation of the entanglement entropy,
as a slice of the closed string (the sphere) crossing the horizon
can only be observed as an open string by a Rindler observer outside
the event horizon (right image of figure (\ref{fig:SU})). Thus, Susskind
and Uglum\textquoteright s conjecture suggests that both black hole
entropy and entanglement entropy can be calculated using string theory,
allowing for the natural determination of their corresponding quantum
states. Recently, Ahmadain and Wall provided a proof of the closed
string perspective of the conjecture using the sphere diagram in off-shell
closed string theory, yielding the result $A/4G_{N}$ \cite{Ahmadain:2022tew,Ahmadain:2022eso}.
However, proving the open string perspective of the conjecture remains
a challenge, as the replica trick introduces a conical singularity
that breaks the conformal symmetry of the string worldsheet.

The goal of this paper is to explore the relationship between the
open string worldsheet and the Ryu-Takayanagi surface. If such a relationship
exists, it could serve as a guiding principle for proving the open
string perspective of Susskind and Uglum\textquoteright s conjecture
in the framework of bosonic string theory, ultimately allowing us
to compute the entanglement entropy. With this framework in place,
we can revisit the black hole information paradox from a new angle.

\begin{figure}[h]
\begin{centering}

\tikzset{every picture/.style={line width=0.65pt}} 

\begin{tikzpicture}[x=0.65pt,y=0.65pt,yscale=-1,xscale=1]

\draw [line width=1.5]    (146,67.08) -- (146,24.08) ;
\draw [shift={(146,21.08)}, rotate = 90] [color={rgb, 255:red, 0; green, 0; blue, 0 }  ][line width=1.5]    (14.21,-4.28) .. controls (9.04,-1.82) and (4.3,-0.39) .. (0,0) .. controls (4.3,0.39) and (9.04,1.82) .. (14.21,4.28)   ;
\draw   (91.54,120.04) .. controls (91.54,89.96) and (115.92,65.58) .. (146,65.58) .. controls (176.08,65.58) and (200.46,89.96) .. (200.46,120.04) .. controls (200.46,150.12) and (176.08,174.5) .. (146,174.5) .. controls (115.92,174.5) and (91.54,150.12) .. (91.54,120.04) -- cycle ;
\draw [line width=1.5]  [dash pattern={on 5.63pt off 4.5pt}]  (146,67.08) -- (146,176.08) ;
\draw [line width=1.5]    (146,176.08) -- (146,215.08) ;
\draw [color={rgb, 255:red, 74; green, 144; blue, 226 }  ,draw opacity=1 ][line width=1.5]    (91.54,120.58) .. controls (96.76,132.08) and (190.76,134.08) .. (200.46,120.58) ;
\draw [color={rgb, 255:red, 74; green, 144; blue, 226 }  ,draw opacity=1 ][line width=1.5]  [dash pattern={on 5.63pt off 4.5pt}]  (91.54,120.58) .. controls (112.76,104.08) and (198.76,113.08) .. (200.46,120.58) ;

\draw [color={rgb, 255:red, 74; green, 144; blue, 226 }  ,draw opacity=1 ][line width=1.5]    (108.54,157.49) .. controls (112.14,165.43) and (176.98,166.8) .. (183.67,157.49) ;
\draw [color={rgb, 255:red, 74; green, 144; blue, 226 }  ,draw opacity=1 ][line width=1.5]  [dash pattern={on 5.63pt off 4.5pt}]  (108.54,157.49) .. controls (123.18,146.11) and (182.5,152.32) .. (183.67,157.49) ;

\draw [color={rgb, 255:red, 74; green, 144; blue, 226 }  ,draw opacity=1 ][line width=1.5]    (108.54,82.49) .. controls (112.14,90.43) and (176.98,91.8) .. (183.67,82.49) ;
\draw [color={rgb, 255:red, 74; green, 144; blue, 226 }  ,draw opacity=1 ][line width=1.5]  [dash pattern={on 5.63pt off 4.5pt}]  (108.54,82.49) .. controls (123.18,71.11) and (182.5,77.32) .. (183.67,82.49) ;

\draw [line width=1.5]    (439,67.08) -- (439,24.08) ;
\draw [shift={(439,21.08)}, rotate = 90] [color={rgb, 255:red, 0; green, 0; blue, 0 }  ][line width=1.5]    (14.21,-4.28) .. controls (9.04,-1.82) and (4.3,-0.39) .. (0,0) .. controls (4.3,0.39) and (9.04,1.82) .. (14.21,4.28)   ;
\draw   (384.54,120.04) .. controls (384.54,89.96) and (408.92,65.58) .. (439,65.58) .. controls (469.08,65.58) and (493.46,89.96) .. (493.46,120.04) .. controls (493.46,150.12) and (469.08,174.5) .. (439,174.5) .. controls (408.92,174.5) and (384.54,150.12) .. (384.54,120.04) -- cycle ;
\draw [line width=1.5]  [dash pattern={on 5.63pt off 4.5pt}]  (439,67.08) -- (439,176.08) ;
\draw [line width=1.5]    (439,176.08) -- (439,215.08) ;
\draw [color={rgb, 255:red, 208; green, 2; blue, 27 }  ,draw opacity=1 ][line width=1.5]    (438.83,174.01) .. controls (450.31,169) and (452.24,76.9) .. (439,67.08) ;
\draw [color={rgb, 255:red, 208; green, 2; blue, 27 }  ,draw opacity=1 ][line width=1.5]    (438.83,174.01) .. controls (422.79,152.72) and (431.54,67.16) .. (439,65.58) ;
\draw [color={rgb, 255:red, 208; green, 2; blue, 27 }  ,draw opacity=1 ][line width=1.5]    (438.12,173.37) .. controls (479.64,168.81) and (488.23,75.71) .. (439.72,65.37) ;
\draw [color={rgb, 255:red, 208; green, 2; blue, 27 }  ,draw opacity=1 ][line width=1.5]    (438.12,173.37) .. controls (378.97,151.44) and (412.67,66.65) .. (439.72,65.37) ;

\draw (52,243) node [anchor=north west][inner sep=0.75pt]   [align=left] { \textbf{Closed string perspective}};
\draw (357,243) node [anchor=north west][inner sep=0.75pt]   [align=left] { \textbf{Open string perspective}};
\draw (169,24) node [anchor=north west][inner sep=0.75pt]   [align=left] {Horizon};
\draw (464,23) node [anchor=north west][inner sep=0.75pt]   [align=left] {Horizon};

\end{tikzpicture}
\par\end{centering}
\caption{\label{fig:SU}The left image of the figure illustrates the closed
string perspective of Susskind and Uglum's setup. The blue circle
represents the closed string, whose initial and final states interact
with the horizon. On the other hand, the right image depicts the open
string, as shown by the red lines. The endpoints of the open string
are fixed on the horizon. The target space can be factorized into
two Hilbert spaces, describing the open strings inside and outside
the event horizon: $\mathcal{H}\subseteq\mathcal{H}_{out}\otimes\mathcal{H}_{in}$,
and therefore gives the statistical interpretation of entanglement
entropy \cite{Ahmadain:2022eso}.}
\end{figure}

In summary, the key issue is how to apply the replica trick to define
the entanglement entropy of the open string worldsheet. Let us now
provide a preliminary discussion on how to apply our correspondence
to this task. In our recent work \cite{Jiang:2024akz}, we proposed
an alternative approach to purification in CFT$_{2}$. Rather than
purifying the mixed state by introducing auxiliary systems, our method
involves subtracting the undetectable regions of the system, as illustrated
in the left image of figure (\ref{fig:SU app}). The remaining regions,
which are disjoint but mutually complementary, form a pure entangled
state $\psi_{AB}$, satisfying $A^{C}=B$. Using the entanglement
Hamiltonian, we computed the von Neumann entropy $S_{\mathrm{vN}}\left(A:B\right)$
for this configuration and demonstrated that $S_{\mathrm{vN}}\left(A:B\right)=E_{W}\left(A:B\right)$
on a time-slice of bulk AdS$_{3}$ (middle image of figure (\ref{fig:SU app})).
This result is well-defined in the framework of the replica trick.
Based on the correspondence between the entanglement wedge (middle
image of figure (\ref{fig:SU app})) and string worldsheet (right
image of figure (\ref{fig:SU app})), it becomes straightforward to
apply the same definition to study the open string entanglement entropy,
as depicted in the right image of figure (\ref{fig:SU app}).

\begin{figure}[H]
\begin{centering}

\tikzset{every picture/.style={line width=0.55pt}} 

\begin{tikzpicture}[x=0.55pt,y=0.55pt,yscale=-1,xscale=1]

\draw  [fill={rgb, 255:red, 74; green, 144; blue, 226 }  ,fill opacity=0.3 ] (63.89,135.94) .. controls (63.89,92.66) and (98.97,57.58) .. (142.25,57.58) .. controls (185.52,57.58) and (220.61,92.66) .. (220.61,135.94) .. controls (220.61,179.21) and (185.52,214.29) .. (142.25,214.29) .. controls (98.97,214.29) and (63.89,179.21) .. (63.89,135.94) -- cycle ;
\draw  [fill={rgb, 255:red, 255; green, 255; blue, 255 }  ,fill opacity=1 ] (95.5,135.94) .. controls (95.5,110.12) and (116.43,89.18) .. (142.25,89.18) .. controls (168.07,89.18) and (189,110.12) .. (189,135.94) .. controls (189,161.76) and (168.07,182.69) .. (142.25,182.69) .. controls (116.43,182.69) and (95.5,161.76) .. (95.5,135.94) -- cycle ;
\draw    (27.48,140.07) -- (269.73,140.07) ;
\draw [shift={(271.73,140.07)}, rotate = 180] [color={rgb, 255:red, 0; green, 0; blue, 0 }  ][line width=0.75]    (10.93,-3.29) .. controls (6.95,-1.4) and (3.31,-0.3) .. (0,0) .. controls (3.31,0.3) and (6.95,1.4) .. (10.93,3.29)   ;
\draw    (142.25,252.45) -- (142.25,13.08) ;
\draw [shift={(142.25,11.08)}, rotate = 90] [color={rgb, 255:red, 0; green, 0; blue, 0 }  ][line width=0.75]    (10.93,-3.29) .. controls (6.95,-1.4) and (3.31,-0.3) .. (0,0) .. controls (3.31,0.3) and (6.95,1.4) .. (10.93,3.29)   ;
\draw [color={rgb, 255:red, 245; green, 166; blue, 35 }  ,draw opacity=1 ][line width=2.25]    (63.89,139.68) -- (95.68,139.68) ;
\draw [color={rgb, 255:red, 208; green, 2; blue, 27 }  ,draw opacity=1 ][line width=2.25]    (95.68,139.68) -- (189,139.68) ;
\draw [color={rgb, 255:red, 126; green, 211; blue, 33 }  ,draw opacity=1 ][line width=2.25]    (189,139.68) -- (221.08,139.68) ;
\draw    (142.34,139.68) -- (104.07,70.85) ;
\draw [shift={(103.09,69.1)}, rotate = 60.92] [color={rgb, 255:red, 0; green, 0; blue, 0 }  ][line width=0.75]    (10.93,-3.29) .. controls (6.95,-1.4) and (3.31,-0.3) .. (0,0) .. controls (3.31,0.3) and (6.95,1.4) .. (10.93,3.29)   ;
\draw    (142.34,139.68) -- (161.22,98.06) ;
\draw [shift={(162.05,96.24)}, rotate = 114.4] [color={rgb, 255:red, 0; green, 0; blue, 0 }  ][line width=0.75]    (10.93,-3.29) .. controls (6.95,-1.4) and (3.31,-0.3) .. (0,0) .. controls (3.31,0.3) and (6.95,1.4) .. (10.93,3.29)   ;
\draw   (292,138.39) -- (307.12,138.39) -- (307.12,133.08) -- (317.2,143.7) -- (307.12,154.32) -- (307.12,149.01) -- (292,149.01) -- cycle ;
\draw    (450.54,145.32) -- (450.54,15.45) ;
\draw [shift={(450.54,13.45)}, rotate = 90] [color={rgb, 255:red, 0; green, 0; blue, 0 }  ][line width=0.75]    (10.93,-3.29) .. controls (6.95,-1.4) and (3.31,-0.3) .. (0,0) .. controls (3.31,0.3) and (6.95,1.4) .. (10.93,3.29)   ;
\draw    (335.21,145.86) -- (570.81,145.86) ;
\draw  [draw opacity=0][line width=1.5]  (348.87,145.75) .. controls (348.87,145.58) and (348.87,145.41) .. (348.87,145.24) .. controls (348.92,89.09) and (394.48,43.61) .. (450.63,43.66) .. controls (506.78,43.71) and (552.25,89.26) .. (552.21,145.41) .. controls (552.21,145.56) and (552.2,145.71) .. (552.2,145.86) -- (450.54,145.32) -- cycle ; \draw  [color={rgb, 255:red, 80; green, 227; blue, 194 }  ,draw opacity=1 ][line width=1.5]  (348.87,145.75) .. controls (348.87,145.58) and (348.87,145.41) .. (348.87,145.24) .. controls (348.92,89.09) and (394.48,43.61) .. (450.63,43.66) .. controls (506.78,43.71) and (552.25,89.26) .. (552.21,145.41) .. controls (552.21,145.56) and (552.2,145.71) .. (552.2,145.86) ;  
\draw [color={rgb, 255:red, 245; green, 166; blue, 35 }  ,draw opacity=1 ][line width=1.5]    (348.87,145.75) -- (398.82,145.75) ;
\draw [color={rgb, 255:red, 144; green, 19; blue, 254 }  ,draw opacity=1 ][line width=1.5]    (502.25,145.86) -- (552.2,145.86) ;
\draw [color={rgb, 255:red, 208; green, 2; blue, 27 }  ,draw opacity=1 ][line width=2.25]  [dash pattern={on 6.75pt off 4.5pt}]  (450.42,43.27) -- (450.42,93.58) ;
\draw  [draw opacity=0][line width=1.5]  (552.2,145.86) .. controls (552.28,146.68) and (552.32,147.5) .. (552.32,148.33) .. controls (552.35,176.61) and (507,199.59) .. (451.02,199.64) .. controls (395.04,199.7) and (349.63,176.82) .. (349.6,148.54) .. controls (349.6,147.2) and (349.7,145.87) .. (349.9,144.56) -- (450.96,148.44) -- cycle ; \draw  [color={rgb, 255:red, 74; green, 144; blue, 226 }  ,draw opacity=1 ][line width=1.5]  (552.2,145.86) .. controls (552.28,146.68) and (552.32,147.5) .. (552.32,148.33) .. controls (552.35,176.61) and (507,199.59) .. (451.02,199.64) .. controls (395.04,199.7) and (349.63,176.82) .. (349.6,148.54) .. controls (349.6,147.2) and (349.7,145.87) .. (349.9,144.56) ;  
\draw  [draw opacity=0][dash pattern={on 5.63pt off 4.5pt}][line width=1.5]  (502.82,146.42) .. controls (502.87,146.84) and (502.89,147.26) .. (502.89,147.69) .. controls (502.91,161.35) and (479.58,172.45) .. (450.8,172.48) .. controls (422.01,172.51) and (398.66,161.46) .. (398.65,147.8) .. controls (398.65,147.11) and (398.71,146.43) .. (398.82,145.75) -- (450.77,147.74) -- cycle ; \draw  [color={rgb, 255:red, 74; green, 144; blue, 226 }  ,draw opacity=1 ][dash pattern={on 5.63pt off 4.5pt}][line width=1.5]  (502.82,146.42) .. controls (502.87,146.84) and (502.89,147.26) .. (502.89,147.69) .. controls (502.91,161.35) and (479.58,172.45) .. (450.8,172.48) .. controls (422.01,172.51) and (398.66,161.46) .. (398.65,147.8) .. controls (398.65,147.11) and (398.71,146.43) .. (398.82,145.75) ;  
\draw  [draw opacity=0][dash pattern={on 5.63pt off 4.5pt}][line width=1.5]  (398.49,144.23) .. controls (399.93,132.97) and (422.67,123.99) .. (450.52,123.96) .. controls (477.85,123.93) and (500.28,132.53) .. (502.48,143.5) -- (450.54,145.32) -- cycle ; \draw  [color={rgb, 255:red, 74; green, 144; blue, 226 }  ,draw opacity=1 ][dash pattern={on 5.63pt off 4.5pt}][line width=1.5]  (398.49,144.23) .. controls (399.93,132.97) and (422.67,123.99) .. (450.52,123.96) .. controls (477.85,123.93) and (500.28,132.53) .. (502.48,143.5) ;  
\draw  [draw opacity=0][dash pattern={on 5.63pt off 4.5pt}][line width=1.5]  (349.6,148.37) .. controls (349.77,121.7) and (395.06,100.06) .. (450.91,100) .. controls (506.21,99.95) and (551.19,121.06) .. (552.3,147.36) -- (450.96,148.44) -- cycle ; \draw  [color={rgb, 255:red, 74; green, 144; blue, 226 }  ,draw opacity=1 ][dash pattern={on 5.63pt off 4.5pt}][line width=1.5]  (349.6,148.37) .. controls (349.77,121.7) and (395.06,100.06) .. (450.91,100) .. controls (506.21,99.95) and (551.19,121.06) .. (552.3,147.36) ;  
\draw  [draw opacity=0][line width=1.5]  (398.82,145.75) .. controls (398.82,145.59) and (398.82,145.44) .. (398.82,145.28) .. controls (398.85,116.72) and (422.02,93.58) .. (450.58,93.61) .. controls (478.96,93.63) and (501.98,116.51) .. (502.25,144.81) -- (450.54,145.32) -- cycle ; \draw  [color={rgb, 255:red, 80; green, 227; blue, 194 }  ,draw opacity=1 ][line width=1.5]  (398.82,145.75) .. controls (398.82,145.59) and (398.82,145.44) .. (398.82,145.28) .. controls (398.85,116.72) and (422.02,93.58) .. (450.58,93.61) .. controls (478.96,93.63) and (501.98,116.51) .. (502.25,144.81) ;  
\draw    (459.14,54.19) -- (459.14,66.18) ;
\draw [shift={(459.14,69.18)}, rotate = 270] [fill={rgb, 255:red, 0; green, 0; blue, 0 }  ][line width=0.08]  [draw opacity=0] (8.93,-4.29) -- (0,0) -- (8.93,4.29) -- cycle    ;
\draw    (459.14,84.09) -- (459.14,72.18) ;
\draw [shift={(459.14,69.18)}, rotate = 90] [fill={rgb, 255:red, 0; green, 0; blue, 0 }  ][line width=0.08]  [draw opacity=0] (8.93,-4.29) -- (0,0) -- (8.93,4.29) -- cycle    ;

\draw [color={rgb, 255:red, 208; green, 2; blue, 27 }  ,draw opacity=1 ][line width=2.25]    (398.82,145.75) -- (502.82,145.75) ;
\draw    (748.54,143.32) -- (748.54,13.45) ;
\draw [shift={(748.54,11.45)}, rotate = 90] [color={rgb, 255:red, 0; green, 0; blue, 0 }  ][line width=0.75]    (10.93,-3.29) .. controls (6.95,-1.4) and (3.31,-0.3) .. (0,0) .. controls (3.31,0.3) and (6.95,1.4) .. (10.93,3.29)   ;
\draw    (633.21,143.86) -- (868.81,143.86) ;
\draw  [draw opacity=0][line width=1.5]  (646.87,143.75) .. controls (646.87,143.58) and (646.87,143.41) .. (646.87,143.24) .. controls (646.92,87.09) and (692.48,41.61) .. (748.63,41.66) .. controls (804.78,41.71) and (850.25,87.26) .. (850.21,143.41) .. controls (850.21,143.56) and (850.2,143.71) .. (850.2,143.86) -- (748.54,143.32) -- cycle ; \draw  [color={rgb, 255:red, 80; green, 227; blue, 194 }  ,draw opacity=1 ][line width=1.5]  (646.87,143.75) .. controls (646.87,143.58) and (646.87,143.41) .. (646.87,143.24) .. controls (646.92,87.09) and (692.48,41.61) .. (748.63,41.66) .. controls (804.78,41.71) and (850.25,87.26) .. (850.21,143.41) .. controls (850.21,143.56) and (850.2,143.71) .. (850.2,143.86) ;  
\draw [color={rgb, 255:red, 208; green, 2; blue, 27 }  ,draw opacity=1 ][line width=2.25]  [dash pattern={on 6.75pt off 4.5pt}]  (748.42,41.27) -- (748.42,91.58) ;
\draw  [draw opacity=0][line width=1.5]  (850.2,143.86) .. controls (850.28,144.68) and (850.32,145.5) .. (850.32,146.33) .. controls (850.35,174.61) and (805,197.59) .. (749.02,197.64) .. controls (693.04,197.7) and (647.63,174.82) .. (647.6,146.54) .. controls (647.6,145.2) and (647.7,143.87) .. (647.9,142.56) -- (748.96,146.44) -- cycle ; \draw  [color={rgb, 255:red, 74; green, 144; blue, 226 }  ,draw opacity=1 ][line width=1.5]  (850.2,143.86) .. controls (850.28,144.68) and (850.32,145.5) .. (850.32,146.33) .. controls (850.35,174.61) and (805,197.59) .. (749.02,197.64) .. controls (693.04,197.7) and (647.63,174.82) .. (647.6,146.54) .. controls (647.6,145.2) and (647.7,143.87) .. (647.9,142.56) ;  
\draw  [draw opacity=0][dash pattern={on 5.63pt off 4.5pt}][line width=1.5]  (800.82,144.42) .. controls (800.87,144.84) and (800.89,145.26) .. (800.89,145.69) .. controls (800.91,159.35) and (777.58,170.45) .. (748.8,170.48) .. controls (720.01,170.51) and (696.66,159.46) .. (696.65,145.8) .. controls (696.65,145.11) and (696.71,144.43) .. (696.82,143.75) -- (748.77,145.74) -- cycle ; \draw  [color={rgb, 255:red, 74; green, 144; blue, 226 }  ,draw opacity=1 ][dash pattern={on 5.63pt off 4.5pt}][line width=1.5]  (800.82,144.42) .. controls (800.87,144.84) and (800.89,145.26) .. (800.89,145.69) .. controls (800.91,159.35) and (777.58,170.45) .. (748.8,170.48) .. controls (720.01,170.51) and (696.66,159.46) .. (696.65,145.8) .. controls (696.65,145.11) and (696.71,144.43) .. (696.82,143.75) ;  
\draw  [draw opacity=0][dash pattern={on 5.63pt off 4.5pt}][line width=1.5]  (696.49,142.23) .. controls (697.93,130.97) and (720.67,121.99) .. (748.52,121.96) .. controls (775.85,121.93) and (798.28,130.53) .. (800.48,141.5) -- (748.54,143.32) -- cycle ; \draw  [color={rgb, 255:red, 74; green, 144; blue, 226 }  ,draw opacity=1 ][dash pattern={on 5.63pt off 4.5pt}][line width=1.5]  (696.49,142.23) .. controls (697.93,130.97) and (720.67,121.99) .. (748.52,121.96) .. controls (775.85,121.93) and (798.28,130.53) .. (800.48,141.5) ;  
\draw  [draw opacity=0][dash pattern={on 5.63pt off 4.5pt}][line width=1.5]  (647.6,146.37) .. controls (647.77,119.7) and (693.06,98.06) .. (748.91,98) .. controls (804.21,97.95) and (849.19,119.06) .. (850.3,145.36) -- (748.96,146.44) -- cycle ; \draw  [color={rgb, 255:red, 74; green, 144; blue, 226 }  ,draw opacity=1 ][dash pattern={on 5.63pt off 4.5pt}][line width=1.5]  (647.6,146.37) .. controls (647.77,119.7) and (693.06,98.06) .. (748.91,98) .. controls (804.21,97.95) and (849.19,119.06) .. (850.3,145.36) ;  
\draw  [draw opacity=0][line width=1.5]  (696.82,143.75) .. controls (696.82,143.59) and (696.82,143.44) .. (696.82,143.28) .. controls (696.85,114.72) and (720.02,91.58) .. (748.58,91.61) .. controls (776.96,91.63) and (799.98,114.51) .. (800.25,142.81) -- (748.54,143.32) -- cycle ; \draw  [color={rgb, 255:red, 80; green, 227; blue, 194 }  ,draw opacity=1 ][line width=1.5]  (696.82,143.75) .. controls (696.82,143.59) and (696.82,143.44) .. (696.82,143.28) .. controls (696.85,114.72) and (720.02,91.58) .. (748.58,91.61) .. controls (776.96,91.63) and (799.98,114.51) .. (800.25,142.81) ;  
\draw    (757.14,52.19) -- (757.14,64.18) ;
\draw [shift={(757.14,67.18)}, rotate = 270] [fill={rgb, 255:red, 0; green, 0; blue, 0 }  ][line width=0.08]  [draw opacity=0] (8.93,-4.29) -- (0,0) -- (8.93,4.29) -- cycle    ;
\draw    (757.14,82.09) -- (757.14,70.18) ;
\draw [shift={(757.14,67.18)}, rotate = 90] [fill={rgb, 255:red, 0; green, 0; blue, 0 }  ][line width=0.08]  [draw opacity=0] (8.93,-4.29) -- (0,0) -- (8.93,4.29) -- cycle    ;

\draw (75.5,143.25) node [anchor=north west][inner sep=0.75pt]    {$x_{2}$};
\draw (193.17,142.7) node [anchor=north west][inner sep=0.75pt]    {$x_{3}$};
\draw (225.93,142.7) node [anchor=north west][inner sep=0.75pt]    {$x_{4}$};
\draw (43.44,142.7) node [anchor=north west][inner sep=0.75pt]    {$x_{1}$};
\draw (73.48,115.7) node [anchor=north west][inner sep=0.75pt]    {$\textcolor[rgb]{0.96,0.65,0.14}{A}$};
\draw (198.01,114.77) node [anchor=north west][inner sep=0.75pt]    {$\textcolor[rgb]{0.49,0.83,0.13}{B}$};
\draw (156.81,115.7) node [anchor=north west][inner sep=0.75pt]    {$\textcolor[rgb]{0.82,0.01,0.11}{C}$};
\draw (234.55,113.83) node [anchor=north west][inner sep=0.75pt]    {$\tau =0$};
\draw (89.17,41.71) node [anchor=north west][inner sep=0.75pt]    {$r_{2}$};
\draw (162.17,73.53) node [anchor=north west][inner sep=0.75pt]    {$r_{1}$};
\draw (92,272) node [anchor=north west][inner sep=0.75pt]   [align=left] {\textbf{EE in CFT}};
\draw (414.97,58.47) node [anchor=north west][inner sep=0.75pt]    {$\textcolor[rgb]{0.82,0.01,0.11}{E}\textcolor[rgb]{0.82,0.01,0.11}{_{W}}$};
\draw (370.05,126.55) node [anchor=north west][inner sep=0.75pt]    {$\textcolor[rgb]{0.96,0.65,0.14}{A}$};
\draw (514.52,126.55) node [anchor=north west][inner sep=0.75pt]    {$\textcolor[rgb]{0.49,0.83,0.13}{B}$};
\draw (460.39,128.41) node [anchor=north west][inner sep=0.75pt]    {$\textcolor[rgb]{0.82,0.01,0.11}{C}$};
\draw (388,274) node [anchor=north west][inner sep=0.75pt]   [align=left] {\textbf{EWCS in AdS}};
\draw (601.39,132) node [anchor=north west][inner sep=0.75pt]   [align=left] {{\Large \textbf{=}}};
\draw (707.97,56.47) node [anchor=north west][inner sep=0.75pt]    {$\textcolor[rgb]{0.82,0.01,0.11}{L}\textcolor[rgb]{0.82,0.01,0.11}{_{AB}}$};
\draw (663,273) node [anchor=north west][inner sep=0.75pt]   [align=left] {\textbf{Open string worldsheet}};
\draw (556.16,156) node [anchor=north west][inner sep=0.75pt]   [align=left] {CFT};
\draw (555.16,33) node [anchor=north west][inner sep=0.75pt]   [align=left] {AdS};
\draw (855.16,153) node [anchor=north west][inner sep=0.75pt]   [align=left] {D-brane};
\draw (833,34) node [anchor=north west][inner sep=0.75pt]   [align=left] {\textcolor[rgb]{0.31,0.89,0.76}{open string}};

\end{tikzpicture}
\par\end{centering}
\caption{\label{fig:SU app}The left image illustrates our recent work on the
entanglement of purification. In this approach, we subtract the region
$C$ and the regions outside of $A\cup B\cup C$. The remaining regions
$A$ and $B$ are disjoint but mutually complementary, and their entanglement
entropy is well-defined. Holographically, the result $S_{\mathrm{vN}}\left(A:B\right)$
agrees with the EWCS $E_{W}\left(A:B\right)$, as shown in the middle
image. The right image is derived from the hyperbolic string vertices,
which share the same geometry as the middle image. This paper confirms
that we can use the framework presented in the left image to calculate
the open string entanglement entropy depicted in the right image.}
\end{figure}

After obtaining the entanglement entropy, and considering the string
worldsheet interactions previously discussed, this configuration yields
a cylindrical structure, agreeing with Susskind and Uglum\textquoteright s
picture, as illustrated in Fig. (\ref{fig:SU app 2}). The boundaries
of the open string cylinder correspond to D-branes compactified around
the horizon (the entangling surface). These boundary states can be
derived explicitly in string theory, providing a direct means to verify
Susskind and Uglum\textquoteright s conjecture about the open string.
Furthermore, this framework establishes the equivalence between black
hole entropy (closed string perspective) and entanglement entropy
(open string perspective), showing that it originates from the phase
transition of the entanglement wedge or the interaction of the string
worldsheet. This result is also consistent with our observations in
the realization of ER=EPR \cite{Jiang:2024xqz}.

\begin{figure}[h]
\begin{centering}

\tikzset{every picture/.style={line width=0.65pt}} 

\begin{tikzpicture}[x=0.65pt,y=0.65pt,yscale=-1,xscale=1]

\draw    (122.86,137.82) -- (122.86,22.45) ;
\draw [shift={(122.86,20.45)}, rotate = 90] [color={rgb, 255:red, 0; green, 0; blue, 0 }  ][line width=0.75]    (10.93,-3.29) .. controls (6.95,-1.4) and (3.31,-0.3) .. (0,0) .. controls (3.31,0.3) and (6.95,1.4) .. (10.93,3.29)   ;
\draw    (20.21,138.3) -- (229.9,138.3) ;
\draw  [draw opacity=0][line width=1.5]  (32.37,138.2) .. controls (32.37,138.05) and (32.37,137.9) .. (32.37,137.74) .. controls (32.41,87.77) and (72.96,47.29) .. (122.93,47.34) .. controls (172.91,47.38) and (213.38,87.93) .. (213.34,137.9) .. controls (213.34,138.03) and (213.34,138.16) .. (213.34,138.3) -- (122.86,137.82) -- cycle ; \draw  [color={rgb, 255:red, 80; green, 227; blue, 194 }  ,draw opacity=1 ][line width=1.5]  (32.37,138.2) .. controls (32.37,138.05) and (32.37,137.9) .. (32.37,137.74) .. controls (32.41,87.77) and (72.96,47.29) .. (122.93,47.34) .. controls (172.91,47.38) and (213.38,87.93) .. (213.34,137.9) .. controls (213.34,138.03) and (213.34,138.16) .. (213.34,138.3) ;  
\draw   (253.96,103.12) -- (276.39,103.12) -- (276.39,97.94) -- (291.35,108.3) -- (276.39,118.66) -- (276.39,113.48) -- (253.96,113.48) -- cycle ;
\draw    (411.4,137.48) -- (411.4,22.11) ;
\draw [shift={(411.4,20.11)}, rotate = 90] [color={rgb, 255:red, 0; green, 0; blue, 0 }  ][line width=0.75]    (10.93,-3.29) .. controls (6.95,-1.4) and (3.31,-0.3) .. (0,0) .. controls (3.31,0.3) and (6.95,1.4) .. (10.93,3.29)   ;
\draw    (308.76,137.95) -- (518.44,137.95) ;
\draw  [draw opacity=0][line width=1.5]  (320.74,138.04) .. controls (320.74,137.97) and (320.74,137.91) .. (320.74,137.84) .. controls (320.76,112.44) and (330.81,91.86) .. (343.18,91.87) .. controls (355.54,91.88) and (365.54,112.43) .. (365.55,137.78) -- (343.14,137.86) -- cycle ; \draw  [color={rgb, 255:red, 80; green, 227; blue, 194 }  ,draw opacity=1 ][line width=1.5]  (320.74,138.04) .. controls (320.74,137.97) and (320.74,137.91) .. (320.74,137.84) .. controls (320.76,112.44) and (330.81,91.86) .. (343.18,91.87) .. controls (355.54,91.88) and (365.54,112.43) .. (365.55,137.78) ;  
\draw  [draw opacity=0][line width=1.5]  (457.08,138.22) .. controls (457.08,138.15) and (457.08,138.08) .. (457.08,138.01) .. controls (457.1,112.61) and (467.15,92.03) .. (479.52,92.04) .. controls (491.88,92.06) and (501.88,112.6) .. (501.88,137.95) -- (479.48,138.03) -- cycle ; \draw  [color={rgb, 255:red, 80; green, 227; blue, 194 }  ,draw opacity=1 ][line width=1.5]  (457.08,138.22) .. controls (457.08,138.15) and (457.08,138.08) .. (457.08,138.01) .. controls (457.1,112.61) and (467.15,92.03) .. (479.52,92.04) .. controls (491.88,92.06) and (501.88,112.6) .. (501.88,137.95) ;  
\draw  [draw opacity=0][line width=1.5]  (501.51,135.18) .. controls (501.58,135.91) and (501.61,136.65) .. (501.61,137.39) .. controls (501.64,162.56) and (461.27,183) .. (411.45,183.05) .. controls (361.62,183.11) and (321.21,162.74) .. (321.19,137.57) .. controls (321.19,136.38) and (321.28,135.2) .. (321.45,134.02) -- (411.4,137.48) -- cycle ; \draw  [color={rgb, 255:red, 74; green, 144; blue, 226 }  ,draw opacity=1 ][line width=1.5]  (501.51,135.18) .. controls (501.58,135.91) and (501.61,136.65) .. (501.61,137.39) .. controls (501.64,162.56) and (461.27,183) .. (411.45,183.05) .. controls (361.62,183.11) and (321.21,162.74) .. (321.19,137.57) .. controls (321.19,136.38) and (321.28,135.2) .. (321.45,134.02) ;  
\draw  [draw opacity=0][dash pattern={on 5.63pt off 4.5pt}][line width=1.5]  (364.43,140.53) .. controls (365.71,130.5) and (385.94,122.51) .. (410.73,122.48) .. controls (435.06,122.46) and (455.02,130.11) .. (456.98,139.88) -- (410.75,141.5) -- cycle ; \draw  [color={rgb, 255:red, 74; green, 144; blue, 226 }  ,draw opacity=1 ][dash pattern={on 5.63pt off 4.5pt}][line width=1.5]  (364.43,140.53) .. controls (365.71,130.5) and (385.94,122.51) .. (410.73,122.48) .. controls (435.06,122.46) and (455.02,130.11) .. (456.98,139.88) ;  
\draw  [draw opacity=0][dash pattern={on 5.63pt off 4.5pt}][line width=1.5]  (457.73,136.3) .. controls (457.77,136.67) and (457.79,137.05) .. (457.79,137.43) .. controls (457.8,149.59) and (437.04,159.47) .. (411.42,159.5) .. controls (385.8,159.52) and (365.02,149.69) .. (365.01,137.53) .. controls (365.01,136.91) and (365.06,136.31) .. (365.17,135.7) -- (411.4,137.48) -- cycle ; \draw  [color={rgb, 255:red, 74; green, 144; blue, 226 }  ,draw opacity=1 ][dash pattern={on 5.63pt off 4.5pt}][line width=1.5]  (457.73,136.3) .. controls (457.77,136.67) and (457.79,137.05) .. (457.79,137.43) .. controls (457.8,149.59) and (437.04,159.47) .. (411.42,159.5) .. controls (385.8,159.52) and (365.02,149.69) .. (365.01,137.53) .. controls (365.01,136.91) and (365.06,136.31) .. (365.17,135.7) ;  
\draw  [draw opacity=0][dash pattern={on 5.63pt off 4.5pt}][line width=1.5]  (320.54,141.44) .. controls (320.69,117.7) and (360.99,98.44) .. (410.7,98.39) .. controls (459.92,98.34) and (499.95,117.13) .. (500.94,140.54) -- (410.75,141.5) -- cycle ; \draw  [color={rgb, 255:red, 74; green, 144; blue, 226 }  ,draw opacity=1 ][dash pattern={on 5.63pt off 4.5pt}][line width=1.5]  (320.54,141.44) .. controls (320.69,117.7) and (360.99,98.44) .. (410.7,98.39) .. controls (459.92,98.34) and (499.95,117.13) .. (500.94,140.54) ;  
\draw  [draw opacity=0][line width=1.5]  (479.56,91.11) .. controls (478.93,104.2) and (448.08,114.78) .. (410.1,114.82) .. controls (372.97,114.85) and (342.63,104.81) .. (340.69,92.13) -- (410.08,90.77) -- cycle ; \draw  [color={rgb, 255:red, 80; green, 227; blue, 194 }  ,draw opacity=1 ][line width=1.5]  (479.56,91.11) .. controls (478.93,104.2) and (448.08,114.78) .. (410.1,114.82) .. controls (372.97,114.85) and (342.63,104.81) .. (340.69,92.13) ;  
\draw  [draw opacity=0][line width=1.5]  (339.13,93) .. controls (348.79,84.59) and (377.09,78.48) .. (410.49,78.45) .. controls (441.53,78.41) and (468.18,83.64) .. (479.56,91.11) -- (410.51,99.3) -- cycle ; \draw  [color={rgb, 255:red, 80; green, 227; blue, 194 }  ,draw opacity=1 ][line width=1.5]  (339.13,93) .. controls (348.79,84.59) and (377.09,78.48) .. (410.49,78.45) .. controls (441.53,78.41) and (468.18,83.64) .. (479.56,91.11) ;  
\draw  [draw opacity=0][line width=1.5]  (213.34,138.3) .. controls (213.41,139.02) and (213.44,139.76) .. (213.44,140.5) .. controls (213.47,165.67) and (173.1,186.11) .. (123.28,186.17) .. controls (73.46,186.22) and (33.05,165.86) .. (33.02,140.69) .. controls (33.02,139.49) and (33.11,138.31) .. (33.29,137.14) -- (123.23,140.59) -- cycle ; \draw  [color={rgb, 255:red, 74; green, 144; blue, 226 }  ,draw opacity=1 ][line width=1.5]  (213.34,138.3) .. controls (213.41,139.02) and (213.44,139.76) .. (213.44,140.5) .. controls (213.47,165.67) and (173.1,186.11) .. (123.28,186.17) .. controls (73.46,186.22) and (33.05,165.86) .. (33.02,140.69) .. controls (33.02,139.49) and (33.11,138.31) .. (33.29,137.14) ;  
\draw  [draw opacity=0][dash pattern={on 5.63pt off 4.5pt}][line width=1.5]  (169.39,138.79) .. controls (169.43,139.17) and (169.45,139.55) .. (169.45,139.93) .. controls (169.46,152.09) and (148.7,161.96) .. (123.08,161.99) .. controls (97.46,162.02) and (76.69,152.18) .. (76.67,140.02) .. controls (76.67,139.41) and (76.72,138.8) .. (76.83,138.2) -- (123.06,139.98) -- cycle ; \draw  [color={rgb, 255:red, 74; green, 144; blue, 226 }  ,draw opacity=1 ][dash pattern={on 5.63pt off 4.5pt}][line width=1.5]  (169.39,138.79) .. controls (169.43,139.17) and (169.45,139.55) .. (169.45,139.93) .. controls (169.46,152.09) and (148.7,161.96) .. (123.08,161.99) .. controls (97.46,162.02) and (76.69,152.18) .. (76.67,140.02) .. controls (76.67,139.41) and (76.72,138.8) .. (76.83,138.2) ;  
\draw  [draw opacity=0][dash pattern={on 5.63pt off 4.5pt}][line width=1.5]  (76.53,136.85) .. controls (77.81,126.82) and (98.05,118.83) .. (122.84,118.81) .. controls (147.16,118.78) and (167.12,126.44) .. (169.09,136.2) -- (122.86,137.82) -- cycle ; \draw  [color={rgb, 255:red, 74; green, 144; blue, 226 }  ,draw opacity=1 ][dash pattern={on 5.63pt off 4.5pt}][line width=1.5]  (76.53,136.85) .. controls (77.81,126.82) and (98.05,118.83) .. (122.84,118.81) .. controls (147.16,118.78) and (167.12,126.44) .. (169.09,136.2) ;  
\draw  [draw opacity=0][dash pattern={on 5.63pt off 4.5pt}][line width=1.5]  (33.02,140.53) .. controls (33.17,116.79) and (73.47,97.54) .. (123.19,97.49) .. controls (172.4,97.43) and (212.44,116.23) .. (213.43,139.63) -- (123.23,140.59) -- cycle ; \draw  [color={rgb, 255:red, 74; green, 144; blue, 226 }  ,draw opacity=1 ][dash pattern={on 5.63pt off 4.5pt}][line width=1.5]  (33.02,140.53) .. controls (33.17,116.79) and (73.47,97.54) .. (123.19,97.49) .. controls (172.4,97.43) and (212.44,116.23) .. (213.43,139.63) ;  
\draw  [draw opacity=0][line width=1.5]  (76.83,138.2) .. controls (76.83,138.06) and (76.83,137.92) .. (76.83,137.78) .. controls (76.85,112.36) and (97.47,91.77) .. (122.9,91.79) .. controls (148.15,91.82) and (168.64,112.17) .. (168.88,137.36) -- (122.86,137.82) -- cycle ; \draw  [color={rgb, 255:red, 80; green, 227; blue, 194 }  ,draw opacity=1 ][line width=1.5]  (76.83,138.2) .. controls (76.83,138.06) and (76.83,137.92) .. (76.83,137.78) .. controls (76.85,112.36) and (97.47,91.77) .. (122.9,91.79) .. controls (148.15,91.82) and (168.64,112.17) .. (168.88,137.36) ;  
\draw    (645.4,185.08) -- (645.4,23.08) ;
\draw [shift={(645.4,21.08)}, rotate = 90] [color={rgb, 255:red, 0; green, 0; blue, 0 }  ][line width=0.75]    (10.93,-3.29) .. controls (6.95,-1.4) and (3.31,-0.3) .. (0,0) .. controls (3.31,0.3) and (6.95,1.4) .. (10.93,3.29)   ;
\draw  [draw opacity=0][line width=1.5]  (673.35,142.09) .. controls (673.37,142.32) and (673.38,142.56) .. (673.38,142.79) .. controls (673.39,150.8) and (660.55,157.3) .. (644.7,157.32) .. controls (628.85,157.33) and (616,150.86) .. (615.99,142.85) .. controls (615.99,142.47) and (616.02,142.09) .. (616.07,141.72) -- (644.68,142.82) -- cycle ; \draw  [color={rgb, 255:red, 74; green, 144; blue, 226 }  ,draw opacity=1 ][line width=1.5]  (673.35,142.09) .. controls (673.37,142.32) and (673.38,142.56) .. (673.38,142.79) .. controls (673.39,150.8) and (660.55,157.3) .. (644.7,157.32) .. controls (628.85,157.33) and (616,150.86) .. (615.99,142.85) .. controls (615.99,142.47) and (616.02,142.09) .. (616.07,141.72) ;  
\draw  [draw opacity=0][dash pattern={on 5.63pt off 4.5pt}][line width=1.5]  (616.07,141.72) .. controls (616.12,134.17) and (628.94,128.04) .. (644.75,128.03) .. controls (660.41,128.01) and (673.14,133.99) .. (673.46,141.44) -- (644.77,141.74) -- cycle ; \draw  [color={rgb, 255:red, 74; green, 144; blue, 226 }  ,draw opacity=1 ][dash pattern={on 5.63pt off 4.5pt}][line width=1.5]  (616.07,141.72) .. controls (616.12,134.17) and (628.94,128.04) .. (644.75,128.03) .. controls (660.41,128.01) and (673.14,133.99) .. (673.46,141.44) ;  
\draw  [draw opacity=0][line width=1.5]  (674.35,65.09) .. controls (674.37,65.32) and (674.38,65.56) .. (674.38,65.79) .. controls (674.39,73.8) and (661.55,80.3) .. (645.7,80.32) .. controls (629.85,80.33) and (617,73.86) .. (616.99,65.85) .. controls (616.99,65.47) and (617.02,65.09) .. (617.07,64.72) -- (645.68,65.82) -- cycle ; \draw  [color={rgb, 255:red, 74; green, 144; blue, 226 }  ,draw opacity=1 ][line width=1.5]  (674.35,65.09) .. controls (674.37,65.32) and (674.38,65.56) .. (674.38,65.79) .. controls (674.39,73.8) and (661.55,80.3) .. (645.7,80.32) .. controls (629.85,80.33) and (617,73.86) .. (616.99,65.85) .. controls (616.99,65.47) and (617.02,65.09) .. (617.07,64.72) ;  
\draw  [draw opacity=0][line width=1.5]  (617.07,64.72) .. controls (617.12,57.17) and (629.94,51.04) .. (645.75,51.03) .. controls (661.6,51.01) and (674.46,57.14) .. (674.47,64.71) .. controls (674.47,65.48) and (674.33,66.23) .. (674.08,66.97) -- (645.77,64.74) -- cycle ; \draw  [color={rgb, 255:red, 74; green, 144; blue, 226 }  ,draw opacity=1 ][line width=1.5]  (617.07,64.72) .. controls (617.12,57.17) and (629.94,51.04) .. (645.75,51.03) .. controls (661.6,51.01) and (674.46,57.14) .. (674.47,64.71) .. controls (674.47,65.48) and (674.33,66.23) .. (674.08,66.97) ;  
\draw [color={rgb, 255:red, 80; green, 227; blue, 194 }  ,draw opacity=1 ][line width=1.5]    (616.07,141.72) .. controls (595.48,126.08) and (594.48,82.08) .. (617.07,64.72) ;
\draw [color={rgb, 255:red, 80; green, 227; blue, 194 }  ,draw opacity=1 ][line width=1.5]    (673.35,142.09) .. controls (696.48,118.08) and (691.48,79.08) .. (674.08,66.97) ;
\draw    (132.14,55.19) -- (132.14,67.18) ;
\draw [shift={(132.14,70.18)}, rotate = 270] [fill={rgb, 255:red, 0; green, 0; blue, 0 }  ][line width=0.08]  [draw opacity=0] (8.93,-4.29) -- (0,0) -- (8.93,4.29) -- cycle    ;
\draw    (132.14,85.09) -- (132.14,73.18) ;
\draw [shift={(132.14,70.18)}, rotate = 90] [fill={rgb, 255:red, 0; green, 0; blue, 0 }  ][line width=0.08]  [draw opacity=0] (8.93,-4.29) -- (0,0) -- (8.93,4.29) -- cycle    ;

\draw (550.39,82) node [anchor=north west][inner sep=0.75pt]   [align=left] {{\Large \textbf{=}}};
\draw (237.77,52.95) node [anchor=north west][inner sep=0.75pt]   [align=left] {\begin{minipage}[lt]{49.77pt}\setlength\topsep{0pt}
\begin{center}
string \\interaction
\end{center}

\end{minipage}};
\draw (667,18) node [anchor=north west][inner sep=0.75pt]   [align=left] {horizon};

\end{tikzpicture}
\par\end{centering}
\caption{\label{fig:SU app 2} The first two images are identical to figure
(\ref{fig:4 open-close}), which illustrates the interaction between
two open string disks, forming a cylinder. This cylinder is consistent
with Susskind and Uglum's open string picture (figure (\ref{fig:SU})),
where the dark blue boundaries represent the D-brane surrounding the
horizon (entangling surface).}
\end{figure}

\section{Discussion and conclusion}

In summary, we have established connections between string scattering
and the evolution of RT surfaces in the context of multipartite entanglement.
For open string field theory, we demonstrated that open string scattering
corresponds to the evolution of the entanglement wedge, and the width
of the open string scattering distance relates to the EWCS. In open-closed
string field theory, by taking a slice of open-closed string vertices
described by hyperbolic open string vertices, we observed that disk-disk
interaction corresponds to hyperbolic open string scattering. The
circumference of its waist cross section corresponds to the reflected
entropy.

Some remarks and future works are as follows:
\begin{itemize}
\item In SFT, the calculation of string scattering amplitudes using marked
Riemann surfaces requires attaching semi-infinite flat strips or cylinders
to open or closed boundaries of vertices. These structures can be
conformally mapped to the punctured disk or semi-disk. However, in
holographic entanglement entropy, hyperbolic strips and cylinders
play a crucial role, as they can be glued to the boundaries of Y-piece
or hexagon. The connections between hyperbolic string vertices and
entanglement entropy motivates us to investigate whether building
blocks of string field theory can be constructed using hyperbolic
strips and cylinders.
\item Based on the relationships between string scattering and the evolution
of an RT surface, it is conceivable to explore whether the entanglement
entropy can be derived from the off-shell string amplitude, or vice
versa.
\end{itemize}
\vspace{5mm}

\noindent {\bf Acknowledgements} 
This work is supported in part by NSFC (Grants No. 12105191, No. 12275183, and No. 12275184).

\end{document}